\pdfoutput=1
\documentclass[10pt,twocolumn,letterpaper]{article}
\usepackage[pagenumbers]{licaudit}
\usepackage{tikz}
\usetikzlibrary{arrows.meta,positioning}
\usepackage{cuted}
\usepackage{caption}
\usepackage{booktabs}
\usepackage{multirow}
\usepackage{amsmath,amssymb,amsthm}
\usepackage{graphicx}
\usepackage{xspace}
\usepackage{siunitx}
\usepackage{enumitem}
\usepackage{microtype}
\setlist{nosep,leftmargin=*}
\newcounter{remark}

\AtBeginDocument{\crefname{remark}{Remark}{Remarks}\Crefname{remark}{Remark}{Remarks}}
\newcounter{proposition}
\newenvironment{proposition}[1][]{\par\smallskip\noindent\refstepcounter{proposition}\textbf{Property~\theproposition}\if\relax\detokenize{#1}\relax\else\ (#1)\fi.\ \itshape\ignorespaces}{\par\smallskip}
\AtBeginDocument{\crefname{proposition}{Property}{Properties}\Crefname{proposition}{Property}{Properties}}
\newcommand{\sbd}{\underline{\Delta}}

\newcommand{\BD}{\mathrm{BD}}
\DeclareRobustCommand{\sref}[1]{\ref*{S-#1}}
\DeclareRobustCommand{\mref}[1]{\ref*{M-#1}}
\makeatletter
\renewcommand\paragraph{\@startsection{paragraph}{4}{\z@}{2.2ex \@plus 0.8ex \@minus .3ex}{-1em}{\normalfont\normalsize\bfseries}}
\makeatother

\usepackage{longtable}
\usepackage{array}
\newcolumntype{P}[1]{>{\raggedright\arraybackslash}p{#1}}
\usepackage{float}
\usepackage{placeins}
\usepackage{needspace}
\usepackage{etoolbox}
\usepackage{balance}
\newcounter{licbibitem}
\makeatletter
\AtBeginDocument{\let\lic@bibitem\bibitem
  \renewcommand{\bibitem}{\stepcounter{licbibitem}\ifnum\value{licbibitem}=56\balance\fi\lic@bibitem}}
\makeatother
\definecolor{linkblue}{rgb}{0.21,0.49,0.74}
\usepackage[pagebackref,breaklinks,colorlinks,allcolors=linkblue]{hyperref}
\newcommand{\licAuthorBlock}{%
Haisheng Fu$^{1}$\qquad Jie Liang$^{2}$\qquad Runyu Yang$^{3}$\qquad Feng Ding$^{3}$\\
Xin Dong$^{2,4}$\qquad Zixiao Jin$^{5}$\qquad Jingning Han$^{6}$\\[0.4em]
{\normalsize $^{1}$University of British Columbia\qquad $^{2}$Eastern Institute of Technology, Ningbo}\\
{\normalsize $^{3}$Simon Fraser University\qquad $^{4}$University of Science and Technology of China}\\
{\normalsize $^{5}$China University of Geosciences\qquad $^{6}$Google}}
\newcommand{\licPdfAuthors}{Haisheng Fu, Jie Liang, Runyu Yang, Feng Ding, Xin Dong, Zixiao Jin, Jingning Han}
\newcommand{\licCodeURL}{https://github.com/fengyurenpingsheng/lic-audit}
\newcommand{\licAcknowledgments}{This research was enabled in part by support provided by the Digital Research Alliance of Canada (alliancecan.ca). We used ChatGPT, Claude, and Codex for feedback on the presentation, for drafting and revising the text, and for assistance with code; the authors verified all sources, code, analyses, and text and take full responsibility for this paper.}
\newcommand{\licCodeLink}{\url{\licCodeURL}}

\hypersetup{pdftitle={LIC-Audit: How Comparable Are Reported Gains in Learned Image Compression?},pdfauthor={\licPdfAuthors}}
\usepackage[capitalize]{cleveref}
\usepackage{xurl}\Urlmuskip=0mu plus 0.3mu

\newcommand{\absSubMax}{1.12}

\newcommand{\anchorLocalC}{4.6}
\newcommand{\anchorLocalK}{2.0}

\newcommand{\anchorSpreadC}{3.56}
\newcommand{\anchorSpreadK}{1.50}
\newcommand{\anchorSpreadT}{1.06}

\newcommand{\claimCmicT}{\mbox{$-$21.34}}

\newcommand{\claimHpcmT}{\mbox{$-$22.20}}

\newcommand{\claimMedGapC}{0.89}
\newcommand{\claimMedGapK}{0.66}
\newcommand{\claimMedGapT}{0.93}

\newcommand{\claimPairsC}{55}
\newcommand{\claimPairsK}{91}
\newcommand{\claimPairsT}{66}
\newcommand{\claimRevC}{10}
\newcommand{\claimRevK}{4}

\newcommand{\claimRevT}{5}
\newcommand{\claimTauC}{0.64}
\newcommand{\claimTauK}{0.91}
\newcommand{\claimTauT}{0.85}

\newcommand{\convConsMed}{0.53}
\newcommand{\convMisMax}{17.4}
\newcommand{\convMisMed}{8.5}

\newcommand{\dCmicTAbs}{2.33}

\newcommand{\dGlicTAbs}{1.00}

\newcommand{\dHpcmVsDcaeC}{3.1}

\newcommand{\dirWinMax}{0.32}

\newcommand{\driftMaxAll}{0.14}

\newcommand{\effAboveHiMin}{18.8}

\newcommand{\effHypMax}{44}
\newcommand{\effHypMin}{27}

\newcommand{\frontKXX}{+2.58}

\newcommand{\frontKXXIII}{\mbox{$-$13.96}}

\newcommand{\frontKXXV}{\mbox{$-$18.95}}

\newcommand{\gapVtmTwelve}{1.33}
\newcommand{\gapVtmTwentyOne}{1.47}
\newcommand{\gapsBelowAnchorPhraseC}{all 12}
\newcommand{\gapsBelowAnchorPhraseK}{12 of the 13}

\newcommand{\nCells}{76}
\newcommand{\nClaimMethods}{49}
\newcommand{\nClaimRows}{198}

\newcommand{\nConvDoc}{6}
\newcommand{\nConvPairs}{29}

\newcommand{\nConvPartial}{4}
\newcommand{\nConvUndoc}{66}

\newcommand{\nCrossAll}{10}

\newcommand{\nCrossSmall}{5}
\newcommand{\nCurves}{353}

\newcommand{\nDrift}{4}

\newcommand{\nFamTests}{44}

\newcommand{\nGapVtmTwelve}{9}

\newcommand{\nGapVtmTwentyOne}{3}

\newcommand{\nIntChanged}{4}

\newcommand{\nMainTierA}{32}

\newcommand{\nNeededOne}{317}

\newcommand{\nNullPairs}{28}

\newcommand{\nNumericClaims}{175}

\newcommand{\nProvDerived}{30}

\newcommand{\nProvPaired}{1}

\newcommand{\nProvSemantic}{322}
\newcommand{\nPsnrRgbCurves}{306}

\newcommand{\nRefPairs}{1{,}596}

\newcommand{\nReposCloned}{77}
\newcommand{\nReposUsed}{36}

\newcommand{\nUnstatedAgg}{231}

\newcommand{\nValClaims}{14}

\newcommand{\nVerbatimRepos}{4}

\newcommand{\nullAboveHiHi}{31.0}

\newcommand{\nullBelowHiHi}{0.5}

\newcommand{\nullNonRefBoot}{7.9}
\newcommand{\nullNonRefOr}{5.3}
\newcommand{\nullR}{400}

\newcommand{\qarvDirect}{3.6}
\newcommand{\qarvFixed}{4.6}
\newcommand{\qarvMediatedAbs}{2.1}

\newcommand{\ratioCC}{1.72}
\newcommand{\ratioCCMax}{2.2}
\newcommand{\ratioCCMin}{1.1}
\newcommand{\refMaxDev}{3\times10^{-4}}

\newcommand{\sigmaCrossMax}{13.5}
\newcommand{\sigmaCrossMin}{5.5}
\newcommand{\sigmaCrossSmallMax}{9.6}

\newcommand{\sigmaHi}{10.91}

\newcommand{\sigmaHypMax}{18.8}

\newcommand{\sigmaRef}{6.35}
\newcommand{\sigmaRefHi}{6.65}
\newcommand{\sigmaRefLo}{5.49}

\newcommand{\sparseMaxT}{0.32}

\newcommand{\tabThreeAttMax}{12}
\newcommand{\tabThreeAttMin}{11}
\newcommand{\tabThreeValidMax}{9}
\newcommand{\tabThreeValidMin}{3}

\newcommand{\transDisC}{2}
\newcommand{\transDisK}{16}
\newcommand{\transDisT}{4}

\newcommand{\transMaxK}{10.4}

\newcommand{\transPairsC}{276}
\newcommand{\transPairsK}{496}
\newcommand{\transPairsT}{190}
\newcommand{\valCloseMax}{0.21}
\newcommand{\valMax}{0.86}

\newcommand{\vtmNineOneDiff}{1.50}

\newcommand{\winC}{32.4--37.4}

\newcommand{\winK}{30.5--36.6}

\newcommand{\winT}{32.5--38.1}

\newcommand{\evCloseNK}{62}
\newcommand{\evFixedFlipsK}{2}
\newcommand{\evOverlapFlipsK}{9}

\newcommand{\evCloseNT}{25}
\newcommand{\evFixedFlipsT}{0}
\newcommand{\evOverlapFlipsT}{3}

\newcommand{\evCloseNC}{33}
\newcommand{\evFixedFlipsC}{1}
\newcommand{\evOverlapFlipsC}{2}

\newcommand{\evDocPairs}{4}
\newcommand{\evRawMin}{12.1}
\newcommand{\evRawMax}{13.6}
\newcommand{\evSelHolmMin}{2.6}
\newcommand{\evSelHolmMax}{4.4}

\newcommand{\rmGpuHours}{3.0}

\newcommand{\rmRecords}{3{,}224}

\newcommand{\rmGpuHoursNew}{2.4}

\newcommand{\rmRecordsNew}{2{,}899}

\newcommand{\atEstConvMax}{2.19}
\newcommand{\atEstConvMin}{0.66}

\newcommand{\atLalicConvMax}{0.53}
\newcommand{\atLalicConvMin}{0.30}

\newcommand{\idCloseBdMax}{0.16}

\newcommand{\idGlicRateMax}{1.3}

\newcommand{\idNextDbMin}{0.034}

\newcommand{\idTcmDbMax}{0.28}
\newcommand{\idTcmRateMax}{4.2}

\newcommand{\rrCkpts}{30}

\newcommand{\rrGlicTOwnAbs}{1.43}

\newcommand{\rrGpuHours}{5.3}

\newcommand{\rrOwnKCMax}{+5.71}
\newcommand{\rrOwnKCMin}{+1.57}
\newcommand{\rrPadSpreadMax}{2.8}

\newcommand{\rrRecords}{10{,}410}

\newcommand{\rrSetups}{five}
\newcommand{\rrShiftMax}{2.43}
\newcommand{\rrShiftMed}{0.79}
\newcommand{\rrShiftMin}{0.31}

\newcommand{\rrStreams}{10{,}535}

\newcommand{\rnAdvHigh}{3}

\newcommand{\rnAdvPninety}{8}

\newcommand{\rnAdvRef}{7}

\newcommand{\rnAdvTotal}{14}
\newcommand{\rnBandKH}{36.6--37.0}

\newcommand{\rnBandReversals}{2}
\newcommand{\rnBandTL}{32.2--32.5}

\newcommand{\rnCalCovPninety}{98.2}

\newcommand{\rnCalFalseMed}{9.3}

\newcommand{\rnCalFalsePninety}{1.6}

\newcommand{\rnCalFalseThree}{12.0}

\newcommand{\rnCalFalseThreeMax}{30}
\newcommand{\rnCalGatePass}{18}

\newcommand{\rnCalM}{24}
\newcommand{\rnCalOuter}{500}
\newcommand{\rnCalPairs}{30}

\newcommand{\rnChwPninetyT}{0.41}

\newcommand{\rnCmicPninetyT}{0.65}

\newcommand{\rnCmicTOwn}{$-$1.28}
\newcommand{\rnCmicTOwnHi}{$-$0.96}
\newcommand{\rnCmicTOwnLo}{$-$1.60}
\newcommand{\rnCmicTRep}{+1.24}

\newcommand{\rnCmicTSh}{+0.95}

\newcommand{\rnDecHpcm}{0.85}
\newcommand{\rnDecMax}{1.07}
\newcommand{\rnDecMaxName}{GLIC}
\newcommand{\rnDecMin}{0.71}
\newcommand{\rnDecMinName}{SAAF and DCAE}

\newcommand{\rnIutAdvSame}{14}
\newcommand{\rnJkAgree}{88}
\newcommand{\rnJkCases}{92}

\newcommand{\rnJointBootExclC}{9}

\newcommand{\rnJointFirstFreqK}{100}

\newcommand{\rnJointMembers}{6}
\newcommand{\rnJointPairs}{15}

\newcommand{\rnKHGlic}{$-$0.37}

\newcommand{\rnLowGapMax}{1.6}
\newcommand{\rnLowGapMin}{1.1}
\newcommand{\rnLowestCMax}{31.2}
\newcommand{\rnLowestCMin}{30.8}
\newcommand{\rnLowestKMax}{29.3}
\newcommand{\rnLowestKMin}{28.9}
\newcommand{\rnLowestTMax}{31.4}
\newcommand{\rnLowestTMin}{31.0}

\newcommand{\rnMlicvTwoMaxAbs}{3.18}
\newcommand{\rnMlicvTwoMinAbs}{1.14}

\newcommand{\rnMlicvTwoPlusMaxAbs}{2.52}
\newcommand{\rnMlicvTwoPlusMinAbs}{0.62}

\newcommand{\rnPairAbstJkResolved}{23}
\newcommand{\rnPairCases}{63}
\newcommand{\rnPairGateAbst}{23}

\newcommand{\rnPairs}{21}
\newcommand{\rnRegChangedMax}{3}

\newcommand{\rnRegTotal}{30}
\newcommand{\rnRho}{0.74}

\newcommand{\rnSigHigh}{10.91}
\newcommand{\rnSigMaxV}{6.0}

\newcommand{\rnSigMedV}{3.3}
\newcommand{\rnSigMinV}{1.0}
\newcommand{\rnSigN}{45}
\newcommand{\rnSigPninety}{5.37}
\newcommand{\rnSigRef}{6.35}
\newcommand{\rnTFGlicDcae}{+0.96}

\newcommand{\rnTLGlicDcae}{$-$4.22}

\newcommand{\rnTabAbst}{22}
\newcommand{\rnTabAbstResolved}{21}

\newcommand{\rnTecFailObs}{3.6}

\newcommand{\rnTecGlicHi}{+2.44}
\newcommand{\rnTecGlicLo}{+0.42}

\newcommand{\rnTecMargin}{0.34}

\newcommand{\rgCmpBase}{Cool-chic 3.4}
\newcommand{\rgCmpEst}{$-$6.83}
\newcommand{\rgCmpHi}{$-$5.47}
\newcommand{\rgCmpLo}{$-$8.25}

\newcommand{\rgCmpTest}{Cool-chic 5.0}
\newcommand{\rgExEst}{1.57}
\newcommand{\rgExHi}{2.20}
\newcommand{\rgExLo}{0.89}
\newcommand{\rgImgOne}{227}
\newcommand{\rgImgOneRef}{317}

\newcommand{\rgLimC}{1.64}
\newcommand{\rgLimK}{2.15}

\newcommand{\rgLimT}{1.05}

\newcommand{\rgNbelow}{18}
\newcommand{\rgNbelowRes}{12}
\newcommand{\rgNcmp}{45}

\newcommand{\rgSigHi}{10.91}
\newcommand{\rgSigMed}{3.30}
\newcommand{\rgSigNinety}{5.37}
\newcommand{\rgSigRef}{6.35}

\newcommand{\cfDecAgreeMax}{2.1}

\newcommand{\cfDecSpearman}{0.94}

\newcommand{\cfDivCmic}{+1.00}
\newcommand{\cfDivCmicHi}{+1.64}
\newcommand{\cfDivCmicLo}{+0.28}
\newcommand{\cfDivFirst}{95}
\newcommand{\cfDivGlic}{+0.97}

\newcommand{\cfDivHi}{36.1}
\newcommand{\cfDivLo}{31.8}

\newcommand{\cfDivPairs}{21}
\newcommand{\cfDivPairsRes}{18}
\newcommand{\cfDivRestMax}{15.4}
\newcommand{\cfDivRestMin}{3.4}
\newcommand{\cfGpuHours}{5.8}
\newcommand{\cfGpuHoursAll}{16.5}
\newcommand{\cfIdentCompared}{6{,}105}

\newcommand{\cfIdentMaxDb}{0.005}
\newcommand{\cfIdentMaxRel}{0.12}

\newcommand{\cfIdentStream}{5{,}967}

\newcommand{\cfLalicRelMax}{2.3}
\newcommand{\cfLalicRelMin}{0.7}

\newcommand{\cfMlicBetterMax}{1.8}
\newcommand{\cfMlicBetterMin}{1.2}
\newcommand{\cfMlicDbMax}{0.12}

\newcommand{\cfMlicOwnC}{+4.96}

\newcommand{\cfMlicOwnK}{+4.85}

\newcommand{\cfMlicOwnT}{+4.39}

\newcommand{\cfMlicRateErrMax}{0.5}

\newcommand{\cfMlicRelLeadMax}{8.9}
\newcommand{\cfMlicRelLeadMin}{2.5}

\newcommand{\cfMlicShiftFiveMax}{4.5}
\newcommand{\cfMlicShiftFiveMin}{1.5}
\newcommand{\cfMlicShiftMax}{4.5}
\newcommand{\cfMlicShiftMin}{1.7}

\newcommand{\cfMsMedMs}{1.05}
\newcommand{\cfMsMedPsnr}{2.54}
\newcommand{\cfMsSame}{23}
\newcommand{\cfMsTot}{24}
\newcommand{\cfRecords}{11{,}660}
\newcommand{\cfRecordsAll}{28{,}193}

\newcommand{\cfRuns}{176}
\newcommand{\cfTcmlDbMax}{0.06}
\newcommand{\cfTcmlGainMax}{0.08}
\newcommand{\cfTcmlGainMin}{0.07}
\newcommand{\cfTcmlRateErrMax}{1.2}

\newcommand{\cfTcmlSaveMax}{3.7}
\newcommand{\cfTcmlSaveMin}{2.9}

\newcommand{\hoByCodec}{HPCM-Large 6, CMIC\_AuxT 14, GLIC 11, SAAF 15, DCAE 15, LALIC 11}
\newcommand{\hoExchHi}{91}
\newcommand{\hoExchLo}{82}

\newcommand{\hoLowCodec}{HPCM-Large}
\newcommand{\hoLowCov}{6}
\newcommand{\hoLowMedEff}{2.81}
\newcommand{\hoLowN}{15}

\newcommand{\hoN}{90}
\newcommand{\hoPairs}{45}
\newcommand{\hoPninetyCov}{80}
\newcommand{\hoPninetyFalse}{0}
\newcommand{\hoPninetyMiss}{14}

\newcommand{\hoRestMedEff}{1.73}

\newcommand{\hoThreeFalse}{4}

\newcommand{\simHolm}{9}

\newcommand{\simSim}{8}

\newcommand{\vtmNoVersionN}{23}
\newcommand{\vtmVersionsClaims}{117}
\newcommand{\vtmVersionsList}{9.1, 12.1, 17.0, 18.0, 21.0, 22.0, and 28.3}
\newcommand{\vtmVersionsN}{7}

\newcommand{\vtmVersionsYear}{2025}
\newcommand{\vtmVersionsYearN}{5}

\DeclareRobustCommand{\sref}[1]{\ref{#1}}
\DeclareRobustCommand{\mref}[1]{\ref{#1}}

\title{LIC-Audit: How Comparable Are Reported Gains\\ in Learned Image Compression?}
\author{\licAuthorBlock\\[0.5em]{\normalsize Code: \licCodeLink}}
\begin{document}
\maketitle
\begin{abstract}
Learned image compression (LIC) papers often report BD-rate improvements of only a few percent.
These improvements are measured against VTM reference software whose version keeps changing: the numeric claims we audit name \vtmVersionsN\ versions, \vtmVersionsYearN\ in \vtmVersionsYear\ papers alone.
But how comparable are these numbers across papers, and what evidence should future papers give for such small gains?
We introduce LIC-Audit, an analysis of \nCurves\ released rate--distortion curves from \nReposUsed\ repositories, and a controlled re-evaluation of eight public codec releases with \cfRecordsAll\ encode--decode records.
We identify three problems.
First, different VTM reference runs change BD-rates by up to \anchorSpreadC\ percentage points, and on CLIC~Pro published numbers order \claimRevC\ of \claimPairsC\ method pairs differently from a common computation; comparing codecs directly over a fixed quality range removes this dependence.
Second, releases use different and mostly undocumented conventions for counting bits, computing PSNR, and padding images.
Under consistent measurement, the GLIC method's \dGlicTAbs\% lead over the HPCM method on Tecnick becomes a \rrGlicTOwnAbs\% deficit, changing only the padding reverses the comparison of HPCM and CMIC, and most gaps narrow when MS-SSIM replaces PSNR.
Third, average curves hide how much results vary from image to image, so whether a small gain is statistically significant depends on the assumed variation.
An optimistic assumption calls \rnCalFalseThree\% of comparisons between equally good codecs significant on test sets as small as Kodak, instead of 5\%; a cautious one leaves \rgNbelow\ of \rgNcmp\ differences between recent codecs undecided, yet per-image results show \rgNbelowRes\ of them to be significant.
Beyond auditing past results, LIC-Audit provides a versioned evaluation protocol, a public leaderboard website with a scoring tool, source-linked benchmark data, per-image measurements, and reporting guidelines.
By fixing how BD-rates are computed and measured and what evidence a gain needs, LIC-Audit aims to help future LIC papers make comparable and verifiable claims of progress.
\end{abstract}

\section{Introduction}
\label{sec:intro}
\begin{figure*}[t]
  \centering
  \includegraphics[width=\linewidth]{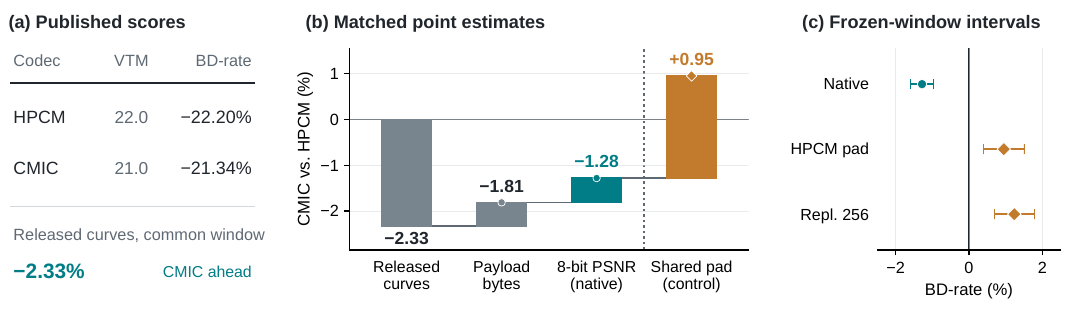}
  \caption{\textbf{Why two reported gains do not settle a comparison.}
  \textbf{a},~HPCM and CMIC report BD-rates against different VTM runs on Tecnick.
  \textbf{b},~Direct CMIC-versus-HPCM BD-rate on the frozen window: from the released curves, after switching both codecs to entropy-payload bits, then to 8-bit PSNR under native padding (the primary comparison), and with shared padding (a diagnostic intervention). The first step also absorbs small checkpoint/curve residuals.
  \textbf{c},~Paired 95\% intervals on the same window (delete-one jackknife, \cref{sec:protocol}). Negative values favor CMIC. Native padding favors CMIC, both padding controls favor HPCM, and all three intervals exclude zero.}
  \label{fig:overview}
\end{figure*}

Several recent learned image compression (LIC) methods report better rate--distortion performance than VVC~\cite{bross2021vvc} on standard test sets, and new methods typically report BD-rate~\cite{bjontegaard2001vcegm33} savings of only a few percent over the previous best.
Readers rank methods by comparing these percentages across papers.
Yet a BD-rate is a comparison under one evaluation setup, not a property of a codec.
It depends on the reference run, the quality range, how bits are counted, how PSNR is computed, and which images are tested.
Even the reference software keeps changing: the numeric claims we audit name \vtmVersionsN\ VTM versions, \vtmVersionsYearN\ in \vtmVersionsYear\ papers alone, and \vtmNoVersionN\ name VVC without any version.
When gains are this small, such choices can decide which method looks best, and every new paper that compares with published numbers inherits them.

\Cref{fig:overview} shows a concrete case.
On Tecnick~\cite{asuni2014testimages}, HPCM~\cite{li2025hpcm} reports \claimHpcmT\% against VTM-22.0 and CMIC~\cite{chen2026cmic} reports \claimCmicT\% against VTM-21.0, so HPCM appears better.
Comparing their released curves directly over a common quality range instead favors CMIC by \dCmicTAbs\%.
Measuring both public checkpoints with the same bit counting and PSNR reduces CMIC's advantage, and changing only how images are padded reverses it.
The reported numbers are not necessarily incorrect, but they do not establish a fair direct comparison.

We ask: \emph{how comparable are reported gains in LIC, and what evidence does a small gain need?}
LIC-Audit separates three questions that one percentage mixes together.
\textbf{Computation:} does an ordering survive a common reference and quality range?
\textbf{Measurement:} do two curves count the same bits and evaluate the same pixels?
\textbf{Evidence:} is a difference larger than what another choice of test images could produce?

We first trace \nCurves\ released curves and \nClaimRows\ quoted claims to pinned source files and compare codecs directly over one frozen PSNR window per dataset, which makes comparisons independent of the VTM run.
Released VTM runs alone disagree by up to \anchorSpreadC\ BD-rate points, and published numbers order \claimRevC\ of \claimPairsC\ CLIC~Pro~\cite{toderici2020clic} pairs differently from a common computation.
We then re-run eight public releases with common measurement definitions (\cfRecordsAll\ encode--decode records).
No release states all its measurement rules, and the six that we can reproduce follow \rrSetups\ different profiles.
Under common definitions, every recent release falls \rrShiftMin--\rrShiftMax\ points further behind HPCM-Large than its released curve suggests; HPCM-Large leads on Kodak~\cite{franzen1999kodak} and CLIC~Pro and has the lowest measured BD-rate on an additional DIV2K~\cite{agustsson2017div2k} set, CMIC\_AuxT leads on Tecnick, where padding alone reverses their order, and most gaps narrow under MS-SSIM.
Finally, we ask what dataset-average (aggregate) curves can show without per-image results.
Recent codec pairs vary less from image to image than older ones, but assuming a variability near their median is still too optimistic: on 24-image test sets it calls \rnCalFalseThree\% of comparisons between equally good codecs significant instead of 5\%, so aggregate-only conclusions use a cautious, calibrated assumption.

These findings matter for every future LIC paper, not only for the codecs we test.
As gains shrink, measurement choices and test-set sampling become as large as the gains themselves, so a printed BD-rate alone no longer shows progress.
LIC-Audit therefore gives future papers a reusable protocol: a new method's released curve is placed among all scored curves without new VTM runs, a stated measurement profile makes it comparable with matched measurements, and per-image records provide paired evidence.
At the calibrated variability, aggregate curves alone cannot support differences below about \rgLimK\% on Kodak, \rgLimC\% on CLIC~Pro, and \rgLimT\% on Tecnick, which tells authors when a claim needs per-image records or a larger test set.
The LIC-Audit website provides a public leaderboard and a simple tool that produces this evidence for a new method.
Our contributions are:
\begin{itemize}[leftmargin=*,topsep=1pt,itemsep=0pt,parsep=0pt]
\item a source-linked corpus of released LIC results, showing how reference runs and quality ranges change published rankings (\cref{sec:ledger,sec:findings});
\item a matched re-measurement of eight releases on four test sets that identifies undocumented measurement rules and quantifies their effects (\cref{sec:remeasure});
\item an evidence framework, calibrated on codec pairs with zero true difference, that separates measured from assumed image-to-image variability (\cref{sec:protocol,sec:progress});
\item a versioned protocol, a public leaderboard and scoring tool, and reporting guidelines that tell future papers how to compute, measure, and support a claimed gain, with test-set planning numbers (\cref{sec:discussion}).
\end{itemize}
Sections, tables, and figures whose numbers begin with ``S'' are in the supplementary material at the end of this paper.

\section{Related Work}
\label{sec:related}

\paragraph{Learned image compression.}
Nonlinear transform coding~\cite{balle2017end,theis2017lossy,balle2021ntc} has advanced through hyperpriors~\cite{balle2018variational}, autoregressive and mixture entropy models~\cite{minnen2018joint,cheng2020learned,fu2023gllmm}, parallel context models~\cite{he2022elic,jiang2025mlicpp}, and hierarchical prediction~\cite{li2025hpcm}.
Transforms have incorporated attention~\cite{liu2023tcm,li2024ftic}, wavelets~\cite{fu2024weconvene,fu2026chwdta}, and efficient global modeling~\cite{feng2025lalic,chen2026cmic,chen2026glic}.
Per-image fitting forms a separate line~\cite{ladune2023coolchic,kim2024c3,ladune2026coolchic5}.
We do not attribute gains to components; we ask whether reported gains are comparable.

\paragraph{Evaluation platforms and test conditions.}
Video coding standards compare proposals under common test conditions that fix sequences, anchors, and the BD-rate computation~\cite{bossen2020jvett2010}.
For LIC, CompressAI~\cite{begaint2020compressai}, Hu \etal~\cite{hu2022benchmark}, OpenDIC~\cite{gao2024opendic}, UI-LIC~\cite{nolen2026uilic}, the MSU benchmark~\cite{rylov2024msulic}, USTC-TD~\cite{li2026ustctd}, and JPEG~AI~\cite{ascenso2023jpegai,esenlik2025jpegai} provide platforms or common protocols that make measurements comparable for the models they run.
LIC-Audit complements them: it examines the numbers that papers already release, including methods without runnable weights, connects each number to its source, and re-runs public checkpoints to test measurement explanations.

\paragraph{BD-rate and reliable comparison.}
BD-rate~\cite{bjontegaard2001vcegm33,bjontegaard2008vcegai11} averages the rate difference between two rate--distortion curves over a quality range.
Herglotz \etal~\cite{herglotz2022beyond,herglotz2024bible}, Barman \etal~\cite{barman2024bdtutorial}, and Hang \etal~\cite{hang2024rethinkingbd} study its interpolation and sampling; Kele{\c s} \etal~\cite{keles2021psnr} and Yilmaz \etal~\cite{yilmaz2024averaging} examine averaging.
Studies in other fields show how unreported evaluation choices inflate apparent progress~\cite{musgrave2020metric,ferraridacrema2019progress,lucic2018gans,parmar2022cleanfid,bouthillier2021variance,agarwal2021precipice,theis2025clicprogress}, and versioned BLEU signatures~\cite{post2018call} show how a field can standardize a familiar score.
We combine these ideas with the paired bootstrap~\cite{koehn2004bootstrap}, power analysis~\cite{card2020power}, and multiverse analysis~\cite{steegen2016multiverse,simonsohn2020specification}, and apply them to LIC claims and checkpoint measurements.

\section{The Released Record}
\label{sec:ledger}
\begin{figure*}[t]
  \centering
  \includegraphics[width=\linewidth]{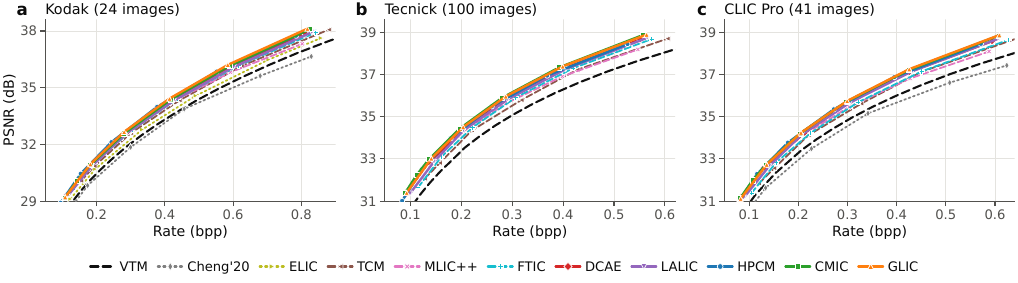}
  \caption{\textbf{The released record as papers plot it.}
  Released rate--distortion curves of representative MSE-optimized methods~\cite{cheng2020learned,he2022elic,liu2023tcm,jiang2023mlicppws,li2024ftic,lu2025dcae,feng2025lalic,li2025hpcm,chen2026cmic,chen2026glic}, read from pinned source repositories without re-measurement (sources: Tab.~\sref{tab:s_sources}); these are the curves that we score, and the PSNR axes start at 29, 31, and 31\,dB.
  ELIC uses a third-party transcription of official values.
  Each VTM curve is one released run (\textbf{a}, \textbf{b}: VTM-18.0 of lossy-vae~\cite{duan2023qresvae}; \textbf{c}: VTM-19.1 of LotteryCodec~\cite{wu2025lotterycodec}); such runs disagree by up to \anchorSpreadC\ points (\cref{sec:anchors}).}
  \label{fig:rd}
\end{figure*}

\paragraph{Scope.}
We study MSE-optimized RGB codecs evaluated by PSNR on Kodak (24 images)~\cite{franzen1999kodak}, Tecnick (100 images of $1200\times1200$ pixels)~\cite{asuni2014testimages}, and the 41-image professional validation set of CLIC~2020 (CLIC~Pro)~\cite{toderici2020clic}.
Only \nReposUsed\ of \nReposCloned\ cloned repositories contain usable numeric curves; results given only as plots or PDF tables are excluded.
This filter may favor particular groups or codec families, so our counts describe the recoverable record, not the whole field.
The scored methods include author releases of \nMainTierA\ methods and third-party transcriptions of official ELIC~\cite{he2022elic} and S2CFormer~\cite{chen2025s2cformer} values (\cref{fig:rd}).

\paragraph{Harvesting and verification.}
Parsers extract every value and its metadata; no rate--distortion point is typed by hand.
Each curve stores its repository, commit, file, dataset label, stated conventions, and provenance tier: author release (A), transcription (B), or third-party run (C).
A verifier checks that the paired (bpp, PSNR) values and the method, dataset, and variant occur at the pinned source.
Of the \nCurves\ curves, \nProvSemantic\ are verified directly, \nProvDerived\ are rebuilt from released per-image rows, and \nProvPaired\ has an unconfirmed dataset label and is excluded from rankings.
Stored hashes detect any later change of a source file (Sec.~\sref{sec:s_prov_audit}).
We also record \nClaimRows\ quoted statements from \nClaimMethods\ papers (\nNumericClaims\ numeric), with their exact text and location.

\paragraph{Traceability is not measurement compatibility.}
Knowing where a curve comes from does not tell us how it was measured.
Among \nCells\ scored method--dataset cells, only \nConvDoc\ document both bit accounting and PSNR averaging, \nConvPartial\ document one, and \nConvUndoc\ neither.
The fully documented cells yield only \evDocPairs\ within-dataset pairs, too few to compare recent methods, so \cref{sec:remeasure} returns to public weights.

\paragraph{Numerical checks.}
Our BD engine agrees with Herglotz \etal's reference implementation~\cite{herglotz2024bible} within $\refMaxDev$ points over \nRefPairs\ curve pairs.
It reproduces \nValClaims\ author-stated claims from their released anchors within \valCloseMax\ points, except one at \valMax\ (Tab.~\sref{tab:s_validation}).
These checks confirm the arithmetic, not the underlying measurement conventions.

\section{Audit Protocol}
\label{sec:protocol}

We first make the computation identical for all curves and then ask which conclusions the evidence supports.
Neither step can repair an unknown measurement convention; \cref{sec:remeasure} tests conventions directly with public checkpoints.

\paragraph{Direct BD-rate.}
A released curve lists operating points $(r_i,q_i)$: rate in bits per pixel (bpp) and PSNR.
Let $L_m(q)$ interpolate $\log_{10}r$ along codec $m$'s upper envelope with piecewise-cubic Hermite interpolation (PCHIP), as in the JVET test conditions~\cite{bossen2020jvett2010}.
Over a quality window $W=[q_0,q_1]$ covered by the curve, the average log-rate is
\begin{equation}
 D_W(m)=\frac{1}{q_1-q_0}\int_{q_0}^{q_1}L_m(q)\,dq.
 \label{eq:level}
\end{equation}
The BD-rate of codec $b$ against codec $a$ is $\BD_W(a\!\to\!b)=10^{D_W(b)-D_W(a)}-1$.
We report $\Delta_W=100\,\BD_W$ in percent; negative values mean that $b$ needs less rate.
Differences between such percentages are given in percentage points (``points'').
We compare codecs with each other directly rather than subtracting their scores against VTM.

\paragraph{One frozen window per dataset.}
For each dataset, we intersect the PSNR ranges of all fixed-rate curves with at least five released points and round inward to 0.1\,dB.
This gives $W_d$ = \winK\ (Kodak), \winT\ (Tecnick), and \winC\,dB (CLIC~Pro).
A curve that does not span $W_d$ receives no score; we never extrapolate or silently narrow the window.
For display only, a canonical score compares each curve with the average log-rate of independently released VTM curves.
Windows, references, numerical rules, and input hashes are versioned together.
A new entry must cover the existing window and cannot change an existing score; a different window defines a new benchmark version.

\begin{proposition}[Reference cancellation on a shared window]
\label{prop:fixed}
For curves spanning the same $W$ under one interpolant and point rule,
$1+\BD_W(a\!\to\!c)=(1+\BD_W(a\!\to\!b))(1+\BD_W(b\!\to\!c))$.
Changing a common reference multiplies all $1+\BD$ scores by the same positive factor, so rankings and direct comparisons do not change.
\end{proposition}
The identity follows by subtracting average log-rates in Eq.~\eqref{eq:level}.
Windows chosen separately for each pair lack this property.

\paragraph{Protocol sensitivity.}
We recompute every comparison with three interpolants and three point subsets on $W_d$, and with PCHIP on predeclared alternative windows, counting only variants valid for both codecs.
We abstain when the main comparison lacks coverage, and call a result unstable when valid variants disagree in sign.
Otherwise, $\Delta_{\rm least}$ is the valid effect closest to zero: a conservative summary of these computations, not a confidence bound.

\paragraph{Evidence from per-image records.}
When both codecs provide per-image results, a paired bootstrap resamples the same test images for both, rebuilds both curves, and recomputes the BD-rate on the unchanged $W_d$.
If more than 1\% of resampled curves no longer span the window, we use the delete-one jackknife~\cite{efron1981jackknife} with a $t_{n-1}$ interval instead.
Where both apply, the two give the same decision in \rnJkAgree\ of \rnJkCases\ comparisons.
For several releases on one dataset, simultaneous intervals over all pairs use the maximum $|t|$ over shared replicates~\cite{westfall1993resampling}.
These intervals describe how results depend on the choice of test images, not on training seeds.

\paragraph{Evidence without per-image records.}
Most papers release only dataset-average curves, which hide image-to-image variability.
For them, we use the working model $\hat\Delta\sim\mathcal N(\Delta,\sigma^2/n_d)$, where $n_d$ is the number of test images and $\sigma$ is an assumed variability in points.
A difference is resolved, \ie, statistically significant at the 5\% level, if $|\Delta_{\rm least}|>1.96\sigma/\sqrt{n_d}$, that is, if $\sigma$ is below the break-even scale
\begin{equation}
 \sigma^\star=|\Delta_{\rm least}|\sqrt{n_d}/1.96.
 \label{eq:sigmastar}
\end{equation}
\begin{proposition}[Conditional margin]
\label{prop:supported}
The part of the difference that remains beyond this uncertainty is
\begin{equation}
 \sbd(\sigma)=\operatorname{sign}(\Delta_{\rm least})
 \max\{0,|\Delta_{\rm least}|-1.96\sigma/\sqrt{n_d}\}.
 \label{eq:supported}
\end{equation}
\end{proposition}
We call it a \emph{conditional margin} because it depends on the assumed $\sigma$; a zero margin is inconclusive.
For the \nFamTests\ historical frontier tests, we also report the Holm-adjusted~\cite{holm1979simple} margin $\sbd_{\rm FW}$ (proofs: Sec.~\sref{sec:s_supported}).

\paragraph{Which variability to assume.}
For a measured pair, $\sigma_{\rm pair}=\sqrt{n_d}\operatorname{SD}_{\rm boot}(\Delta)$.
The historical reference $\sigma_{\rm ref}=\sigmaRef$ is the median over four pairs of older learned codecs (\sigmaRefLo--\sigmaRefHi); the stress value $\sigma_{\rm hi}=\sigmaHi$ scales it by \ratioCC, the median CLIC~Pro/Kodak ratio for the same codec pairs.
The \rnSigN\ comparisons among six re-measured 2025--26 releases give a recent reference: their 90th percentile is $\hat\sigma_{90}=\rnSigPninety$ (\cref{sec:split}).
All three are assumptions for an unmeasured pair, not estimates of its variability.

\section{What Released Results Show}
\label{sec:findings}

\begin{table}[t]
  \centering
  \caption{\textbf{How much each choice moves a BD-rate} (points).
  Top: median change of a method's score against one released VTM run (VTM run: largest disagreement between two runs).
  Middle: median change of a direct comparison between two methods.
  Bottom: 95\% half-width of a paired difference due to test-image sampling, observed in image bootstraps of Cool-chic versions~\cite{ladune2023coolchic} or modeled at an assumed $\sigma$ (--: no per-image results).
  Rows are sensitivities, not confidence intervals, and are not additive.}
  \label{tab:variation}
  \small
  \setlength{\tabcolsep}{5pt}
  \begin{tabular}{@{}l@{\hspace{8pt}}S[table-format=1.2]S[table-format=1.2]S[table-format=1.2]@{}}
\toprule
& {Kodak} & {Tecnick} & {CLIC~Pro} \\
\midrule
\multicolumn{4}{@{}l}{\textit{Absolute score against a released VTM run}} \\
\quad VTM run (largest pair) & 1.50 & 1.06 & 3.56 \\
\quad Pairwise vs.\ fixed window & 0.43 & 0.44 & 0.54 \\
\quad Interpolation & 0.08 & 0.68 & 0.10 \\
\quad Point subset & 1.12 & 0.89 & 0.80 \\
\addlinespace[1pt]
\multicolumn{4}{@{}l}{\textit{Direct comparison on one shared protocol}} \\
\quad Interpolation & 0.10 & 0.06 & 0.11 \\
\quad Point subset (where valid) & 0.01 & 0.00 & 0.03 \\
\quad Window rule & 0.26 & 0.20 & 0.32 \\
\addlinespace[1pt]
\multicolumn{4}{@{}l}{\textit{Test-set sampling, 95\% half-width}} \\
\quad Observed (Cool-chic versions) & 1.10 & {--} & 1.50 \\
\quad Modeled at $\sigma_{\text{ref}}$ & 2.54 & 1.24 & 1.94 \\
\quad Modeled at $\sigma_{\text{hi}}$ & 4.36 & 2.14 & 3.34 \\
\bottomrule
\end{tabular}

\end{table}

\subsection{References and windows change the ordering}
\label{sec:anchors}
\label{sec:variation}
\label{sec:multiverse}
The VTM run used as reference is the largest single source of variation.
Released VTM runs on the same images disagree by up to \anchorSpreadK, \anchorSpreadT, and \anchorSpreadC\ BD-rate points on Kodak, Tecnick, and CLIC~Pro, and even the sign of their difference changes along the rate range.
Two Kodak runs both labeled VTM-9.1 differ by \vtmNineOneDiff\ points.
These spreads exceed \gapsBelowAnchorPhraseK\ gaps between neighboring Kodak methods in \cref{tab:leaderboard} and \gapsBelowAnchorPhraseC\ on CLIC~Pro.
The CompressAI VTM-9.1 Kodak curve~\cite{begaint2020compressai} appears verbatim in \nVerbatimRepos\ repositories and matches three of \nDrift\ Kodak BD-rates labeled ``VTM-12.1'' within \driftMaxAll\ points~\cite{lu2025dcae,li2024ftic,han2024cca}, so a version label may name copied values rather than the software that was run.

\Cref{tab:variation} separates these effects.
Scores against one VTM run change by up to \anchorSpreadC\ points with the run and by medians up to \absSubMax\ with the point subset, whereas direct comparisons on one shared protocol change by medians of at most \dirWinMax\ points.
Window choice matters as well.
With fixed reference, interpolant, and point subset, order reversals among close pairs (gap below two points) fall from \evOverlapFlipsK\ to \evFixedFlipsK\ of \evCloseNK\ on Kodak, \evOverlapFlipsT\ to \evFixedFlipsT\ of \evCloseNT\ on Tecnick, and \evOverlapFlipsC\ to \evFixedFlipsC\ of \evCloseNC\ on CLIC~Pro when pairwise overlap windows are replaced by the frozen window.
With pairwise windows, ordering methods by their scores against one VTM run contradicts their direct comparison for \transDisK\ of \transPairsK\ Kodak pairs, \transDisT\ of \transPairsT\ Tecnick pairs, and \transDisC\ of \transPairsC\ CLIC~Pro pairs, by up to \transMaxK\ points.
Frozen windows remove this inconsistency (\cref{prop:fixed}), at a price: they omit the lowest rates, where our rankings do not apply.

\subsection{Averaging is often unstated and matters}
\label{sec:conventions}
How per-image values are averaged is rarely documented: only two audited groups (lossy-vae~\cite{duan2023qresvae} and Cool-chic~\cite{ladune2023coolchic}) state it, and \nUnstatedAgg\ of \nPsnrRgbCurves\ PSNR curves do not.
On released per-image records, the common alternatives (PSNR of the mean MSE, pixel-weighted bpp, or both) change a direct BD-rate by a median of \convConsMed\ points when applied to both curves, but by a median of \convMisMed\ and up to \convMisMax\ points when applied to only one (\nConvPairs\ cases), as when a baseline was averaged differently.
Re-scoring therefore aligns computations, but undocumented measurements remain provisional in \cref{tab:leaderboard}.

\begin{table}[t]
  \centering
  \caption{\textbf{Released-curve scores.}
  Fourteen methods since 2022, ordered by Kodak: fixed-window BD-rate (\%) against the consensus VTM reference, computed from each released curve with its own conventions.
  E/F: the best-fitting tested rates are estimated/written, both with unrounded PSNR; these scores are diagnostic, not a measurement-compatible ranking. Unmarked does not mean verified; HPCM-Large matches written bits and 8-bit PSNR.
  Bold: lowest score. $^\dagger$preprint; $^\ddagger$transcription. Sensitivities: Tabs.~\sref{tab:s_lb_kodak}--\sref{tab:s_lb_clic_pv}.}
  \label{tab:leaderboard}
  \small
  \setlength{\tabcolsep}{3.2pt}\renewcommand{\arraystretch}{1.04}
  \begin{tabular}{@{}lc rrr@{}}
\toprule
Method & Year & Kodak & Tecnick & CLIC~Pro \\
\midrule
HPCM~\cite{li2025hpcm} & 2025 & \textbf{-18.95} & -21.38 & \textbf{-19.26} \\
GLIC~\cite{chen2026glic} & 2026 & -18.60$^{\mathrm{E}}$ & -22.17$^{\mathrm{E}}$ & -18.91$^{\mathrm{E}}$ \\
ChWDTA$^\dagger$~\cite{fu2026chwdta} & 2026 & -17.29 & -22.99 & -18.38 \\
CMIC~\cite{chen2026cmic} & 2026 & -17.13$^{\mathrm{F}}$ & \textbf{-23.21}$^{\mathrm{E}}$ & -18.90$^{\mathrm{E}}$ \\
SAAF~\cite{ma2026saaf} & 2026 & -16.33$^{\mathrm{E}}$ & -20.63$^{\mathrm{E}}$ & -16.85$^{\mathrm{E}}$ \\
DCAE~\cite{lu2025dcae} & 2025 & -16.03$^{\mathrm{E}}$ & -20.06$^{\mathrm{E}}$ & -16.69$^{\mathrm{E}}$ \\
S2CFormer$^\dagger$$^\ddagger$~\cite{chen2025s2cformer} & 2025 & -15.23 & \multicolumn{1}{c}{--} & \multicolumn{1}{c}{--} \\
LALIC~\cite{feng2025lalic} & 2025 & -14.57$^{\mathrm{F}}$ & -18.20$^{\mathrm{F}}$ & -15.70$^{\mathrm{F}}$ \\
FTIC~\cite{li2024ftic} & 2024 & -14.06 & -15.11 & -11.20 \\
MLIC++~\cite{jiang2023mlicppws,jiang2025mlicpp} & 2023 & -13.96 & -16.41 & -13.74 \\
CCA~\cite{han2024cca} & 2024 & -13.34 & -14.50 & -12.03 \\
MLIC+~\cite{jiang2023mlic} & 2023 & -12.02 & -15.61 & -13.50 \\
TCM~\cite{liu2023tcm} & 2023 & -11.37 & -11.37 & -9.10 \\
SegPIC~\cite{liu2024segpic} & 2024 & -8.80 & \multicolumn{1}{c}{--} & -9.38 \\
\bottomrule
\end{tabular}

\end{table}

\subsection{Published and re-scored rankings disagree}
\label{sec:claims}
Author-stated BD-rates differ from our canonical scores by a median of \claimMedGapK, \claimMedGapT, and \claimMedGapC\ points on the three sets.
The reference explains part of this gap: \nGapVtmTwelve\ claims labeled VTM-12.1 are on average \gapVtmTwelve\ points stronger than our scores, consistent with the CompressAI VTM-9.1 curve, whereas \nGapVtmTwentyOne\ VTM-21.0 claims are \gapVtmTwentyOne\ points weaker.
Ordering methods by their published numbers instead of their re-scored curves reverses \claimRevK\ of \claimPairsK\ Kodak, \claimRevT\ of \claimPairsT\ Tecnick, and \claimRevC\ of \claimPairsC\ CLIC~Pro pairs (Kendall $\tau$: \claimTauK, \claimTauT, \claimTauC).
A new paper that compares with released curves on the frozen windows avoids this inconsistency without running VTM again.

\begin{figure}[t]
  \centering
  \includegraphics[width=\linewidth]{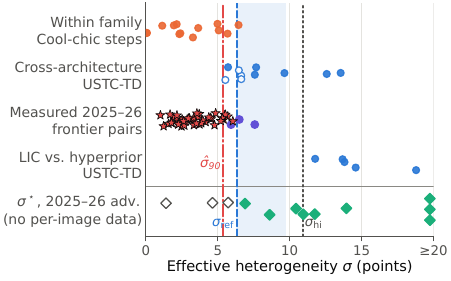}
  \caption{\textbf{Image-to-image variability.} Measured $\sigma_{\text{pair}}$ of learned-codec pairs (open circles: four calibration pairs; stars: all \rnSigN\ comparisons among six re-measured 2025--26 releases; dash-dot: their 90th percentile $\hat\sigma_{90}$) and the break-even scale $\sigma^\star$ of the released-curve advances in \cref{tab:frontier} (filled: non-zero conditional margin; values above 20 at the edge). Shaded: $\sigma_{\text{ref}}$ over usable bootstrap windows. A claim is resolved under the scale model if the true $\sigma$ lies left of its $\sigma^\star$.}
  \label{fig:sampling}
\end{figure}

\subsection{How much do test images matter?}
\label{sec:sampling}
Where per-image results exist, the uncertainty of a paired difference shrinks roughly as $1/\sqrt{n}$ with the number of images, at a pair-dependent scale $\sigma_{\text{pair}}$ (\cref{fig:sampling}).
The \nCrossAll\ cross-architecture USTC-TD pairs range from \sigmaCrossMin\ to \sigmaCrossMax: \nCrossSmall\ pairs with effects below 12\% reach \sigmaCrossSmallMax, while \effHypMin--\effHypMax\% steps over the hyperprior model reach \sigmaHypMax.
The \rnSigN\ comparisons among the six re-measured 2025--26 releases lie between \rnSigMinV\ and \rnSigMaxV, below $\sigma_{\text{ref}}$.
Four older calibration pairs thus do not justify carrying $\sigma_{\text{ref}}$ over to new architectures, and $\sigma_{\text{hi}}$ is a stress value, not a universal bound.
At $\sigma_{\text{ref}}$, detecting a one-point difference with 80\% power would need about \nNeededOne\ images (half-widths: \cref{tab:variation}).

\paragraph{Null controls.}
Dividing a codec's rates by $1+\beta$, where $\beta$ is its full-sample BD-rate, creates a pair with exactly zero effect but the original variability $\sigma_0$.
On \nNullPairs\ such pairs, the $\sigma_{\text{hi}}$ rule falsely resolves \nullBelowHiHi\% of resampled test sets when $\sigma_0\le\sigma_{\text{hi}}$, but \nullAboveHiHi\% when $\sigma_0$ is larger (all such pairs differ by \effAboveHiMin\% or more); with the true $\sigma_0$ the rate is \nullNonRefOr\%, close to the nominal 5\%.
Picking the best codec on each resampled set and testing the others against it falsely resolves \evRawMin--\evRawMax\% of draws without and \evSelHolmMin--\evSelHolmMax\% with Holm adjustment (controls: Sec.~\sref{sec:s_evidence}).

\section{What Aggregate Curves Can Support}
\label{sec:progress}

\subsection{Released-curve trends}
\label{sec:trends}
Among the included Kodak curves, the best canonical score improves from \frontKXX\% in 2020 to \frontKXXIII\% in 2023 and \frontKXXV\% in 2025.
ELIC~\cite{he2022elic}, MLIC++~\cite{jiang2023mlicppws}, and HPCM~\cite{li2025hpcm} mark the large steps.
No 2026 curve beats HPCM on Kodak or CLIC~Pro, but CMIC~\cite{chen2026cmic}, GLIC~\cite{chen2026glic}, and ChWDTA~\cite{fu2026chwdta} release lower-rate curves on Tecnick.

\begin{table}[t]
 \centering
 \caption{\textbf{Conditional evidence at two assumed variabilities.}
 For each dataset, $\Delta$ is the direct BD-rate (\%) against the best method of earlier years; the next two columns give the Holm-adjusted conditional margin $\sbd_{\rm FW}$ at the measured $\hat\sigma_{90}=\rnSigPninety$ and at the historical $\sigma_{\rm ref}=\rnSigRef$, computed from the least favorable valid effect over \nFamTests\ tests.
 Negative: rate saving; positive: rate cost; bold: non-zero saving; zero: inconclusive.
 Neither column is an observed confidence bound.
 $^\dagger$Preprint; $^\S$weights also re-measured in \cref{sec:remeasure}.}
 \label{tab:frontier}
 \scriptsize\setlength{\tabcolsep}{1.3pt}
 \resizebox{\linewidth}{!}{\begin{tabular}{@{}l rrr rrr rrr@{}}
\toprule
 & \multicolumn{3}{c}{Kodak} & \multicolumn{3}{c}{Tecnick} & \multicolumn{3}{c}{CLIC~Pro} \\
\cmidrule(lr){2-4}\cmidrule(lr){5-7}\cmidrule(lr){8-10}
Method & $\Delta$ & $5.37$ & $6.35$ & $\Delta$ & $5.37$ & $6.35$ & $\Delta$ & $5.37$ & $6.35$ \\
\midrule
FTIC~\cite{li2024ftic} '24 & $-$0.11 & 0 & 0 & +1.56 & 0 & 0 & +2.95 & 0 & 0 \\
CCA~\cite{han2024cca} '24 & +0.72 & 0 & 0 & +2.28 & +0.63 & +0.34 & +1.99 & 0 & 0 \\
HPCM$^\S$~\cite{li2025hpcm} '25 & $-$5.69 & \textbf{$-$2.18} & \textbf{$-$1.56} & $-$5.95 & \textbf{$-$4.05} & \textbf{$-$3.74} & $-$6.39 & \textbf{$-$3.69} & \textbf{$-$3.22} \\
DCAE$^\S$~\cite{lu2025dcae} '25 & $-$2.29 & 0 & 0 & $-$4.36 & \textbf{$-$2.50} & \textbf{$-$2.19} & $-$3.42 & \textbf{$-$0.79} & \textbf{$-$0.33} \\
LALIC$^\S$~\cite{feng2025lalic} '25 & $-$0.60 & 0 & 0 & $-$2.14 & \textbf{$-$0.06} & 0 & $-$2.27 & 0 & 0 \\
GLIC$^\S$~\cite{chen2026glic} '26 & +0.43 & 0 & 0 & $-$1.00 & 0 & 0 & +0.43 & 0 & 0 \\
CMIC$^\S$~\cite{chen2026cmic} '26 & +2.24 & 0 & 0 & $-$2.33 & \textbf{$-$0.65} & \textbf{$-$0.35} & +0.45 & 0 & 0 \\
ChWDTA$^\dagger$~\cite{fu2026chwdta} '26 & +2.04 & 0 & 0 & $-$2.05 & \textbf{$-$0.41} & \textbf{$-$0.12} & +1.08 & 0 & 0 \\
SAAF$^\S$~\cite{ma2026saaf} '26 & +3.23 & 0 & 0 & +0.96 & 0 & 0 & +2.98 & 0 & 0 \\
\bottomrule
\end{tabular}
}
\end{table}

\subsection{The answer depends on the assumed variability}
\label{sec:split}
Of \rnAdvTotal\ released-curve advances among the \nFamTests\ frozen comparisons, \rnAdvPninety\ keep a non-zero conditional saving at $\hat\sigma_{90}=\rnSigPninety$, \rnAdvRef\ at $\sigma_{\rm ref}=\rnSigRef$, and \rnAdvHigh\ at $\sigma_{\rm hi}=\rnSigHigh$ (\cref{tab:frontier}); testing against every earlier method instead of the best one changes none of these advances (Sec.~\sref{sec:s_frontier}), whereas window shifts and interior-point removal can change decisions (Tab.~\sref{tab:s_interior}).
At $\hat\sigma_{90}$, the Tecnick advances of CMIC and ChWDTA keep \rnCmicPninetyT\% and \rnChwPninetyT\%, and that of GLIC none; at the stress scale, no 2026 advance keeps a margin.
This outcome reflects the assumed scale, not evidence that recent progress is absent.

Which scale is appropriate cannot be read from released curves, so we calibrated the scale rules where the answer is known.
Measured scales vary with the pair and the dataset and grow with the size of the effect (Spearman $\rho=\rnRho$ over all \rnPairCases\ matched comparisons).
For the \rnCalPairs\ frontier pairs, we drew \rnCalM-image test sets (the size of Kodak) from Tecnick and CLIC~Pro and built pairs with zero true difference as in \cref{sec:sampling}: assuming $\sigma=3$ wrongly declares a difference on \rnCalFalseThree\% of these test sets (up to \rnCalFalseThreeMax\% for one pair), the measured median on \rnCalFalseMed\%, and $\hat\sigma_{90}$ on only \rnCalFalsePninety\%, while its intervals still contain the full-set effect in \rnCalCovPninety\% of draws.
A borrowed scale must also transfer to a codec that it was not estimated from.
When each codec in turn is left out and $\hat\sigma_{90}$ is computed from the other codecs' pairs, it makes \hoPninetyFalse\ false and \hoPninetyMiss\ missed decisions in the \hoN\ comparisons of the left-out codecs, whereas $\sigma=3$ makes \hoThreeFalse\ false ones (Tab.~\sref{tab:s_holdout}).
For a new paper without per-image records, $\hat\sigma_{90}$ is therefore a conservative working assumption, not a guarantee.

\paragraph{Observed conclusions remain separate.}
These results concern aggregate curves under assumed variability.
The matched measurements in \cref{sec:remeasure} give image-resampling evidence on all frozen windows, and they, not an assumed scale, decide our conclusions about the tested checkpoints.

\section{Matched Checkpoint Evaluation}
\label{sec:remeasure}
\begin{figure*}[t]
  \centering
  \includegraphics[width=\linewidth]{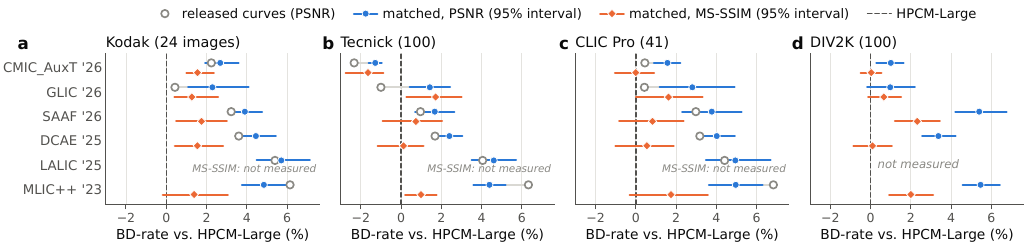}
  \caption{\textbf{Matched comparisons against HPCM-Large.}
  Direct BD-rate (\%) on the frozen windows (\textbf{a}--\textbf{c}) and the DIV2K window (\textbf{d}); positive values mean more rate.
  Open circles: released curves. Filled markers: matched measurements (payload bytes, native padding) with PSNR (blue) or MS-SSIM (orange) of the same decoded images, and paired 95\% intervals (image bootstrap, or the delete-one jackknife where more than 1\% of bootstrap replicates fail coverage).
  TCM-S, far behind on every set, is omitted; all values and the padding controls are in Tab.~\sref{tab:s_releases}.}
  \label{fig:matched}
\end{figure*}
\begin{table}[t]
  \centering
  \caption{\textbf{Measurement profiles of the eight releases}: best-fitting tested rules (no release states all of them; LALIC states written rates).
  Padding: zero or replicate, centered (c) or right/bottom (rb), to a multiple of 64--256.
  Params: author-reported, millions. Dec.: median seconds per CLIC~Pro image on one H100 MIG 3g.40gb slice.
  $^a$Bytes on Kodak. $^b$Same rates as released, higher PSNR.}
  \label{tab:releases}
  \scriptsize
  \setlength{\tabcolsep}{3pt}
  \begin{tabular}{@{}l r l l l r@{}}
\toprule
Release & Params & Bits & PSNR & Padding & Dec. \\
\midrule
HPCM-Large~\cite{li2025hpcm} '25 & 89.71 & written & 8-bit & zero c 256 & 0.85 \\
CMIC\_AuxT~\cite{chen2026cmic} '26 & -- & estimate$^a$ & unrounded & repl. rb 64 & 0.96 \\
GLIC~\cite{chen2026glic} '26 & 67.20 & estimate & unrounded & repl. rb 128 & 1.07 \\
SAAF~\cite{ma2026saaf} '26 & 123 & estimate & unrounded & zero c 128 & 0.71 \\
DCAE~\cite{lu2025dcae} '25 & 119.2 & estimate & unrounded & zero c 128 & 0.71 \\
LALIC~\cite{feng2025lalic} '25 & 63.24 & written & unrounded & zero c 128 & 0.83 \\
MLIC++~\cite{jiang2023mlicppws} '23 & -- & \multicolumn{2}{l}{none; weights beat curve$^b$} & zero rb 64 & 0.74 \\
TCM-S~\cite{liu2023tcm} '23 & 44.96 & \multicolumn{2}{l}{none matches} & zero c 128 & 0.51 \\
\bottomrule
\end{tabular}

\end{table}
Source verification shows where a curve comes from, not how it was measured, so we re-ran eight public releases with common definitions: the six 2025--26 releases HPCM-Large, CMIC\_AuxT, DCAE~\cite{lu2025dcae}, GLIC~\cite{chen2026glic}, LALIC~\cite{feng2025lalic}, and SAAF~\cite{ma2026saaf} (the published methods of \cref{tab:frontier} with runnable code and at least six public MSE weights), and two 2023 baselines, the small TCM~\cite{liu2023tcm} (TCM-S) and MLIC++~\cite{jiang2023mlicppws}.
All pairs of the first seven are compared on Kodak, Tecnick, and CLIC~Pro (\rnPairCases\ comparisons); statements about recent progress use the \rnJointPairs\ pairs of the six recent releases (\rgNcmp\ comparisons); MLIC++, measured last, is compared with all others.
For every weight, we wrote and decoded every bitstream and stored payload and file bytes, rate estimates, PSNR before and after 8-bit rounding, MS-SSIM~\cite{wang2003msssim}, and timings (\cfRecordsAll\ records, about \cfGpuHoursAll\ GPU-hours; Sec.~\sref{sec:s_batches}).
Matched comparisons count payload bytes over the original image area and compute 8-bit RGB PSNR on the cropped reconstruction, with each codec's native padding (the padding its released code applies); two diagnostic controls give both codecs HPCM's padding or right/bottom replicate padding to multiples of 256.
Repeated runs were bitwise identical, and a final re-encoding reproduced \cfIdentStream\ of \cfIdentCompared\ bitstreams exactly.

\paragraph{Different releases fit different measurement profiles.}
For six releases, one tested rule fits the released points best (\cref{tab:releases}): the other PSNR rounding misses by at least \idNextDbMin\,dB, and rate definitions separate clearly except for DCAE's three-decimal values.
HPCM-Large matches written bytes and 8-bit PSNR, LALIC written bytes and unrounded PSNR, and CMIC (Tecnick, CLIC~Pro), GLIC, SAAF, and DCAE estimated rates and unrounded PSNR; with padding, these form \rrSetups\ profiles.
A best fit does not uniquely identify the authors' setup: residual BD-rate differences are within \idCloseBdMax\% for DCAE, LALIC, and SAAF, and GLIC differs by up to \idGlicRateMax\% in rate at two points.
No tested rule reproduces TCM-S (up to \idTcmRateMax\% in rate and \idTcmDbMax\,dB) or MLIC++, whose public weights match the released rates within \cfMlicRateErrMax\% but exceed their PSNR by up to \cfMlicDbMax\,dB, outperforming the published curves by \cfMlicBetterMin--\cfMlicBetterMax\%.

\paragraph{Common definitions change the comparison.}
With common definitions and native padding, every 2025--26 release needs more rate than HPCM-Large on Kodak and CLIC~Pro, from \rrOwnKCMin\% (CMIC, CLIC~Pro) to \rrOwnKCMax\% (LALIC, Kodak), all resolved, also under either shared padding (\cref{fig:matched}).
On Tecnick, HPCM-Large's highest point lies only \rnTecMargin\,dB above the window edge, so \rnTecFailObs\% of bootstrap replicates no longer span the window; the jackknife decides all \rnTabAbst\ such cells, and \rnTabAbstResolved\ are significant.
CMIC\_AuxT's BD-rate against HPCM-Large is \rnCmicTOwn\% with native padding (95\%: \rnCmicTOwnLo\ to \rnCmicTOwnHi) but \rnCmicTSh\% and \rnCmicTRep\% under the two shared paddings: padding alone reverses the comparison.
GLIC's \dGlicTAbs\% released-curve lead becomes a \rrGlicTOwnAbs\% deficit (\rnTecGlicLo\ to \rnTecGlicHi).
Every 2025--26 release falls \rrShiftMin--\rrShiftMax\ points further behind HPCM-Large than its released curve suggests (median \rrShiftMed; open vs.\ filled circles).
Their leads over MLIC++ shrink by \cfMlicShiftFiveMin--\cfMlicShiftFiveMax\ points, partly because its weights beat its released curves: MLIC++ is then not resolved from DCAE and SAAF on Kodak and CLIC~Pro and is level with LALIC on all three sets (Tab.~\sref{tab:s_cf_mlic_pairs}).
Padding alone moves a matched comparison by up to \rrPadSpreadMax\ points, so a new method should state its padding with its measurement profile.

\paragraph{All pairs and joint ordering.}
Over the \rnJointPairs\ pairs of the six 2025--26 releases, simultaneous 95\% intervals place HPCM-Large first on Kodak and CLIC~Pro and CMIC\_AuxT first on Tecnick, and every comparison with the leader excludes zero; the leader ranks first in \rnJointFirstFreqK\% of bootstrap replicates on each set.
Below the leader, orderings are only partly resolved, \eg, only \rnJointBootExclC\ of \rnJointPairs\ pairs are significant under simultaneous intervals on CLIC~Pro (Tabs.~\sref{tab:s_r19_pairs}--\sref{tab:s_r19_joint}).

\paragraph{A fourth test set and a second quality measure.}
The 100 DIV2K validation images~\cite{agustsson2017div2k} enter none of our other analyses, although some releases may have trained on DIV2K images.
Their window (\cfDivLo--\cfDivHi\,dB) was fixed before any comparison by the rule of our resampling analyses: the common range of the measured curves, each inset by one point and rounded inward to 0.1\,dB.
HPCM-Large needs the least rate and ranks first in \cfDivFirst\% of bootstrap replicates; against it, CMIC\_AuxT's BD-rate is \cfDivCmic\% (95\%: \cfDivCmicLo\ to \cfDivCmicHi), GLIC's \cfDivGlic\% (unresolved), and the others' +\cfDivRestMin\ to +\cfDivRestMax\%.
All models are MSE-optimized, so MS-SSIM on the same reconstructions is a sensitivity analysis, but it matters: comparisons with HPCM-Large keep their direction in \cfMsSame\ of \cfMsTot\ cases, yet the gaps of CMIC\_AuxT, GLIC, SAAF, and DCAE shrink from a median of \cfMsMedPsnr\ to \cfMsMedMs\ points, and HPCM-Large, CMIC\_AuxT, and DCAE are not resolved from one another on CLIC~Pro and DIV2K.
A ranking under PSNR therefore need not hold under another quality measure.
LALIC is not included in the DIV2K and MS-SSIM measurements, and FTIC~\cite{li2024ftic} is not measured (Sec.~\sref{sec:s_cf}).

\paragraph{Rankings depend on the rate regime.}
The frozen windows omit the lowest rates, where the public checkpoints end (mean PSNRs of \rnLowestKMin--\rnLowestKMax, \rnLowestTMin--\rnLowestTMax, and \rnLowestCMin--\rnLowestCMax\,dB on the three sets).
Bands next to the windows that all resampled curves reach exist only on Kodak (\rnBandKH\,dB) and Tecnick (\rnBandTL\,dB): in the first, GLIC no longer trails HPCM-Large (\rnKHGlic\%, unresolved); in the second, DCAE overtakes GLIC (\rnTLGlicDcae\%, against \rnTFGlicDcae\% on the frozen window) and SAAF (Tab.~\sref{tab:s_r19_regimes}).

\paragraph{Size and decoding time are separate axes.}
Decoding a CLIC~Pro image takes \rnDecMin\,s (\rnDecMinName) to \rnDecMax\,s (\rnDecMaxName), and \rnDecHpcm\,s for HPCM-Large, so its rate advantage does not come with the slowest decoder; implementations differ in optimization, so these times are context, not a ranking (\cref{tab:releases}).

\section{Implications for Future LIC Papers}
\label{sec:discussion}

\paragraph{Padding belongs to the codec.}
We recommend native padding for the primary comparison: the alignment a deployed codec requires costs bits, which should be counted over the original image area, and distortion should be measured on the decoded image cropped to its original size.
Shared padding is a useful diagnostic, but it changes the operating conditions and is not necessarily fairer.

\paragraph{Measurement differences are not errors.}
The inferred profiles are the best fit among tested rules, not a unique reconstruction, and their authors have not confirmed them.
The audit establishes no error in any paper: legitimate measurement choices produce different comparisons, and author clarifications can enter later benchmark versions.

\paragraph{Using LIC-Audit for a new method.}
LIC-Audit is a protocol, not a new metric: as versioned signatures did for BLEU~\cite{post2018call}, it keeps BD-rate but fixes how it is computed, measured, and supported.
Authors can adopt it in steps (Tab.~\sref{tab:s_r20_ladder}).
A released curve spanning the frozen windows is placed among all scored curves without new VTM runs; a stated profile (written bytes, 8-bit RGB PSNR, padding, averaging) makes it measurement-compatible; per-image records give paired intervals.
Differences below $1.96\sigma/\sqrt{n_d}$ cannot be shown from aggregate curves alone (\cref{eq:supported}); at $\hat\sigma_{90}$ this limit is \rgLimK\%, \rgLimT\%, and \rgLimC\% on Kodak, Tecnick, and CLIC~Pro.
Of the \rgNcmp\ comparisons among the 2025--26 releases, \rgNbelow\ fall below this limit, yet \rgNbelowRes\ of them are significant with per-image results (\simSim\ also when all pairs of a dataset are tested jointly).
Claims should also report rate regimes, several datasets, baseline sources, and complexity (checklist: Sec.~\sref{sec:s_card}).
The LIC-Audit website, linked from the code repository (\licCodeLink), turns a method's curves or per-image records into this report in the browser, offers a prompt with which an AI assistant can extract curves and their sources from a paper or repository, and adds a submitted curve to the leaderboard only after its numbers are found at the stated repository, commit, and file.

\paragraph{What this means for future research.}
\emph{Test sets:} Kodak's 24 images cannot support gains of about 2\% from aggregate curves, so claims of this size need larger sets with per-image results (detecting a one-point gain with 80\% power needs about \rgImgOne\ images at $\hat\sigma_{90}$).
\emph{Low rates:} public checkpoints rarely reach the lowest rates, where orderings change; releasing low-rate models would extend the comparable range.
\emph{Regimes and measures:} the best release differs between rate regimes and between datasets, and MS-SSIM narrows the gaps that PSNR shows, so a method can lead in one setting only.
\emph{Padding and complexity:} alignment costs bits, and the decoding times of the 2025--26 releases differ by up to $1.5\times$ on one GPU, so both belong in the design trade-off rather than in footnotes.

\paragraph{Limitations.}
The corpus is selected by numeric-release availability and covers MSE-optimized codecs; it does not represent all published methods, and perceptual metrics need their own calibrated protocol.
Availability can hide strong methods: bridging the claims of MLICv2~\cite{jiang2025mlicv2} through its authors' own MLIC++ results places it \rnMlicvTwoMinAbs--\rnMlicvTwoMaxAbs\% behind HPCM, but its test-time-refined variant \rnMlicvTwoPlusMinAbs--\rnMlicvTwoPlusMaxAbs\% ahead (Tab.~\sref{tab:s_r19_mlicv2}).
Re-measurement covers eight releases and inference only, and its rankings concern public checkpoints of different sizes, not architectures in general; ELIC has no author-released weights, TCM-L a public weight at one rate only, and FTIC is not measured.
Image resampling on small test sets does not establish population-wide or training-seed robustness, and aggregate-only margins assume a Gaussian scale model calibrated on six recent releases.

\section{Conclusion}
\label{sec:conclusion}
Reported BD-rate gains in learned image compression are reproducible numbers, but they are often not comparable claims of progress: reference runs and quality windows change rankings inferred from published percentages, and bit accounting, PSNR rounding, and padding change the measurements themselves.
Direct comparisons on a fixed range do not change when the VTM reference software changes, so they stay comparable across papers and over time.
Under common definitions, HPCM-Large has the lowest measured rate among the tested recent releases on Kodak, CLIC~Pro, and DIV2K, and CMIC\_AuxT on Tecnick, where padding alone reverses their order; orderings also change with the rate regime and the quality measure, and aggregate curves alone cannot show whether a small difference exceeds image-to-image variability.
These findings reach beyond the codecs examined here: as gains shrink, consistent measurement and adequate evidence decide whether a reported improvement is progress or a difference in evaluation.
LIC-Audit fixes the comparison range and the measurement conventions as a versioned protocol and states what evidence a gain needs; we hope that future LIC papers will use such practices to report comparable and verifiable evidence of progress rather than isolated BD-rate improvements.

\section*{Acknowledgments}
\licAcknowledgments
{
    \footnotesize
    \bibliographystyle{ieeenat_fullname}
    \bibliography{refs}

\begin{thebibliography}{71}
\providecommand{\natexlab}[1]{#1}
\providecommand{\url}[1]{\texttt{#1}}
\expandafter\ifx\csname urlstyle\endcsname\relax
  \providecommand{\doi}[1]{doi: #1}\else
  \providecommand{\doi}{doi: \begingroup \urlstyle{rm}\Url}\fi

\bibitem[Agarwal et~al.(2021)Agarwal, Schwarzer, Castro, Courville, and
  Bellemare]{agarwal2021precipice}
Rishabh Agarwal, Max Schwarzer, Pablo~Samuel Castro, Aaron Courville, and
  Marc~G. Bellemare.
\newblock Deep reinforcement learning at the edge of the statistical precipice.
\newblock In \emph{Advances in Neural Information Processing Systems
  (NeurIPS)}, 2021.

\bibitem[Agustsson and Timofte(2017)]{agustsson2017div2k}
Eirikur Agustsson and Radu Timofte.
\newblock {NTIRE} 2017 challenge on single image super-resolution: Dataset and
  study.
\newblock In \emph{Proceedings of the IEEE Conference on Computer Vision and
  Pattern Recognition Workshops (CVPRW)}, pages 126--135, 2017.

\bibitem[Ascenso et~al.(2023)Ascenso, Alshina, and Ebrahimi]{ascenso2023jpegai}
Jo{\~a}o Ascenso, Elena Alshina, and Touradj Ebrahimi.
\newblock The {JPEG AI} standard: Providing efficient human and machine visual
  data consumption.
\newblock \emph{IEEE MultiMedia}, 30\penalty0 (1):\penalty0 100--111, 2023.

\bibitem[Asuni and Giachetti(2014)]{asuni2014testimages}
Nicola Asuni and Andrea Giachetti.
\newblock {TESTIMAGES}: A large-scale archive for testing visual devices and
  basic image processing algorithms.
\newblock In \emph{Smart Tools and Apps for Graphics -- Eurographics Italian
  Chapter Conference (STAG)}, pages 63--70. The Eurographics Association, 2014.

\bibitem[Ball{\'e} et~al.(2017)Ball{\'e}, Laparra, and
  Simoncelli]{balle2017end}
Johannes Ball{\'e}, Valero Laparra, and Eero~P. Simoncelli.
\newblock End-to-end optimized image compression.
\newblock In \emph{International Conference on Learning Representations
  (ICLR)}, 2017.

\bibitem[Ball{\'e} et~al.(2018)Ball{\'e}, Minnen, Singh, Hwang, and
  Johnston]{balle2018variational}
Johannes Ball{\'e}, David Minnen, Saurabh Singh, Sung~Jin Hwang, and Nick
  Johnston.
\newblock Variational image compression with a scale hyperprior.
\newblock In \emph{International Conference on Learning Representations
  (ICLR)}, 2018.

\bibitem[Ball{\'e} et~al.(2021)Ball{\'e}, Chou, Minnen, Singh, Johnston,
  Agustsson, Hwang, and Toderici]{balle2021ntc}
Johannes Ball{\'e}, Philip~A. Chou, David Minnen, Saurabh Singh, Nick Johnston,
  Eirikur Agustsson, Sung~Jin Hwang, and George Toderici.
\newblock Nonlinear transform coding.
\newblock \emph{IEEE Journal of Selected Topics in Signal Processing}, 15,
  2021.

\bibitem[Barman et~al.(2024)Barman, Martini, and Reznik]{barman2024bdtutorial}
Nabajeet Barman, Maria~G. Martini, and Yuriy Reznik.
\newblock {Bj{\o}ntegaard} delta ({BD}): A tutorial overview of the metric,
  evolution, challenges, and recommendations.
\newblock arXiv:2401.04039, 2024.

\bibitem[B{\'e}gaint et~al.(2020)B{\'e}gaint, Racap{\'e}, Feltman, and
  Pushparaja]{begaint2020compressai}
Jean B{\'e}gaint, Fabien Racap{\'e}, Simon Feltman, and Akshay Pushparaja.
\newblock {CompressAI}: A {PyTorch} library and evaluation platform for
  end-to-end compression research.
\newblock arXiv:2011.03029, 2020.

\bibitem[Berger(1982)]{berger1982iut}
Roger~L. Berger.
\newblock Multiparameter hypothesis testing and acceptance sampling.
\newblock \emph{Technometrics}, 24\penalty0 (4):\penalty0 295--300, 1982.

\bibitem[Bj{\o}ntegaard(2001)]{bjontegaard2001vcegm33}
Gisle Bj{\o}ntegaard.
\newblock Calculation of average {PSNR} differences between {RD}-curves.
\newblock VCEG Document VCEG-M33, ITU-T SG16 Q.6 Video Coding Experts Group
  (VCEG), Austin, TX, USA, 2001.
\newblock 13th VCEG Meeting, 2--4 April 2001.

\bibitem[Bj{\o}ntegaard(2008)]{bjontegaard2008vcegai11}
Gisle Bj{\o}ntegaard.
\newblock Improvements of the {BD-PSNR} model.
\newblock VCEG Document VCEG-AI11, ITU-T SG16 Q.6 Video Coding Experts Group
  (VCEG), Berlin, Germany, 2008.

\bibitem[Bossen et~al.(2020)]{bossen2020jvett2010}
Frank Bossen et~al.
\newblock {VTM} common test conditions and software reference configurations
  for {SDR} video.
\newblock JVET Document JVET-T2010, Joint Video Experts Team (JVET) of ITU-T
  SG16 WP3 and ISO/IEC JTC1/SC29, 2020.
\newblock 20th JVET Meeting, by teleconference, October 2020.

\bibitem[Bouthillier et~al.(2021)Bouthillier, Delaunay, Bronzi, Trofimov,
  Nichyporuk, Szeto, Mohammadi~Sepahvand, Raff, Madan, Voleti, Ebrahimi~Kahou,
  Michalski, Arbel, Pal, Varoquaux, and Vincent]{bouthillier2021variance}
Xavier Bouthillier, Pierre Delaunay, Mirko Bronzi, Assya Trofimov, Brennan
  Nichyporuk, Justin Szeto, Nazanin Mohammadi~Sepahvand, Edward Raff, Kanika
  Madan, Vikram Voleti, Samira Ebrahimi~Kahou, Vincent Michalski, Tal Arbel,
  Chris Pal, Ga{\"e}l Varoquaux, and Pascal Vincent.
\newblock Accounting for variance in machine learning benchmarks.
\newblock In \emph{Proceedings of Machine Learning and Systems (MLSys)}, pages
  747--769, 2021.

\bibitem[Bross et~al.(2021)Bross, Wang, Ye, Liu, Chen, Sullivan, and
  Ohm]{bross2021vvc}
Benjamin Bross, Ye-Kui Wang, Yan Ye, Shan Liu, Jianle Chen, Gary~J. Sullivan,
  and Jens-Rainer Ohm.
\newblock Overview of the versatile video coding ({VVC}) standard and its
  applications.
\newblock \emph{IEEE Transactions on Circuits and Systems for Video
  Technology}, 31\penalty0 (10):\penalty0 3736--3764, 2021.

\bibitem[Card et~al.(2020)Card, Henderson, Khandelwal, Jia, Mahowald, and
  Jurafsky]{card2020power}
Dallas Card, Peter Henderson, Urvashi Khandelwal, Robin Jia, Kyle Mahowald, and
  Dan Jurafsky.
\newblock With little power comes great responsibility.
\newblock In \emph{Proceedings of the 2020 Conference on Empirical Methods in
  Natural Language Processing ({EMNLP})}, pages 9263--9274, 2020.

\bibitem[Chen et~al.(2025)Chen, Li, He, Feng, Wu, Wang, Song, Lu, and
  Zhang]{chen2025s2cformer}
Yunuo Chen, Qian Li, Bing He, Donghui Feng, Ronghua Wu, Qi Wang, Li Song, Guo
  Lu, and Wenjun Zhang.
\newblock {S2CFormer}: Revisiting the {RD}-latency trade-off in
  transformer-based learned image compression.
\newblock arXiv:2502.00700, 2025.

\bibitem[Chen et~al.(2026{\natexlab{a}})Chen, He, Lyu, Hu, Gu, Tian, and
  Lu]{chen2026glic}
Yunuo Chen, Bing He, Zezheng Lyu, Hongwei Hu, Qunshan Gu, Yuan Tian, and Guo
  Lu.
\newblock Adaptive learned image compression with graph neural networks.
\newblock In \emph{Proceedings of the IEEE/CVF Conference on Computer Vision
  and Pattern Recognition (CVPR)}, 2026{\natexlab{a}}.

\bibitem[Chen et~al.(2026{\natexlab{b}})Chen, Lyu, He, Hu, Wang, Tian, Song,
  Zhang, and Lu]{chen2026cmic}
Yunuo Chen, Zezheng Lyu, Bing He, Hongwei Hu, Qi Wang, Yuan Tian, Li Song,
  Wenjun Zhang, and Guo Lu.
\newblock Content-aware {Mamba} for learned image compression.
\newblock In \emph{International Conference on Learning Representations
  (ICLR)}, 2026{\natexlab{b}}.

\bibitem[Cheng et~al.(2020)Cheng, Sun, Takeuchi, and Katto]{cheng2020learned}
Zhengxue Cheng, Heming Sun, Masaru Takeuchi, and Jiro Katto.
\newblock Learned image compression with discretized {Gaussian} mixture
  likelihoods and attention modules.
\newblock In \emph{Proceedings of the IEEE/CVF Conference on Computer Vision
  and Pattern Recognition (CVPR)}, pages 7939--7948, 2020.

\bibitem[Duan et~al.(2023)Duan, Lu, Ma, and Zhu]{duan2023qresvae}
Zhihao Duan, Ming Lu, Zhan Ma, and Fengqing Zhu.
\newblock Lossy image compression with quantized hierarchical {VAEs}.
\newblock In \emph{Proceedings of the IEEE/CVF Winter Conference on
  Applications of Computer Vision (WACV)}, pages 198--207, 2023.

\bibitem[Efron and Stein(1981)]{efron1981jackknife}
Bradley Efron and Charles Stein.
\newblock The jackknife estimate of variance.
\newblock \emph{The Annals of Statistics}, 9\penalty0 (3):\penalty0 586--596,
  1981.

\bibitem[Esenlik et~al.(2025)Esenlik, Wu, Zhang, Wang, Zhang, Zhang, Ascenso,
  and Liu]{esenlik2025jpegai}
Semih Esenlik, Yaojun Wu, Zhaobin Zhang, Ye-Kui Wang, Kai Zhang, Li Zhang,
  Jo{\~a}o Ascenso, and Shan Liu.
\newblock An overview of the {JPEG AI} learning-based image coding standard.
\newblock \emph{IEEE Transactions on Circuits and Systems for Video
  Technology}, 2025.

\bibitem[Feng et~al.(2025)Feng, Cheng, Wang, Wu, Hu, Lu, and
  Song]{feng2025lalic}
Donghui Feng, Zhengxue Cheng, Shen Wang, Ronghua Wu, Hongwei Hu, Guo Lu, and Li
  Song.
\newblock Linear attention modeling for learned image compression.
\newblock In \emph{Proceedings of the IEEE/CVF Conference on Computer Vision
  and Pattern Recognition (CVPR)}, pages 7623--7632, 2025.

\bibitem[Ferrari~Dacrema et~al.(2019)Ferrari~Dacrema, Cremonesi, and
  Jannach]{ferraridacrema2019progress}
Maurizio Ferrari~Dacrema, Paolo Cremonesi, and Dietmar Jannach.
\newblock Are we really making much progress? {A} worrying analysis of recent
  neural recommendation approaches.
\newblock In \emph{Proceedings of the 13th ACM Conference on Recommender
  Systems (RecSys '19)}, Copenhagen, Denmark, 2019. ACM.

\bibitem[Franzen(1999)]{franzen1999kodak}
Rich Franzen.
\newblock {Kodak} lossless true color image suite.
\newblock \url{http://r0k.us/graphics/kodak/}, 1999.
\newblock 24 images from Kodak PhotoCD PCD0992.

\bibitem[Fu et~al.(2023)Fu, Liang, Lin, Li, Akbari, Liang, Zhang, Liu, Tu, and
  Han]{fu2023gllmm}
Haisheng Fu, Feng Liang, Jianping Lin, Bing Li, Mohammad Akbari, Jie Liang,
  Guohe Zhang, Dong Liu, Chengjie Tu, and Jingning Han.
\newblock Learned image compression with {Gaussian-Laplacian-Logistic} mixture
  model and concatenated residual modules.
\newblock \emph{IEEE Transactions on Image Processing}, 32:\penalty0
  2063--2076, 2023.

\bibitem[Fu et~al.(2024)Fu, Liang, Fang, Han, Liang, and
  Zhang]{fu2024weconvene}
Haisheng Fu, Jie Liang, Zhenman Fang, Jingning Han, Feng Liang, and Guohe
  Zhang.
\newblock {WeConvene}: Learned image compression with wavelet-domain
  convolution and entropy model.
\newblock In \emph{Computer Vision -- ECCV 2024}, pages 37--53. Springer, 2024.

\bibitem[Fu et~al.(2026)Fu, Yang, Ding, Zhu, Liang, Li, Fang, and
  Han]{fu2026chwdta}
Haisheng Fu, Runyu Yang, Feng Ding, Siyu Zhu, Jie Liang, Xiaoxiao Li, Zhenman
  Fang, and Jingning Han.
\newblock {ChWDTA}: Channel-wise wavelet-domain transformer attention and
  entropy modeling for learned image compression.
\newblock arXiv:2606.00111, 2026.

\bibitem[Gao et~al.(2024)Gao, Zheng, Zhang, Zheng, Yu, Li, Ye, and
  Zhang]{gao2024opendic}
Wei Gao, Huiming Zheng, Chenhao Zhang, Kaiyu Zheng, Zhuozhen Yu, Yuan Li, Hua
  Ye, and Yongchi Zhang.
\newblock {OpenDIC}: An open-source library and performance evaluation for
  deep-learning-based image compression.
\newblock In \emph{Proceedings of the 32nd ACM International Conference on
  Multimedia (MM '24)}, pages 11202--11205. ACM, 2024.

\bibitem[Han et~al.(2024)Han, Jiang, Li, Deng, Xu, Zhu, and Gu]{han2024cca}
Minghao Han, Shiyin Jiang, Shengxi Li, Xin Deng, Mai Xu, Ce Zhu, and Shuhang
  Gu.
\newblock Causal context adjustment loss for learned image compression.
\newblock In \emph{Advances in Neural Information Processing Systems
  (NeurIPS)}, pages 133231--133253, 2024.

\bibitem[Hang et~al.(2024)Hang, Song, Huang, Jia, Ma, and
  Gao]{hang2024rethinkingbd}
Xinyu Hang, Shenpeng Song, Zhimeng Huang, Chuanmin Jia, Siwei Ma, and Wen Gao.
\newblock Rethinking {Bj{\o}ntegaard} delta for compression efficiency
  evaluation: Are we calculating it precisely and reliably?
\newblock arXiv:2410.12220, 2024.

\bibitem[He et~al.(2022)He, Yang, Peng, Ma, Qin, and Wang]{he2022elic}
Dailan He, Ziming Yang, Weikun Peng, Rui Ma, Hongwei Qin, and Yan Wang.
\newblock {ELIC}: Efficient learned image compression with unevenly grouped
  space-channel contextual adaptive coding.
\newblock In \emph{Proceedings of the IEEE/CVF Conference on Computer Vision
  and Pattern Recognition (CVPR)}, pages 5718--5727, 2022.

\bibitem[Herglotz et~al.(2022)Herglotz, Kr{\"a}nzler, Mons, and
  Kaup]{herglotz2022beyond}
Christian Herglotz, Matthias Kr{\"a}nzler, R. Mons, and Andr{\'e} Kaup.
\newblock Beyond {Bj{\o}ntegaard}: Limits of video compression performance
  comparisons.
\newblock In \emph{Proceedings of the IEEE International Conference on Image
  Processing (ICIP)}, Bordeaux, France, 2022.

\bibitem[Herglotz et~al.(2024)Herglotz, Och, Meyer, Ramasubbu,
  Eicherm{\"u}ller, Kr{\"a}nzler, Brand, Fischer, Nguyen, Regensky, and
  Kaup]{herglotz2024bible}
Christian Herglotz, Hannah Och, Anna Meyer, Geetha Ramasubbu, Lena
  Eicherm{\"u}ller, Matthias Kr{\"a}nzler, Fabian Brand, Kristian Fischer,
  Dat~Thanh Nguyen, Andy Regensky, and Andr{\'e} Kaup.
\newblock The {Bj{\o}ntegaard} bible -- why your way of comparing video codecs
  may be wrong.
\newblock \emph{IEEE Transactions on Image Processing}, 33:\penalty0 987--1001,
  2024.

\bibitem[Holm(1979)]{holm1979simple}
Sture Holm.
\newblock A simple sequentially rejective multiple test procedure.
\newblock \emph{Scandinavian Journal of Statistics}, 6\penalty0 (2):\penalty0
  65--70, 1979.

\bibitem[Hu et~al.(2022)Hu, Yang, Ma, and Liu]{hu2022benchmark}
Yueyu Hu, Wenhan Yang, Zhan Ma, and Jiaying Liu.
\newblock Learning end-to-end lossy image compression: A benchmark.
\newblock \emph{IEEE Transactions on Pattern Analysis and Machine
  Intelligence}, 44\penalty0 (8):\penalty0 4194--4211, 2022.

\bibitem[Jiang and Wang(2023)]{jiang2023mlicppws}
Wei Jiang and Ronggang Wang.
\newblock {MLIC}++: Linear complexity multi-reference entropy modeling for
  learned image compression.
\newblock In \emph{ICML 2023 Workshop on Neural Compression (NCW)}, 2023.

\bibitem[Jiang et~al.(2023)Jiang, Yang, Zhai, Ning, Gao, and
  Wang]{jiang2023mlic}
Wei Jiang, Jiayu Yang, Yongqi Zhai, Peirong Ning, Feng Gao, and Ronggang Wang.
\newblock {MLIC}: Multi-reference entropy model for learned image compression.
\newblock In \emph{Proceedings of the 31st ACM International Conference on
  Multimedia (MM '23)}, 2023.

\bibitem[Jiang et~al.(2025{\natexlab{a}})Jiang, Yang, Zhai, Gao, and
  Wang]{jiang2025mlicpp}
Wei Jiang, Jiayu Yang, Yongqi Zhai, Feng Gao, and Ronggang Wang.
\newblock {MLIC}++: Linear complexity multi-reference entropy modeling for
  learned image compression.
\newblock \emph{ACM Transactions on Multimedia Computing, Communications and
  Applications}, 2025{\natexlab{a}}.

\bibitem[Jiang et~al.(2025{\natexlab{b}})Jiang, Zhai, Yang, Gao, and
  Wang]{jiang2025mlicv2}
Wei Jiang, Yongqi Zhai, Jiayu Yang, Feng Gao, and Ronggang Wang.
\newblock {MLICv2}: Enhanced multi-reference entropy modeling for learned image
  compression.
\newblock arXiv:2504.19119, 2025{\natexlab{b}}.

\bibitem[Kele{\c{s}} et~al.(2021)Kele{\c{s}}, Y{\i}lmaz, Tekalp, Korkmaz, and
  Do{\u{g}}an]{keles2021psnr}
Onur Kele{\c{s}}, M.~Ak{\i}n Y{\i}lmaz, A.~Murat Tekalp, Cansu Korkmaz, and
  Zafer Do{\u{g}}an.
\newblock On the computation of {PSNR} for a set of images or video.
\newblock In \emph{Proceedings of the Picture Coding Symposium (PCS)}. IEEE,
  2021.

\bibitem[Kim et~al.(2024)Kim, Bauer, Theis, Schwarz, and Dupont]{kim2024c3}
Hyunjik Kim, Matthias Bauer, Lucas Theis, Jonathan~Richard Schwarz, and Emilien
  Dupont.
\newblock {C3}: High-performance and low-complexity neural compression from a
  single image or video.
\newblock In \emph{Proceedings of the IEEE/CVF Conference on Computer Vision
  and Pattern Recognition (CVPR)}, pages 9347--9358, 2024.

\bibitem[Koehn(2004)]{koehn2004bootstrap}
Philipp Koehn.
\newblock Statistical significance tests for machine translation evaluation.
\newblock In \emph{Proceedings of the 2004 Conference on Empirical Methods in
  Natural Language Processing}, pages 388--395, 2004.

\bibitem[Ladune et~al.(2023)Ladune, Philippe, Henry, Clare, and
  Leguay]{ladune2023coolchic}
Th{\'e}o Ladune, Pierrick Philippe, F{\'e}lix Henry, Gordon Clare, and Thomas
  Leguay.
\newblock {COOL-CHIC}: Coordinate-based low complexity hierarchical image
  codec.
\newblock In \emph{Proceedings of the IEEE/CVF International Conference on
  Computer Vision (ICCV)}, pages 13515--13522, 2023.

\bibitem[Ladune et~al.(2026)Ladune, Philippe, Jaffuer, Blard, Kervadec, Henry,
  and Clare]{ladune2026coolchic5}
Th{\'e}o Ladune, Pierrick Philippe, Pierre Jaffuer, Th{\'e}ophile Blard,
  Sylvain Kervadec, F{\'e}lix Henry, and Gordon Clare.
\newblock Cool-chic 5.0: Faster encoding and inter-feature entropy modeling for
  overfitted image compression.
\newblock arXiv:2605.02726, 2026.

\bibitem[Li et~al.(2024)Li, Li, Dai, Li, Zou, and Xiong]{li2024ftic}
Han Li, Shaohui Li, Wenrui Dai, Chenglin Li, Junni Zou, and Hongkai Xiong.
\newblock Frequency-aware transformer for learned image compression.
\newblock In \emph{International Conference on Learning Representations
  (ICLR)}, 2024.

\bibitem[Li et~al.(2025)Li, Zhang, Li, and Liu]{li2025hpcm}
Yuqi Li, Haotian Zhang, Li Li, and Dong Liu.
\newblock Learned image compression with hierarchical progressive context
  modeling.
\newblock In \emph{Proceedings of the IEEE/CVF International Conference on
  Computer Vision (ICCV)}, pages 18834--18843, 2025.

\bibitem[Li et~al.(2026)Li, Liao, Tang, Zhang, Li, Bian, Sheng, Feng, Li, Gao,
  Li, Liu, and Wu]{li2026ustctd}
Zhuoyuan Li, Junqi Liao, Chuanbo Tang, Haotian Zhang, Yuqi Li, Yifan Bian,
  Xihua Sheng, Xinmin Feng, Yao Li, Changsheng Gao, Li Li, Dong Liu, and Feng
  Wu.
\newblock {USTC-TD}: A test dataset and benchmark for image and video coding in
  2020s.
\newblock \emph{IEEE Transactions on Multimedia}, 2026.

\bibitem[Liu et~al.(2023)Liu, Sun, and Katto]{liu2023tcm}
Jinming Liu, Heming Sun, and Jiro Katto.
\newblock Learned image compression with mixed {Transformer-CNN} architectures.
\newblock In \emph{Proceedings of the IEEE/CVF Conference on Computer Vision
  and Pattern Recognition (CVPR)}, pages 14388--14397, 2023.

\bibitem[Liu et~al.(2024)Liu, Yang, Bai, Wei, and Zhao]{liu2024segpic}
Yuxi Liu, Wenhan Yang, Huihui Bai, Yunchao Wei, and Yao Zhao.
\newblock Region-adaptive transform with segmentation prior for image
  compression.
\newblock In \emph{Computer Vision -- ECCV 2024}. Springer, 2024.

\bibitem[Lu et~al.(2025)Lu, Zhang, Zhou, Li, Li, and Gu]{lu2025dcae}
Jingbo Lu, Leheng Zhang, Xingyu Zhou, Mu Li, Wen Li, and Shuhang Gu.
\newblock Learned image compression with dictionary-based entropy model.
\newblock In \emph{Proceedings of the IEEE/CVF Conference on Computer Vision
  and Pattern Recognition (CVPR)}, pages 12850--12859, 2025.

\bibitem[Lu et~al.(2022)Lu, Chen, Pu, and Ma]{lu2022tinylic}
Ming Lu, Fangdong Chen, Shiliang Pu, and Zhan Ma.
\newblock High-efficiency lossy image coding through adaptive neighborhood
  information aggregation.
\newblock arXiv:2204.11448, 2022.

\bibitem[Lucic et~al.(2018)Lucic, Kurach, Michalski, Gelly, and
  Bousquet]{lucic2018gans}
Mario Lucic, Karol Kurach, Marcin Michalski, Sylvain Gelly, and Olivier
  Bousquet.
\newblock Are {GANs} created equal? {A} large-scale study.
\newblock In \emph{Advances in Neural Information Processing Systems
  (NeurIPS)}, 2018.

\bibitem[Ma et~al.(2026)Ma, Shi, Sun, Yue, Liu, Wang, and Cai]{ma2026saaf}
Huidong Ma, Xinyan Shi, Hui Sun, Xiaofei Yue, Xiaoguang Liu, Gang Wang, and
  Wentong Cai.
\newblock Learned image compression via sparse attention and adaptive
  frequency.
\newblock In \emph{Proceedings of the IEEE/CVF Conference on Computer Vision
  and Pattern Recognition (CVPR)}, 2026.

\bibitem[Minnen et~al.(2018)Minnen, Ball{\'e}, and Toderici]{minnen2018joint}
David Minnen, Johannes Ball{\'e}, and George~D. Toderici.
\newblock Joint autoregressive and hierarchical priors for learned image
  compression.
\newblock In \emph{Advances in Neural Information Processing Systems
  (NeurIPS)}, 2018.

\bibitem[Musgrave et~al.(2020)Musgrave, Belongie, and Lim]{musgrave2020metric}
Kevin Musgrave, Serge Belongie, and Ser-Nam Lim.
\newblock A metric learning reality check.
\newblock In \emph{Computer Vision -- ECCV 2020}, pages 681--699. Springer,
  2020.

\bibitem[Nolen et~al.(2026)Nolen, Trudeau, and Freeman]{nolen2026uilic}
Nicholas~J. Nolen, Luc Trudeau, and Andrew~C. Freeman.
\newblock {UI-LIC}: A unified framework for evaluating learned image
  compression models.
\newblock arXiv:2606.23545, 2026.

\bibitem[Parmar et~al.(2022)Parmar, Zhang, and Zhu]{parmar2022cleanfid}
Gaurav Parmar, Richard Zhang, and Jun-Yan Zhu.
\newblock On aliased resizing and surprising subtleties in {GAN} evaluation.
\newblock In \emph{Proceedings of the IEEE/CVF Conference on Computer Vision
  and Pattern Recognition (CVPR)}, pages 11410--11420, 2022.

\bibitem[Post(2018)]{post2018call}
Matt Post.
\newblock A call for clarity in reporting {BLEU} scores.
\newblock In \emph{Proceedings of the Third Conference on Machine Translation:
  Research Papers}, pages 186--191, Brussels, Belgium, 2018. Association for
  Computational Linguistics.

\bibitem[Rylov et~al.(2024)Rylov, Kazantsev, and Vatolin]{rylov2024msulic}
Vitaly Rylov, Roman Kazantsev, and Dmitriy Vatolin.
\newblock Learning-based image compression benchmark.
\newblock MSU Graphics \& Media Lab,
  \url{https://videoprocessing.ai/benchmarks/learning-based-image-compression.html},
  2024.

\bibitem[Simonsohn et~al.(2020)Simonsohn, Simmons, and
  Nelson]{simonsohn2020specification}
Uri Simonsohn, Joseph~P. Simmons, and Leif~D. Nelson.
\newblock Specification curve analysis.
\newblock \emph{Nature Human Behaviour}, 4\penalty0 (11):\penalty0 1208--1214,
  2020.

\bibitem[Steegen et~al.(2016)Steegen, Tuerlinckx, Gelman, and
  Vanpaemel]{steegen2016multiverse}
Sara Steegen, Francis Tuerlinckx, Andrew Gelman, and Wolf Vanpaemel.
\newblock Increasing transparency through a multiverse analysis.
\newblock \emph{Perspectives on Psychological Science}, 11\penalty0
  (5):\penalty0 702--712, 2016.

\bibitem[Theis and Acz{\'e}l(2025)]{theis2025clicprogress}
Lucas Theis and Till Acz{\'e}l.
\newblock Measuring the progress of learned image compression.
\newblock Mabyduck blog,
  \url{https://www.mabyduck.com/blog/challenge-on-learned-image-compression-2025/},
  2025.

\bibitem[Theis et~al.(2017)Theis, Shi, Cunningham, and
  Husz{\'a}r]{theis2017lossy}
Lucas Theis, Wenzhe Shi, Andrew Cunningham, and Ferenc Husz{\'a}r.
\newblock Lossy image compression with compressive autoencoders.
\newblock In \emph{International Conference on Learning Representations
  (ICLR)}, 2017.

\bibitem[Toderici et~al.(2020)Toderici, Shi, Timofte, Theis, Ball{\'e},
  Agustsson, Johnston, and Mentzer]{toderici2020clic}
George Toderici, Wenzhe Shi, Radu Timofte, Lucas Theis, Johannes Ball{\'e},
  Eirikur Agustsson, Nick Johnston, and Fabian Mentzer.
\newblock Workshop and challenge on learned image compression ({CLIC2020}).
\newblock CVPR 2020 Workshop, \url{http://www.compression.cc}, 2020.

\bibitem[Wang et~al.(2003)Wang, Simoncelli, and Bovik]{wang2003msssim}
Zhou Wang, Eero~P. Simoncelli, and Alan~C. Bovik.
\newblock Multiscale structural similarity for image quality assessment.
\newblock In \emph{Proceedings of the Thirty-Seventh Asilomar Conference on
  Signals, Systems and Computers}, pages 1398--1402, 2003.

\bibitem[Westfall and Young(1993)]{westfall1993resampling}
Peter~H. Westfall and S.~Stanley Young.
\newblock \emph{Resampling-Based Multiple Testing: Examples and Methods for
  $p$-Value Adjustment}.
\newblock Wiley, New York, 1993.

\bibitem[Wu et~al.(2025)Wu, Chen, Dragotti, and
  G{\"u}nd{\"u}z]{wu2025lotterycodec}
Haotian Wu, Gongpu Chen, Pier~Luigi Dragotti, and Deniz G{\"u}nd{\"u}z.
\newblock {LotteryCodec}: Searching the implicit representation in a random
  network for low-complexity image compression.
\newblock In \emph{Proceedings of the 42nd International Conference on Machine
  Learning (ICML)}, pages 67147--67171, 2025.

\bibitem[Yilmaz et~al.(2024)Yilmaz, Kele{\c{s}}, and
  Tekalp]{yilmaz2024averaging}
M.~Akin Yilmaz, Onur Kele{\c{s}}, and A.~Murat Tekalp.
\newblock The practice of averaging rate-distortion curves over testsets to
  compare learned video codecs can cause misleading conclusions.
\newblock arXiv:2409.08772, 2024.

\bibitem[Zou et~al.(2022)Zou, Song, and Zhang]{zou2022devil}
Renjie Zou, Chunfeng Song, and Zhaoxiang Zhang.
\newblock The devil is in the details: Window-based attention for image
  compression.
\newblock In \emph{Proceedings of the IEEE/CVF Conference on Computer Vision
  and Pattern Recognition (CVPR)}, pages 17492--17501, 2022.

\end{thebibliography}
}
\clearpage
\onecolumn
\setcounter{section}{0}\setcounter{table}{0}\setcounter{figure}{0}
\renewcommand{\thesection}{S\arabic{section}}
\renewcommand{\thetable}{S\arabic{table}}
\renewcommand{\thefigure}{S\arabic{figure}}
\renewcommand{\theHsection}{S\arabic{section}}
\renewcommand{\theHtable}{S\arabic{table}}
\renewcommand{\theHfigure}{S\arabic{figure}}
\pretocmd{\section}{\Needspace*{5\baselineskip}}{}{}
\pretocmd{\subsection}{\Needspace*{4\baselineskip}}{}{}
\renewcommand{\topfraction}{0.95}\renewcommand{\bottomfraction}{0.9}\renewcommand{\textfraction}{0.05}\renewcommand{\floatpagefraction}{0.85}
\setcounter{topnumber}{4}\setcounter{bottomnumber}{4}\setcounter{totalnumber}{8}
\raggedbottom
\makeatletter\@secpenalty=0 \setlength{\@fptop}{0pt}\setlength{\@fpsep}{12pt plus 2fil}\setlength{\@fpbot}{0pt plus 1fil}\makeatother
\setlength{\LTpre}{6pt}\setlength{\LTpost}{6pt}
\begin{center}{\Large\textbf{LIC-Audit: How Comparable Are Reported Gains in Learned Image Compression?}\\[0.5em]Supplementary Material\par}\end{center}
\vspace{0.6em}
\small
This supplement collects what is needed to check the main paper: where the data come from (\cref{sec:s_ledger}), the protocol and its proofs (\cref{sec:s_protocol}), the leaderboards, reference runs, and published claims (\cref{sec:s_anchors}), the frontier tests (\cref{sec:s_frontier}), image-to-image variability and the calibration of the evidence rules (\cref{sec:s_boot}), the matched measurement of public checkpoints (\cref{sec:s_batches}), and how a new method uses the benchmark (\cref{sec:s_r20_guide}).
All rate--distortion and statistical tables are generated by scripts from the record and the analysis outputs. The public code repository (\licCodeLink) contains the benchmark data behind these tables, the reference scoring tool, and the LIC-Audit website (\cref{sec:s_repro}).
Numbers prefixed with ``S'' refer to this supplementary material; unprefixed numbers refer to the main paper.

\section{The released record}
\label{sec:s_ledger}
\paragraph{Sources and tiers.}
\Cref{tab:s_sources} lists every repository that contributes a curve, with its pinned commit and license.
Curves are parsed from JSON, CSV/TSV, Markdown tables and code blocks, Python literals (with \texttt{ast.literal\_eval}), and arrays in README files; per-image releases (Cool-chic, USTC-TD) are averaged by script into dataset curves using the release's own convention.
Tier A curves are released by the method's authors; tier B are third-party transcriptions or plot digitizations of official values (ELIC and S2CFormer enter only this way); tier C are third-party runs and re-implementations, including every VTM anchor.
For methods with several released models we score the strongest variant (\eg, TCM-Large, HPCM-Large); the others stay in the record.
Every curve is reduced to its rate--distortion upper envelope before interpolation, which affects two dense variable-rate releases only.
Cells whose dataset label disagrees with the paper (\eg, a ``professional'' CLIC curve that the paper reports as CLIC~2021 test) are kept but not scored.

\begingroup\footnotesize
\begin{longtable}{@{}p{5.6cm}lllp{0.8cm}p{2.2cm}l@{}}
\caption{Repositories that contribute curves to the record: pinned commit (first 10 hex digits) and its date, number of curves, provenance tiers (A author-released, B third-party transcription or digitization, C third-party run or re-implementation), datasets (K Kodak, T Tecnick, C CLIC~Pro, U USTC-TD; others are kept but not scored), and the license found at the pinned commit (``none found'': no license file, header, or README statement; ``non-standard'': a free-text non-commercial note).}\label{tab:s_sources}\\
\toprule Repository & Commit & Date & Curves & Tiers & Datasets & License \\ \midrule \endfirsthead
\toprule Repository & Commit & Date & Curves & Tiers & Datasets & License \\ \midrule \endhead
\midrule \multicolumn{7}{r}{\emph{continued on next page}} \\ \endfoot
\bottomrule \endlastfoot
\texttt{baoyu2020/\allowbreak{}ShiftLIC} & \texttt{21e52e1579} & 2025-04-17 & 3 & A & K & Apache-2.0 \\
\texttt{BaylorMultimediaLab/\allowbreak{}UI-\allowbreak{}LIC} & \texttt{bbba3bc029} & 2026-08-05 & 1 & B & K & MIT \\
\texttt{CVL-\allowbreak{}UESTC/\allowbreak{}DCAE} & \texttt{e2525a0046} & 2025-10-23 & 3 & A & C,K,T & MIT \\
\texttt{Dezhao-\allowbreak{}Wang/\allowbreak{}Neural-\allowbreak{}Syntax-\allowbreak{}Code} & \texttt{ae3e928a38} & 2022-06-03 & 4 & A & C,K & none found \\
\texttt{duanzhiihao/\allowbreak{}lossy-\allowbreak{}vae} & \texttt{005c3c22e0} & 2024-08-06 & 10 & ABC & C22,K,T & non-standard \\
\texttt{eedavidwu/\allowbreak{}LotteryCodec} & \texttt{4abcc1a429} & 2025-11-03 & 26 & AC & C,K & MIT \\
\texttt{EIDOSLAB/\allowbreak{}StanH} & \texttt{c6a29b6756} & 2025-02-16 & 3 & A & C,K,T & none found \\
\texttt{EmilienDupont/\allowbreak{}coin} & \texttt{a21e50746a} & 2021-04-11 & 1 & B & K & MIT \\
\texttt{EmilienDupont/\allowbreak{}coinpp} & \texttt{55cd43107f} & 2022-07-13 & 1 & B & K & MIT \\
\texttt{EsakaK/\allowbreak{}USTCTD} & \texttt{ae99f3a49c} & 2026-04-09 & 9 & C & U & none found \\
\texttt{fengyurenpingsheng/\allowbreak{}ChWDTA-\allowbreak{}Learned-\allowbreak{}image-\allowbreak{}compression} & \texttt{2f698554f3} & 2026-06-13 & 3 & A & C,K,T & none found \\
\texttt{fengyurenpingsheng/\allowbreak{}WeConvene} & \texttt{7fae6c47e6} & 2025-08-15 & 3 & A & C,K,T & none found \\
\texttt{GityuxiLiu/\allowbreak{}SegPIC-\allowbreak{}for-\allowbreak{}Image-\allowbreak{}Compression} & \texttt{2049d42402} & 2025-03-20 & 2 & A & C,K & none found \\
\texttt{google-\allowbreak{}deepmind/\allowbreak{}c3\_neural\_compression} & \texttt{e63e751964} & 2025-09-18 & 37 & AB & C,K & Apache-2.0 \\
\texttt{Googolxx/\allowbreak{}STF} & \texttt{bbcfe3a48f} & 2023-11-11 & 4 & A & C,K & Apache-2.0 \\
\texttt{huidong-\allowbreak{}ma/\allowbreak{}SAAF} & \texttt{78103edc66} & 2026-06-01 & 3 & A & C,K,T & MIT \\
\texttt{hytu99/\allowbreak{}MSINN-\allowbreak{}VRLIC} & \texttt{ed78e39687} & 2025-03-28 & 6 & A & C,K,T & none found \\
\texttt{InterDigitalInc/\allowbreak{}CompressAI} & \texttt{3698ce739e} & 2026-10-28 & 95 & BC & C20m,C20p$^*$,K & BSD-3-Clear \\
\texttt{JiangWeibeta/\allowbreak{}ELIC} & \texttt{b240a4ca31} & 2024-09-25 & 2 & B & K & Apache-2.0 \\
\texttt{JiangWeibeta/\allowbreak{}MLIC} & \texttt{4aa9d2a512} & 2026-02-21 & 9 & A & C,K,T & Apache-2.0 \\
\texttt{jmliu206/\allowbreak{}LIC\_TCM} & \texttt{6c6d7a3648} & 2024-04-19 & 9 & A & C,K,T & MIT \\
\texttt{LabShuHangGU/\allowbreak{}CCA} & \texttt{c7088d297d} & 2025-04-04 & 3 & A & C,K,T & MIT \\
\texttt{lumingzzz/\allowbreak{}TinyLIC} & \texttt{b9081f7c4d} & 2023-12-04 & 3 & A & C,K,T & Apache-2.0 \\
\texttt{lyq133/\allowbreak{}LIC-\allowbreak{}HPCM} & \texttt{c271cfe241} & 2026-06-03 & 6 & A & C,K,T & MIT \\
\texttt{micmic123/\allowbreak{}QmapCompression} & \texttt{d719568cb7} & 2021-10-29 & 2 & A & K & none found \\
\texttt{microsoft/\allowbreak{}DCVC} & \texttt{cbdae87a54} & 2026-08-25 & 3 & A & K & MIT \\
\texttt{mmSir/\allowbreak{}GainedVAE} & \texttt{e1b9eb2918} & 2023-04-06 & 2 & C & K & none found \\
\texttt{Orange-\allowbreak{}OpenSource/\allowbreak{}Cool-\allowbreak{}Chic} & \texttt{a6fe38a414} & 2026-07-17 & 21 & A & C,K & BSD-3 \\
\texttt{qingshi9974/\allowbreak{}ICLR2024-\allowbreak{}FTIC} & \texttt{e0d488626d} & 2025-03-06 & 3 & A & C,K,T & none found \\
\texttt{sjtu-\allowbreak{}medialab/\allowbreak{}RwkvCompress} & \texttt{d2808464be} & 2025-03-31 & 7 & ABC & C,K,T & MIT \\
\texttt{tensorflow/\allowbreak{}compression} & \texttt{dda8cb4c43} & 2026-04-17 & 56 & ABC & K,T & Apache-2.0 \\
\texttt{tokkiwa/\allowbreak{}S2CFormer} & \texttt{98fd338adb} & 2026-01-27 & 2 & BC & K & Apache-2.0 \\
\texttt{UnoC-\allowbreak{}727/\allowbreak{}CMIC} & \texttt{3b01303292} & 2026-07-14 & 3 & A & C,K,T & none found \\
\texttt{UnoC-\allowbreak{}727/\allowbreak{}GLIC} & \texttt{53e6d4e3af} & 2026-06-21 & 3 & A & C,K,T & MIT \\
\texttt{xyq7/\allowbreak{}InvCompress} & \texttt{cc68eac111} & 2023-02-09 & 3 & A & C,K,T & Apache-2.0 \\
\texttt{ZhengxueCheng/\allowbreak{}Learned-\allowbreak{}Image-\allowbreak{}Compression-\allowbreak{}with-\allowbreak{}GMM-\allowbreak{}and-\allowbreak{}Attention} & \texttt{c13652f7b7} & 2020-04-15 & 2 & A & C,K & none found \\
\end{longtable}
\endgroup

\subsection{Provenance verification and licenses}
\label{sec:s_prov_audit}
A curve is verified directly (\emph{semantic} in \cref{tab:s_prov_audit}) if its full set of (bpp, PSNR) pairs occurs at exactly one paired structure of the pinned source file (two sibling arrays, a Python literal, a list of point records, or two columns of one table), with relative tolerance $10^{-9}$, and if the labels at that location, or the method's own repository, identify its dataset, method, and variant.
Curves averaged from released per-image rows are \emph{derived} and must reproduce the parser's aggregation within $10^{-6}$.
The one \emph{paired} curve, a copy of the CompressAI VTM curve whose file never names its dataset, is not used for scoring.
For every curve the frozen audit stores the locator, the identity evidence, and the SHA-256 of the source file; the tests confirm that swapped arrays, moved metadata, a wrong dataset label, a changed source hash, or a single altered value are all detected (\cref{tab:s_prov_audit}).
Thirteen source repositories carry no standard license (twelve none, one a non-commercial note); the public data give the repository, commit, and file of their curves but not the values, which can be read from the pinned originals.

\begin{table}[htbp]
\centering
\caption{Provenance audit of all \nCurves\ curves by source format and tier.}
\label{tab:s_prov_audit}
\small
\begin{tabular}{@{}lrrrrrr@{}}
\toprule
Source format & Semantic & Derived & Paired & Lexical & Failed & Total \\
\midrule
\texttt{.csv} & 5 & 0 & 0 & 0 & 0 & 5 \\
\texttt{.dat} & 6 & 0 & 0 & 0 & 0 & 6 \\
\texttt{.json} & 145 & 9 & 1 & 0 & 0 & 155 \\
\texttt{.md} & 64 & 0 & 0 & 0 & 0 & 64 \\
\texttt{.py} & 42 & 0 & 0 & 0 & 0 & 42 \\
\texttt{.tsv} & 0 & 21 & 0 & 0 & 0 & 21 \\
\texttt{.txt} & 60 & 0 & 0 & 0 & 0 & 60 \\
\midrule
Tier A (author release) & 150 & 21 & 0 & 0 & 0 & 171 \\
Tier B (transcription) & 50 & 0 & 1 & 0 & 0 & 51 \\
Tier B (digitization) & 3 & 0 & 0 & 0 & 0 & 3 \\
Tier C (third-party run) & 74 & 9 & 0 & 0 & 0 & 83 \\
Tier C (re-implementation) & 45 & 0 & 0 & 0 & 0 & 45 \\
\midrule
All curves & 322 & 30 & 1 & 0 & 0 & 353 \\
\bottomrule
\end{tabular}

\end{table}

\subsection{Numerical checks of stated claims}
\label{sec:s_validation}
\Cref{tab:s_validation} recomputes every author-stated BD-rate whose anchor curve is released, with the pairwise-overlap PCHIP computation that papers use.
The last block recomputes the \nDrift\ Kodak claims labeled ``VTM-12.1'' against the CompressAI VTM-9.1 curve.
\begingroup\footnotesize
\begin{longtable}{@{}p{10.5cm}rrr@{}}
\caption{Recomputation of author-stated BD-rates (\%) whose anchor curves are released (pairwise-overlap PCHIP, as the papers compute them).}\label{tab:s_validation}\\
\toprule Claim (anchor curve released) & Stated & Recomputed & Diff. \\ \midrule \endfirsthead
\toprule Claim (anchor curve released) & Stated & Recomputed & Diff. \\ \midrule \endhead
\midrule \multicolumn{4}{r}{\emph{continued on next page}} \\ \endfoot
\bottomrule \endlastfoot
LALIC Kodak vs released VTM-9.1 & $-$15.26 & $-$15.26 & $-$0.00 \\
LALIC CLIC Pro vs released VTM-9.1 & $-$15.41 & $-$15.41 & $-$0.00 \\
LALIC Tecnick vs released VTM-9.1 & $-$17.63 & $-$17.63 & +0.00 \\
GLIC Kodak vs released VTM-9.1 & $-$19.29 & $-$19.24 & +0.05 \\
GLIC CLIC Pro vs released VTM-9.1 & $-$18.71 & $-$18.87 & $-$0.16 \\
GLIC Tecnick vs released VTM-9.1 & $-$21.69 & $-$21.90 & $-$0.21 \\
ChWDTA Kodak vs released VTM-9.1 & $-$17.82 & $-$17.82 & +0.00 \\
ChWDTA CLIC Pro vs released VTM-9.1 & $-$19.15 & $-$18.29 & +0.86 \\
ChWDTA Tecnick vs released VTM-9.1 & $-$22.56 & $-$22.56 & $-$0.00 \\
QARV mean of Kodak/Tecnick/CLIC22 vs VTM-18.0 & $-$7.26 & $-$7.21 & +0.05 \\
QRes-VAE mean of Kodak/Tecnick/CLIC22 vs VTM-18.0 & $-$3.95 & $-$4.06 & $-$0.11 \\
MLIC CLIC Pro vs released VTM-17.0 & $-$8.79 & $-$8.79 & $-$0.00 \\
MLIC+ CLIC Pro vs released VTM-17.0 & $-$12.56 & $-$12.65 & $-$0.09 \\
MLIC++ CLIC Pro vs released VTM-17.0 & $-$13.08 & $-$13.14 & $-$0.06 \\
Cool-chic 5.0 vs MLIC++ Kodak (README: 5.0.1) & +10.20 & +11.22 & +1.02 \\
Cool-chic 5.0 vs MLIC++ CLIC Pro (README: 5.0.1) & +0.50 & +1.28 & +0.78 \\
Cool-chic 5.0 vs ELIC Kodak (README: 5.0.1) & +0.10 & +1.48 & +1.38 \\
Cool-chic 5.0 vs Cheng'20 Kodak (README: 5.0.1) & $-$10.40 & $-$7.26 & +3.14 \\
Cool-chic 5.0 vs Cheng'20 CLIC Pro (README: 5.0.1) & $-$21.30 & $-$20.71 & +0.59 \\
Cool-chic 5.0 vs 4.2 Kodak (README: 5.0.1) & $-$6.10 & $-$6.96 & $-$0.86 \\
Cool-chic 5.0 vs 4.2 CLIC Pro (README: 5.0.1) & $-$9.90 & $-$9.71 & +0.19 \\
\midrule
\multicolumn{4}{@{}l}{\emph{Kodak claims labeled VTM-12.1, recomputed against the CompressAI VTM-9.1 curve}} \\
DCAE & $-$17.00 & $-$16.98 & +0.02 \\
FTIC & $-$14.50 & $-$14.56 & $-$0.06 \\
CCA & $-$13.87 & $-$13.73 & +0.14 \\
TCM & $-$12.30 & $-$11.73 & +0.57 \\
\end{longtable}
\endgroup

\subsection{MS-SSIM display curves}
\label{sec:s_msssim}
\begin{figure}[htbp]
\centering
\includegraphics[width=0.6\linewidth]{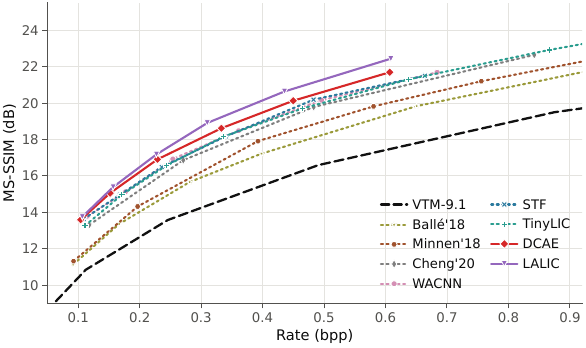}
\caption{\textbf{Released Kodak MS-SSIM curves} of MS-SSIM-optimized models~\cite{balle2018variational,minnen2018joint,cheng2020learned,zou2022devil,lu2022tinylic,lu2025dcae,feng2025lalic} ($-10\log_{10}(1-\text{MS-SSIM})$ in dB; display only). The VTM curve is the VTM-9.1 run of CompressAI~\cite{begaint2020compressai}.}
\label{fig:s_msssim}
\end{figure}

\Cref{fig:s_msssim} draws the released MS-SSIM curves of MS-SSIM-optimized models on Kodak, with their sources in \cref{tab:s_msssim}.
They are display only: they enter no score, ranking, or decision.
Matched MS-SSIM comparisons computed from our own decoded images of the MSE-optimized releases are in Fig.~\mref{fig:matched} and \cref{sec:s_cf}.

\begin{table}[htbp]
\centering
\caption{Sources of the Kodak MS-SSIM display curves (MS-SSIM-optimized models; the VTM-9.1 run is CompressAI's QP sweep).}
\label{tab:s_msssim}
\footnotesize
\begin{tabular}{@{}lrlrP{4.3cm}P{4.8cm}@{}}
\toprule
Curve & Year & Optimized for & Points & Repository @ commit & File \\
\midrule
VTM-9.1 (CompressAI run) & 2020 & n/a (QP anchor) & 8 & \nolinkurl{InterDigitalInc/CompressAI} @\,\texttt{3698ce73} & \nolinkurl{results/image/kodak/vtm.json} \\
Ballé'18 & 2018 & MS-SSIM & 8 & \nolinkurl{tensorflow/compression} @\,\texttt{dda8cb4c} & \nolinkurl{results/image_compression/kodak/MS-SSIM_sRGB_RGB/balle-2018-iclr-opt-msssim.txt} \\
Minnen'18 & 2018 & MS-SSIM & 10 & \nolinkurl{tensorflow/compression} @\,\texttt{dda8cb4c} & \nolinkurl{results/image_compression/kodak/MS-SSIM_sRGB_RGB/minnen-2018-neurips-opt-msssim.txt} \\
Cheng'20 & 2020 & MS-SSIM & 4 & \nolinkurl{ZhengxueCheng/Learned-Image-Compression-with-GMM-and-Attention} @\,\texttt{c13652f7} & \nolinkurl{RDdata/data_Kodak_Proposed_optimized_by_MSSSIM_MSSSIM.dat} \\
WACNN & 2022 & MS-SSIM & 6 & \nolinkurl{Googolxx/STF} @\,\texttt{bbcfe3a4} & \nolinkurl{results/cnn_ms-ssim_Kodak - .json} \\
STF & 2022 & MS-SSIM & 6 & \nolinkurl{Googolxx/STF} @\,\texttt{bbcfe3a4} & \nolinkurl{results/stf_ms-ssim_Kodak.json} \\
TinyLIC & 2022 & MS-SSIM & 8 & \nolinkurl{lumingzzz/TinyLIC} @\,\texttt{b9081f7c} & \nolinkurl{results/ms-ssim/kodak.json} \\
DCAE & 2025 & MS-SSIM & 6 & \nolinkurl{CVL-UESTC/DCAE} @\,\texttt{e2525a00} & \nolinkurl{RD_data.json} \\
LALIC & 2025 & MS-SSIM & 6 & \nolinkurl{sjtu-medialab/RwkvCompress} @\,\texttt{d2808464} & \nolinkurl{results/msssim/kodak-lalic-msssim.json} \\
\bottomrule
\end{tabular}

\end{table}

\section{Protocol details and proofs}
\label{sec:s_protocol}
\paragraph{Windows.} $W_d$ is the intersection of the released PSNR ranges of all fixed-rate models with at least five points, rounded inward to 0.1\,dB: Kodak \winK, Tecnick \winT, and CLIC~Pro \winC\,dB.
Variable-rate releases do not define the window.
\paragraph{Strict coverage.} A level $D_W$ is computed only if the curve, after the upper envelope and the point rule, spans $W$; otherwise the cell has no value. The window is never clipped to the data, and the interpolant never extrapolates.
\paragraph{Consensus reference.} The canonical score averages $D_W$ over the independently released VTM runs (Kodak: four runs; Tecnick: two; CLIC~Pro: three; \cref{tab:s_vtmconf}). Copies of the same run in other repositories are counted once.
\paragraph{Why pairwise windows break transitivity.} With $W_{ab}$ the overlap of $a$ and $b$, comparing $a$ and $b$ through an anchor $r$ combines levels over $W_{ra}$ and $W_{rb}$, which is guaranteed to equal the direct value when $W_{ra}=W_{rb}=W_{ab}$; with unequal windows the two routes can disagree. On Kodak, QARV needs \qarvDirect\% more rate than WACNN in a direct comparison but \qarvMediatedAbs\% less when both are compared through their BD-rates against the VTM-18.0 run; on the fixed window both routes give +\qarvFixed\%.
\paragraph{Protocol family.} A direct comparison is evaluated under nine predeclared variants on $W_d$ (PCHIP, Akima, or cubic interpolation; all points, or dropping the highest or the lowest), and with PCHIP and all points on each of two or three predeclared alternative windows. A variant is valid only if both curves span its window; decisions use the valid variants and report their number.
\paragraph{Image bootstrap and scale.} For a pair with matched per-image results we resample $n$ images with replacement (the same images for both codecs), rebuild both curves, and recompute the direct BD-rate on a window fixed before resampling (the common range of the two curves, inset by one operating point at each end). $\sigma$ is the least-squares slope of the bootstrap SD against $1/\sqrt n$ through the origin, and $\sigma_{\rm pair}=\sqrt n\,\mathrm{SD}$ at the full image count. It is an effective scale of the dataset-level BD-rate, not the SD of per-image BD-rates.
\paragraph{Reference scales.} $\sigma_{\rm ref}=\sigmaRef$ is the median over four consecutive-generation pairs of learned codecs on USTC-TD (Cheng'20/Minnen'18, ELIC/Cheng'20, MLIC++/ELIC, iWave++/ELIC; \sigmaRefLo--\sigmaRefHi). For identical Cool-chic steps, $\sigma$ on CLIC~Pro is \ratioCCMin--\ratioCCMax\ times that on Kodak (median \ratioCC), which defines $\sigma_{\rm hi}=\sigmaHi\approx\ratioCC\,\sigma_{\rm ref}$. Both are frozen in the manifest.
\paragraph{Decision function.} One function produces every status. \emph{Abstained}: the canonical variant is invalid. \emph{Unstable}: valid variants disagree in sign. Otherwise $\Delta_{\rm least}$ is the valid value of smallest magnitude, $\sigma^\star=|\Delta_{\rm least}|\sqrt{n_d}/1.96$, and the status is \emph{robust} ($\sigma^\star>\sigma_{\rm hi}$), \emph{resolved} ($\sigma_{\rm ref}<\sigma^\star\le\sigma_{\rm hi}$), or \emph{unresolved}. With matched per-image results, the status is \emph{resolved (observed)} if the paired interval excludes zero. The prior best of a method of year $y$ is the best canonical score among earlier years (first peer-reviewed publication; preprints by first release).

\subsection{Proofs of the two properties}
\label{sec:s_supported}
\begin{proof}[Property~\mref{prop:fixed} (reference independence)]
Write $D=D_W$. For any two curves that span $W$ under the same interpolant and point rule, $1+\BD_W(a\!\to\!b)=10^{D(b)-D(a)}$, since both levels are integrals of fixed functions over the same interval.
(i) $10^{D(c)-D(a)}=10^{D(b)-D(a)}\,10^{D(c)-D(b)}$.
(ii) $10^{D(m)-D(r')}=10^{D(r)-D(r')}\,10^{D(m)-D(r)}$, and the factor $\kappa=10^{D(r)-D(r')}$ does not depend on $m$; it preserves the order over $m$, and $1+\BD_W(a\!\to\!b)=\bigl(1+\BD_W(r\!\to\!b)\bigr)/\bigl(1+\BD_W(r\!\to\!a)\bigr)$ cancels it.
For the consensus reference, $\sum_j (D-D_j)^2$ over the run levels $D_1,\dots,D_J$ is minimized at their mean $\bar D$, and the score against run $j$ is $10^{D(m)-\bar D}\,10^{\bar D-D_j}$ with $|\bar D-D_j|\le\max_{j'}|D_{j'}-D_j|$.
When $W$ depends on the pair, the levels are integrals over different intervals and (i) fails.
\end{proof}
\begin{proof}[Property~\mref{prop:supported} (conditional margin)]
Let $\delta=\Delta_{\rm least}$, $n=n_d$, $z=1.96$, and $\sigma^\star=|\delta|\sqrt n/z$.
(i) Since $1-\sigma/\sigma^\star=(|\delta|-z\sigma/\sqrt n)/|\delta|$, $\delta(1-\sigma/\sigma^\star)_+=\operatorname{sign}(\delta)\max\{0,|\delta|-z\sigma/\sqrt n\}$, the endpoint of $\delta\pm z\sigma/\sqrt n$ nearest zero when the interval excludes zero and $0$ otherwise. Continuity, monotonicity, $\sbd(0)=\delta$, and vanishing exactly at $\sigma\ge\sigma^\star$ follow from this form.
(ii) The decision function declares robust iff $\sigma^\star>\sigma_{\rm hi}$, resolved iff $\sigma_{\rm ref}<\sigma^\star\le\sigma_{\rm hi}$, and unresolved iff $\sigma^\star\le\sigma_{\rm ref}$; by (i), $\sbd(\sigma)\ne0$ iff $\sigma<\sigma^\star$.
(iii) With the $m$ tests ordered by decreasing $\sigma^\star$ and $z_k=\Phi^{-1}(1-0.025/(m-k+1))$, the Holm-adjusted~\cite{holm1979simple} $p$ of the test of rank $r$ is below 0.05 iff $\sigma<\min_{k\le r}z\,\sigma^\star_{(k)}/z_k=\sigma^\star_{\rm FW}$; the order does not depend on $\sigma$, because every $p$ decreases in $\sigma^\star_{(k)}/\sigma$. Hence $\sbd_{\rm FW}(\sigma)=\delta(1-\sigma/\sigma^\star_{\rm FW})_+\ne0$ iff the adjusted $p$ is below 0.05, and $|\sbd_{\rm FW}|\le|\sbd|$ because $z_r\ge z$.
Finally, any jointly valid variant has $|\delta_v|\ge|\delta|$, so $|\sbd|$ never exceeds that variant's bound.
\end{proof}
Because $\delta$ is the least favorable effect over valid variants and its sampling distribution is not observed for aggregate-only pairs, $\sbd$ is a conditional sensitivity summary: it establishes neither observed coverage for an unmeasured pair nor simultaneous coverage over protocol choices.
An independent script recomputes every margin and recovers every $\sigma^\star_{\rm FW}$ by bisection on the recomputed Holm decisions.

\section{Leaderboards, reference runs, and published claims}
\label{sec:s_anchors}
\subsection{Leaderboards}
\label{sec:s_lb}
\Cref{tab:s_lb_kodak,tab:s_lb_tecnick,tab:s_lb_clic_pv} rank all scored methods.
\emph{Canonical} is the BD-rate (\%) against the consensus VTM reference on the fixed window (PCHIP, all points); \emph{Overlap} is the conventional pairwise-overlap BD-rate against one VTM run, as papers compute it.
\emph{[min, max]} is the range over the valid protocol variants (VTM run, interpolation, window convention, point subset), and \emph{Valid} counts them among those attempted; a variant is invalid when a curve does not span its window or Akima interpolation would need a substitute.
The range is a sensitivity over analyst choices, dominated by the choice of VTM run (\cref{sec:s_vtm}), not a statistical interval.
\emph{Comp.} is the component that the paper presents as its main contribution (E entropy or context model, T transform or backbone, O per-image overfitted codec, R rate control and efficiency) and is descriptive only; \emph{Tier} is the provenance of the curve (\cref{sec:s_ledger}); \emph{Conv.} states which measurement conventions the paper documents (D: rate accounting and aggregation; P: one of them; U: neither).
\emph{Low/Mid/High} are fixed-window BD-rates in each third of the window; n/a marks a curve that does not span the fixed window, and $^\dagger$ a preprint.
\begingroup\footnotesize
\begin{longtable}{@{}lcccc rr lr rrr@{}}
\caption{Full leaderboard on Kodak (fixed window 30.5--36.6\,dB), ordered by canonical score; overlap against the VTM-18.0 run. The columns are explained in \cref{sec:s_lb}.}\label{tab:s_lb_kodak}\\
\toprule Method & Year & Comp. & Tier & Conv. & Canonical & Overlap & [min, max] & Valid & Low & Mid & High \\ \midrule \endfirsthead
\toprule Method & Year & Comp. & Tier & Conv. & Canonical & Overlap & [min, max] & Valid & Low & Mid & High \\ \midrule \endhead
\midrule \multicolumn{12}{r}{\emph{continued on next page}} \\ \endfoot
\bottomrule \endlastfoot
HPCM & 2025 & E & A & U & $-$18.95 & $-$19.39 & [$-$21.0, $-$17.6] & 60/72 & $-$21.3 & $-$18.6 & $-$16.8 \\
GLIC & 2026 & T & A & U & $-$18.60 & $-$18.69 & [$-$19.8, $-$17.2] & 48/72 & $-$20.0 & $-$18.4 & $-$17.4 \\
ChWDTA$^\dagger$ & 2026 & T & A & U & $-$17.29 & $-$17.28 & [$-$18.5, $-$16.0] & 48/72 & $-$19.1 & $-$16.8 & $-$15.9 \\
CMIC & 2026 & T & A & U & $-$17.13 & $-$17.40 & [$-$18.6, $-$16.2] & 60/72 & $-$18.8 & $-$16.8 & $-$15.8 \\
SAAF & 2026 & T & A & U & $-$16.33 & $-$16.81 & [$-$17.9, $-$15.2] & 48/72 & $-$17.5 & $-$16.0 & $-$15.5 \\
DCAE & 2025 & E & A & U & $-$16.03 & $-$16.41 & [$-$17.6, $-$14.7] & 48/72 & $-$17.8 & $-$15.6 & $-$14.6 \\
S2CFormer$^\dagger$ & 2025 & T & B & U & $-$15.23 & $-$15.10 & [$-$16.1, $-$14.0] & 48/72 & $-$16.2 & $-$15.0 & $-$14.5 \\
LALIC & 2025 & T & A & P & $-$14.57 & $-$14.69 & [$-$15.9, $-$13.3] & 48/72 & $-$16.2 & $-$14.4 & $-$13.1 \\
FTIC & 2024 & T & A & U & $-$14.06 & $-$13.99 & [$-$15.3, $-$12.4] & 48/72 & $-$15.9 & $-$13.8 & $-$12.4 \\
MLIC++ & 2023 & E & A & U & $-$13.96 & $-$14.41 & [$-$16.5, $-$12.2] & 48/72 & $-$17.0 & $-$14.3 & $-$10.4 \\
CCA & 2024 & E & A & U & $-$13.34 & $-$13.14 & [$-$14.0, $-$11.4] & 48/72 & $-$15.1 & $-$12.9 & $-$12.0 \\
MLIC+ & 2023 & E & A & U & $-$12.02 & $-$12.49 & [$-$14.3, $-$10.4] & 48/72 & $-$15.4 & $-$11.4 & $-$9.1 \\
TCM & 2023 & T & A & U & $-$11.37 & $-$11.15 & [$-$12.2, $-$10.1] & 48/72 & $-$12.6 & $-$11.1 & $-$10.4 \\
SegPIC & 2024 & T & A & U & $-$8.80 & $-$9.10 & [$-$10.5, $-$7.5] & 48/72 & $-$9.6 & $-$9.2 & $-$7.6 \\
MLIC & 2023 & E & A & U & $-$8.58 & $-$9.11 & [$-$11.0, $-$6.9] & 48/72 & $-$11.7 & $-$8.3 & $-$5.7 \\
WeConvene & 2024 & T & A & U & $-$8.06 & $-$7.75 & [$-$8.9, $-$6.7] & 48/72 & $-$9.1 & $-$8.0 & $-$7.2 \\
ELIC & 2022 & E & B & U & $-$6.22 & $-$6.37 & [$-$7.3, $-$5.0] & 48/72 & $-$7.5 & $-$5.9 & $-$5.2 \\
Cool-chic 5$^\dagger$ & 2026 & O & A & D & $-$5.36 & $-$6.11 & [$-$9.7, $-$2.4] & 48/72 & $-$3.1 & $-$5.3 & $-$7.7 \\
WACNN & 2022 & T & A & U & $-$3.72 & $-$3.89 & [$-$5.2, $-$2.5] & 48/72 & $-$2.8 & $-$4.7 & $-$3.6 \\
STF & 2022 & T & A & U & $-$3.14 & $-$3.46 & [$-$5.1, $-$1.8] & 60/72 & $-$4.1 & $-$3.8 & $-$1.5 \\
TinyLIC$^\dagger$ & 2022 & T & A & U & $-$3.11 & $-$3.08 & [$-$4.1, $-$1.3] & 60/72 & $-$4.7 & $-$3.3 & $-$1.3 \\
MSINN-VR & 2025 & R & A & U & $-$1.35 & $-$4.08 & [$-$5.8, $-$0.4] & 60/72 & +1.1 & $-$1.8 & $-$3.3 \\
InvCompress & 2021 & T & A & U & $-$0.76 & +0.34 & [$-$1.9, +1.2] & 72/72 & +0.0 & $-$0.9 & $-$1.4 \\
Neural-Syntax & 2022 & T & A & U & $-$0.04 & +0.24 & [$-$0.9, +1.7] & 60/72 & $-$0.9 & +0.7 & +0.1 \\
QRes-VAE & 2023 & T & A & D & +0.05 & $-$4.03 & [$-$5.2, +1.0] & 60/72 & +2.1 & +0.3 & $-$2.2 \\
ShiftLIC & 2025 & R & A & U & +0.34 & +0.33 & [$-$0.7, +1.6] & 60/72 & $-$0.9 & +0.4 & +1.6 \\
QARV & 2024 & R & A & D & +0.71 & $-$5.89 & [$-$6.7, +1.6] & 60/72 & +4.5 & +0.4 & $-$2.7 \\
LotteryCodec & 2025 & O & A & U & +2.39 & +0.07 & [$-$2.3, +3.7] & 72/72 & +5.5 & +2.4 & $-$0.7 \\
ChARM & 2020 & E & A & U & +2.58 & +0.83 & [+0.4, +3.7] & 72/72 & +1.5 & +3.3 & +3.0 \\
Cheng'20 & 2020 & E & A & U & +4.23 & +4.25 & [+2.7, +5.2] & 60/72 & +4.3 & +2.0 & +6.5 \\
C3 & 2024 & O & A & U & +5.10 & +5.82 & [+2.7, +8.8] & 72/72 & +9.2 & +4.9 & +1.4 \\
Qmap & 2021 & R & A & P & +15.20 & +18.02 & [+14.2, +19.1] & 60/72 & +17.2 & +14.7 & +13.7 \\
EVC & 2023 & R & A & U & \multicolumn{1}{c}{n/a} & $-$0.43 & \multicolumn{1}{c}{--} & -- & -- & -- & -- \\
STanH & 2025 & R & A & U & \multicolumn{1}{c}{n/a} & +0.49 & \multicolumn{1}{c}{--} & -- & -- & -- & -- \\
\end{longtable}
\endgroup

\begingroup\footnotesize
\begin{longtable}{@{}lcccc rr lr rrr@{}}
\caption{Full leaderboard on Tecnick (fixed window 32.5--38.1\,dB), ordered by canonical score. Overlap: against the VTM-18.0 run; other columns as in \cref{tab:s_lb_kodak}.}\label{tab:s_lb_tecnick}\\
\toprule Method & Year & Comp. & Tier & Conv. & Canonical & Overlap & [min, max] & Valid & Low & Mid & High \\ \midrule \endfirsthead
\toprule Method & Year & Comp. & Tier & Conv. & Canonical & Overlap & [min, max] & Valid & Low & Mid & High \\ \midrule \endhead
\midrule \multicolumn{12}{r}{\emph{continued on next page}} \\ \endfoot
\bottomrule \endlastfoot
CMIC & 2026 & T & A & U & $-$23.21 & $-$23.94 & [$-$24.6, $-$22.7] & 30/36 & $-$24.5 & $-$22.8 & $-$22.3 \\
ChWDTA$^\dagger$ & 2026 & T & A & U & $-$22.99 & $-$23.46 & [$-$24.5, $-$22.0] & 24/36 & $-$24.6 & $-$22.7 & $-$21.6 \\
GLIC & 2026 & T & A & U & $-$22.17 & $-$22.78 & [$-$23.2, $-$21.4] & 24/36 & $-$22.9 & $-$21.8 & $-$21.7 \\
HPCM & 2025 & E & A & U & $-$21.38 & $-$22.25 & [$-$23.0, $-$20.7] & 24/36 & $-$23.1 & $-$21.1 & $-$19.9 \\
SAAF & 2026 & T & A & U & $-$20.63 & $-$21.48 & [$-$21.9, $-$20.1] & 24/36 & $-$20.0 & $-$20.2 & $-$21.7 \\
DCAE & 2025 & E & A & U & $-$20.06 & $-$21.01 & [$-$21.5, $-$19.2] & 24/36 & $-$21.4 & $-$19.2 & $-$19.5 \\
LALIC & 2025 & T & A & P & $-$18.20 & $-$18.54 & [$-$19.5, $-$17.6] & 24/36 & $-$18.0 & $-$18.4 & $-$18.2 \\
MLIC++ & 2023 & E & A & U & $-$16.41 & $-$18.06 & [$-$19.6, $-$15.5] & 24/36 & $-$20.4 & $-$16.2 & $-$12.5 \\
MLIC+ & 2023 & E & A & U & $-$15.61 & $-$16.92 & [$-$18.2, $-$14.7] & 24/36 & $-$18.8 & $-$14.9 & $-$13.1 \\
FTIC & 2024 & T & A & U & $-$15.11 & $-$15.36 & [$-$16.6, $-$14.1] & 24/36 & $-$14.2 & $-$15.1 & $-$16.0 \\
CCA & 2024 & E & A & U & $-$14.50 & $-$14.96 & [$-$16.2, $-$13.8] & 24/36 & $-$14.1 & $-$14.1 & $-$15.3 \\
MLIC & 2023 & E & A & U & $-$11.60 & $-$13.19 & [$-$14.5, $-$10.8] & 24/36 & $-$15.0 & $-$10.8 & $-$8.9 \\
TCM & 2023 & T & A & U & $-$11.37 & $-$11.94 & [$-$12.7, $-$10.4] & 24/36 & $-$10.7 & $-$10.8 & $-$12.6 \\
WeConvene & 2024 & T & A & U & $-$8.34 & $-$8.89 & [$-$9.7, $-$7.7] & 24/36 & $-$8.0 & $-$8.5 & $-$8.6 \\
TinyLIC$^\dagger$ & 2022 & T & A & U & $-$4.03 & $-$4.96 & [$-$5.3, $-$3.0] & 30/36 & $-$5.3 & $-$4.6 & $-$2.1 \\
InvCompress & 2021 & T & A & U & $-$2.28 & $-$1.36 & [$-$3.6, $-$0.1] & 30/36 & $-$1.9 & $-$2.3 & $-$2.6 \\
QARV & 2024 & R & A & D & $-$2.08 & $-$8.88 & [$-$9.8, $-$1.5] & 30/36 & +2.4 & $-$2.0 & $-$6.5 \\
ChARM & 2020 & E & A & U & +0.16 & $-$3.53 & [$-$3.6, +0.8] & 36/36 & $-$1.2 & +1.4 & +0.3 \\
QRes-VAE & 2023 & T & A & D & +0.83 & $-$4.53 & [$-$5.8, +1.5] & 30/36 & +0.9 & +3.0 & $-$1.4 \\
MSINN-VR & 2025 & R & A & U & +1.18 & $-$3.46 & [$-$5.3, +1.8] & 36/36 & +5.3 & +1.0 & $-$2.6 \\
STanH & 2025 & R & A & U & \multicolumn{1}{c}{n/a} & +2.78 & \multicolumn{1}{c}{--} & -- & -- & -- & -- \\
\end{longtable}
\endgroup

\begingroup\footnotesize
\begin{longtable}{@{}lcccc rr lr rrr@{}}
\caption{Full leaderboard on CLIC~Pro (fixed window 32.4--37.4\,dB), ordered by canonical score. Overlap: against the VTM-19.1 run; other columns as in \cref{tab:s_lb_kodak}.}\label{tab:s_lb_clic_pv}\\
\toprule Method & Year & Comp. & Tier & Conv. & Canonical & Overlap & [min, max] & Valid & Low & Mid & High \\ \midrule \endfirsthead
\toprule Method & Year & Comp. & Tier & Conv. & Canonical & Overlap & [min, max] & Valid & Low & Mid & High \\ \midrule \endhead
\midrule \multicolumn{12}{r}{\emph{continued on next page}} \\ \endfoot
\bottomrule \endlastfoot
HPCM & 2025 & E & A & U & $-$19.26 & $-$21.12 & [$-$21.6, $-$17.6] & 45/54 & $-$20.7 & $-$19.0 & $-$18.0 \\
GLIC & 2026 & T & A & U & $-$18.91 & $-$20.79 & [$-$20.8, $-$17.5] & 36/54 & $-$19.2 & $-$18.7 & $-$18.8 \\
CMIC & 2026 & T & A & U & $-$18.90 & $-$20.89 & [$-$21.0, $-$17.5] & 45/54 & $-$19.6 & $-$18.6 & $-$18.4 \\
ChWDTA$^\dagger$ & 2026 & T & A & U & $-$18.38 & $-$20.26 & [$-$20.3, $-$16.9] & 36/54 & $-$19.1 & $-$18.2 & $-$17.9 \\
SAAF & 2026 & T & A & U & $-$16.85 & $-$19.28 & [$-$19.3, $-$15.4] & 36/54 & $-$16.2 & $-$16.4 & $-$17.9 \\
DCAE & 2025 & E & A & U & $-$16.69 & $-$18.89 & [$-$19.1, $-$15.2] & 36/54 & $-$17.6 & $-$16.1 & $-$16.4 \\
LALIC & 2025 & T & A & P & $-$15.70 & $-$17.38 & [$-$17.6, $-$14.0] & 36/54 & $-$15.3 & $-$16.0 & $-$15.8 \\
MLIC++ & 2023 & E & A & U & $-$13.74 & $-$16.31 & [$-$17.5, $-$11.5] & 36/54 & $-$17.5 & $-$13.3 & $-$10.2 \\
MLIC+ & 2023 & E & A & U & $-$13.50 & $-$15.84 & [$-$16.6, $-$11.4] & 36/54 & $-$16.2 & $-$12.8 & $-$11.3 \\
Cool-chic 5$^\dagger$ & 2026 & O & A & D & $-$13.47 & $-$15.74 & [$-$16.6, $-$11.4] & 36/54 & $-$13.1 & $-$13.2 & $-$14.0 \\
CCA & 2024 & E & A & U & $-$12.03 & $-$13.88 & [$-$14.4, $-$10.4] & 36/54 & $-$11.4 & $-$11.7 & $-$12.9 \\
FTIC & 2024 & T & A & U & $-$11.20 & $-$12.86 & [$-$13.3, $-$9.1] & 36/54 & $-$10.7 & $-$11.1 & $-$11.8 \\
MLIC & 2023 & E & A & U & $-$9.43 & $-$12.12 & [$-$13.1, $-$7.4] & 36/54 & $-$12.7 & $-$8.8 & $-$6.6 \\
SegPIC & 2024 & T & A & U & $-$9.38 & $-$11.60 & [$-$11.9, $-$7.8] & 36/54 & $-$10.0 & $-$9.4 & $-$8.8 \\
TCM & 2023 & T & A & U & $-$9.10 & $-$11.31 & [$-$11.7, $-$7.0] & 36/54 & $-$7.8 & $-$9.1 & $-$10.3 \\
LotteryCodec & 2025 & O & A & U & $-$7.12 & $-$9.79 & [$-$10.4, $-$5.3] & 42/54 & $-$6.3 & $-$7.0 & $-$8.0 \\
TinyLIC$^\dagger$ & 2022 & T & A & U & $-$4.29 & $-$6.65 & [$-$6.7, $-$2.5] & 45/54 & $-$4.8 & $-$4.4 & $-$3.6 \\
C3 & 2024 & O & A & U & $-$3.60 & $-$4.90 & [$-$6.3, $-$0.1] & 54/54 & $-$1.4 & $-$3.5 & $-$5.9 \\
Neural-Syntax & 2022 & T & A & U & $-$2.26 & $-$4.77 & [$-$4.8, $-$0.4] & 45/54 & $-$3.1 & $-$1.8 & $-$1.9 \\
WACNN & 2022 & T & A & U & $-$1.28 & $-$3.29 & [$-$3.6, +0.4] & 45/54 & +0.8 & $-$3.2 & $-$1.4 \\
InvCompress & 2021 & T & A & U & $-$0.84 & $-$0.70 & [$-$3.0, +3.8] & 54/54 & +0.3 & $-$1.0 & $-$1.8 \\
STF & 2022 & T & A & U & $-$0.53 & $-$3.22 & [$-$3.3, +1.3] & 45/54 & +0.6 & $-$1.4 & $-$0.8 \\
MSINN-VR & 2025 & R & A & U & +1.80 & $-$1.16 & [$-$2.9, +4.1] & 54/54 & +5.4 & +1.4 & $-$1.3 \\
Cheng'20 & 2020 & E & A & U & +9.52 & +7.46 & [+7.0, +11.7] & 45/54 & +9.3 & +7.6 & +11.7 \\
STanH & 2025 & R & A & U & \multicolumn{1}{c}{n/a} & +1.04 & \multicolumn{1}{c}{--} & -- & -- & -- & -- \\
\end{longtable}
\endgroup

\subsection{Reference runs}
\label{sec:s_vtm}
\begin{figure}[htbp]
  \centering
  \includegraphics[width=0.82\linewidth]{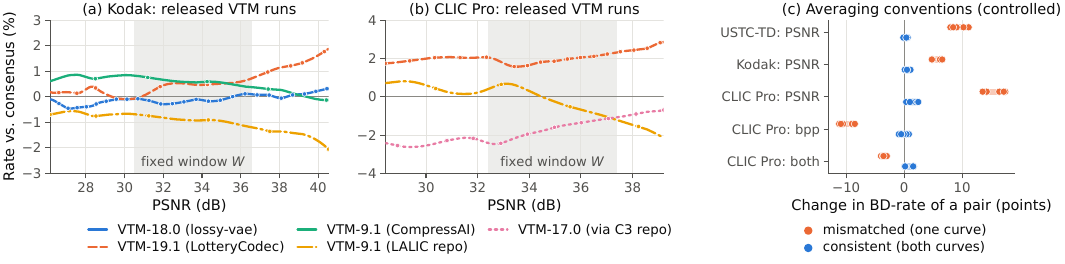}
  \caption{\textbf{Anchor runs and averaging conventions.}
  (a, b) Rate of each released VTM run relative to the consensus of all runs (positive: the run needs more bits); dots mark released points. Inside $W$, runs differ locally by up to \anchorLocalK\ (Kodak) and \anchorLocalC\ (CLIC~Pro) points, and their order changes with rate.
  (c) Change of a between-codec BD-rate on released per-image results when an alternative averaging convention (PSNR of the mean MSE, pixel-weighted bpp, or both) is applied to both curves (consistent) or to the test curve only (mismatched).}
  \label{fig:s_anchors}
\end{figure}

\Cref{fig:s_anchors} shows how far the released VTM runs differ, and \cref{tab:s_anchors} gives the BD-rate between every pair of runs.
Sub-sampling the dense VTM-18.0 run to every fifth QP changes BD-rates by at most \sparseMaxT\ points on Tecnick, so the differences between runs are not interpolation artifacts.
The two Kodak VTM-9.1 runs both document 8-bit 4:4:4 intra coding (\cref{tab:s_vtmconf}) and still differ by \vtmNineOneDiff\ points.
The reference also changes between papers: the \vtmVersionsClaims\ numeric claims that name a VTM version name \vtmVersionsN\ versions (\vtmVersionsList, as stated), \vtmVersionsYearN\ of them in \vtmVersionsYear\ papers alone, and \vtmNoVersionN\ further claims name VVC without a version (claims record in the artifact).

\begin{table}[htbp]\centering\footnotesize
\caption{BD-rate (\%) between independently released VTM runs for each interpolation, on the fixed window (fixed) and on the pairwise overlap (overlap).}\label{tab:s_anchors}
\begin{tabular}{@{}lllrrrrrr@{}}
\toprule Dataset & Reference run & Test run & \multicolumn{2}{c}{PCHIP} & \multicolumn{2}{c}{Akima} & \multicolumn{2}{c}{Cubic} \\ & & & fixed & overlap & fixed & overlap & fixed & overlap \\ \midrule
Kodak & VTM-18.0 (lossy-vae) (36 QPs) & VTM-19.1 (LotteryCodec) (11 QPs) & +0.56 & +0.72 & +0.58 & +0.77 & +0.41 & +0.73 \\
Kodak & VTM-18.0 (lossy-vae) (36 QPs) & VTM-9.1 (CompressAI) (8 QPs) & +0.72 & +0.17 & +0.74 & +0.17 & +0.90 & +0.01 \\
Kodak & VTM-18.0 (lossy-vae) (36 QPs) & VTM-9.1 (LALIC repo) (6 QPs) & $-$0.79 & $-$0.89 & $-$0.80 & $-$0.89 & $-$0.81 & $-$0.96 \\
Kodak & VTM-19.1 (LotteryCodec) (11 QPs) & VTM-9.1 (CompressAI) (8 QPs) & +0.16 & $-$0.17 & +0.15 & $-$0.22 & +0.49 & $-$0.01 \\
Kodak & VTM-19.1 (LotteryCodec) (11 QPs) & VTM-9.1 (LALIC repo) (6 QPs) & $-$1.34 & $-$1.54 & $-$1.38 & $-$1.61 & $-$1.21 & $-$1.60 \\
Kodak & VTM-9.1 (CompressAI) (8 QPs) & VTM-9.1 (LALIC repo) (6 QPs) & $-$1.50 & $-$1.51 & $-$1.53 & $-$1.51 & $-$1.69 & $-$1.75 \\
Tecnick & VTM-18.0 (lossy-vae) (36 QPs) & VTM-9.1 (LALIC repo) (6 QPs) & $-$1.06 & $-$1.25 & $-$1.16 & $-$1.25 & $-$1.75 & $-$1.26 \\
CLIC Pro & VTM-19.1 (LotteryCodec) (19 QPs) & VTM-17.0 (C3 repo) (11 QPs) & $-$3.56 & $-$3.79 & $-$3.55 & $-$3.80 & $-$3.60 & $-$3.68 \\
CLIC Pro & VTM-19.1 (LotteryCodec) (19 QPs) & VTM-9.1 (LALIC repo) (6 QPs) & $-$2.06 & $-$2.57 & $-$2.08 & $-$2.57 & $-$2.15 & $-$2.53 \\
CLIC Pro & VTM-17.0 (C3 repo) (11 QPs) & VTM-9.1 (LALIC repo) (6 QPs) & +1.56 & +1.72 & +1.52 & +1.74 & +1.50 & +1.69 \\
\bottomrule
\end{tabular}
\end{table}

\begin{table}[htbp]
\centering
\caption{Documented configurations of the released VTM runs (from the releasing repository or paper).}
\label{tab:s_vtmconf}
\footnotesize
\begin{tabular}{@{}lllp{8.2cm}@{}}
\toprule
Run & Datasets & QPs & Documented configuration (source) \\
\midrule
VTM-9.1, CompressAI & Kodak & 8 & 8-bit YCbCr 4:4:4, default intra configuration (CompressAI paper, Sec.~5.2) \\
VTM-9.1, LALIC repository & K, T, C & 6 & intra, \texttt{--InputChromaFormat=444}, 8-bit, QP 22--47 (LALIC paper and result files) \\
VTM-17.0, shipped in C3 & CLIC~Pro & 11 & intra, YUV 4:4:4, 8-bit (MLIC papers) \\
VTM-18.0, lossy-vae & K, T & 36 & version and RGB PSNR only (README) \\
VTM-19.1, LotteryCodec & K, C & 11/19 & \texttt{--InputChromaFormat=444}; decoder output bit depth 10 (README) \\
\bottomrule
\end{tabular}
\end{table}

\subsection{Published claims}
\label{sec:s_claims}
\begin{figure}[htbp]
\centering
\includegraphics[width=0.92\linewidth]{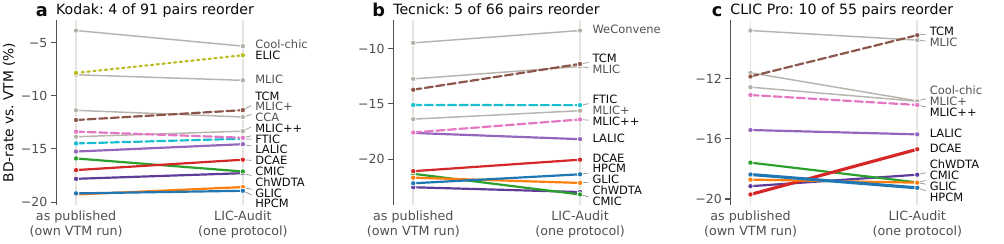}
\caption{\textbf{Published BD-rates do not order learned codecs consistently.} Each line joins a method's BD-rate as its paper states it, against the VTM run that paper chose (left), to its canonical score from the released curve on the frozen window (right). Crossing lines are pairs whose published numbers reverse their order; on CLIC~Pro, DCAE leads as published, but HPCM is \dHpcmVsDcaeC\% better on one protocol (direct BD-rate). Colored: the frontier methods of Tab.~\mref{tab:frontier}.}
\label{fig:s_claims}
\end{figure}

\Cref{fig:s_claims} contrasts the BD-rates that papers state with the canonical scores of their released curves.

\section{Frontier tests}
\label{sec:s_frontier}
Each test compares a method with the best method of earlier years on the released curves (Tab.~\mref{tab:frontier}).
The family has the \nFamTests\ tests of 2024--2026 methods with a stable sign, and every status is Holm-adjusted over it.
\Cref{tab:s_r19_sensitivity} lists the Holm-adjusted conditional margin of every test at five scales: $\sigma=3$, close to the measured median; the measured median and 90th percentile $\hat\sigma_{90}$ of the comparisons among the six matched 2025--26 releases (\cref{sec:s_r19_scale}); the historical $\sigma_{\rm ref}$; and the stress value $\sigma_{\rm hi}$.
The tests use the exact frozen values behind the displayed $\sigma_{\rm ref}$ and $\sigma_{\rm hi}$.

\paragraph{Does choosing the comparator matter?}
An advance claims a saving against every earlier method, so its null hypothesis is the union of the pairwise nulls; rejecting when every pairwise test rejects is a valid intersection--union test without adjustment for having selected the hardest comparator~\cite{berger1982iut}.
Testing each earlier method in this way leaves all \rnIutAdvSame\ advance decisions unchanged at every scale.
A regression claims that some earlier method needs less rate, for which a Bonferroni bound over the earlier methods is valid; it changes at most \rnRegChangedMax\ of \rnRegTotal\ regression decisions, none at $\hat\sigma_{90}$.

\paragraph{Protocol sensitivity.}
Dropping the highest- or lowest-rate point usually leaves $W_d$ uncovered, so only \tabThreeValidMin--\tabThreeValidMax\ of \tabThreeAttMin--\tabThreeAttMax\ predeclared variants are valid for these tests.
\Cref{tab:s_interior} adds variants that keep coverage by construction (one interior point removed, or the window moved by 0.1\,dB); they were not predeclared, leave the frozen decisions unchanged, and change the status of \nIntChanged\ tests.

\begin{table}[htbp]\centering\scriptsize\setlength{\tabcolsep}{3pt}
\caption{Conditional margins $\sbd_{\rm FW}(\sigma)$ (\%) of all \nFamTests\ frozen frontier tests at five scales: $\sigma=3$, the median (3.30) and 90th percentile (5.37) of the 45 measured 2025--26 frontier heterogeneities (\cref{sec:s_r19_scale}), and the frozen $\sigma_{\rm ref}$ and $\sigma_{\rm hi}$ (exact values 6.3461789962842365 and 10.909968429365659; rounded labels). $\Delta$: canonical direct BD-rate against the preceding-year comparator (vs.); $\Delta_{\rm least}$: least favorable valid effect; $\sigma^\star_{\rm FW}$: Holm break-even scale. Negative: saving; positive: rate cost; 0: inconclusive. Last column: hardest earlier comparator in the intersection--union test at $\sigma=\rnSigPninety$ (advances) when it differs from the selected one, and the scales at which a Bonferroni union test over all earlier methods changes a regression decision. All values assume the Gaussian scale model.}\label{tab:s_r19_sensitivity}
\begin{tabular}{@{}lll rrr rrrrr l@{}}
\toprule Dataset & Method & vs. & $\Delta$ & $\Delta_{\rm least}$ & $\sigma^\star_{\rm FW}$ & $3$ & $3.30$ & $5.37$ & $6.35$ & $10.91$ & Selection \\ \midrule
Kodak & C3 '24 & MLIC++ & +22.16 & +20.71 & 31.37 & +18.73 & +18.53 & +17.16 & +16.52 & +13.51 &  \\
Kodak & CCA '24 & MLIC++ & +0.72 & +0.20 & 0.44 & 0 & 0 & 0 & 0 & 0 &  \\
Kodak & FTIC '24 & MLIC++ & $-$0.11 & $-$0.03 & 0.08 & 0 & 0 & 0 & 0 & 0 &  \\
Kodak & QARV '24 & MLIC++ & +17.05 & +15.71 & 24.07 & +13.75 & +13.55 & +12.20 & +11.56 & +8.59 &  \\
Kodak & SegPIC '24 & MLIC++ & +6.00 & +5.49 & 8.71 & +3.60 & +3.41 & +2.11 & +1.49 & 0 &  \\
Kodak & WeConvene '24 & MLIC++ & +6.87 & +6.32 & 9.91 & +4.41 & +4.21 & +2.90 & +2.28 & 0 &  \\
Kodak & DCAE '25 & FTIC & $-$2.29 & $-$2.29 & 3.85 & \textbf{$-$0.50} & \textbf{$-$0.32} & 0 & 0 & 0 &  \\
Kodak & HPCM '25 & FTIC & $-$5.69 & $-$5.58 & 8.81 & \textbf{$-$3.68} & \textbf{$-$3.49} & \textbf{$-$2.18} & \textbf{$-$1.56} & 0 &  \\
Kodak & LALIC '25 & FTIC & $-$0.60 & $-$0.56 & 1.04 & 0 & 0 & 0 & 0 & 0 &  \\
Kodak & LotteryCodec '25 & FTIC & +19.14 & +18.24 & 27.76 & +16.27 & +16.07 & +14.71 & +14.07 & +11.07 &  \\
Kodak & MSINN-VR '25 & FTIC & +14.78 & +14.20 & 21.82 & +12.25 & +12.05 & +10.70 & +10.07 & +7.10 &  \\
Kodak & S2CFormer '25 & FTIC & $-$1.36 & $-$1.34 & 2.39 & 0 & 0 & 0 & 0 & 0 &  \\
Kodak & ShiftLIC '25 & FTIC & +16.75 & +16.57 & 25.26 & +14.60 & +14.40 & +13.04 & +12.40 & +9.41 &  \\
Kodak & CMIC '26 & HPCM & +2.24 & +2.05 & 3.47 & +0.28 & +0.10 & 0 & 0 & 0 & Bonf.\ changes at 3, 3.30 \\
Kodak & ChWDTA '26 & HPCM & +2.04 & +1.83 & 3.19 & +0.11 & 0 & 0 & 0 & 0 & Bonf.\ changes at 3 \\
Kodak & Cool-chic 5 '26 & HPCM & +16.76 & +15.84 & 24.21 & +13.87 & +13.68 & +12.32 & +11.69 & +8.70 &  \\
Kodak & GLIC '26 & HPCM & +0.43 & +0.25 & 0.50 & 0 & 0 & 0 & 0 & 0 &  \\
Kodak & SAAF '26 & HPCM & +3.23 & +2.97 & 4.90 & +1.15 & +0.97 & 0 & 0 & 0 &  \\
Tecnick & CCA '24 & MLIC++ & +2.28 & +2.28 & 7.44 & +1.36 & +1.27 & +0.63 & +0.34 & 0 & Bonf.\ changes at 6.35 \\
Tecnick & FTIC '24 & MLIC++ & +1.56 & +1.56 & 5.20 & +0.66 & +0.57 & 0 & 0 & 0 &  \\
Tecnick & QARV '24 & MLIC++ & +17.14 & +17.08 & 52.59 & +16.11 & +16.01 & +15.34 & +15.02 & +13.54 &  \\
Tecnick & WeConvene '24 & MLIC++ & +9.65 & +9.60 & 29.74 & +8.63 & +8.53 & +7.86 & +7.55 & +6.08 &  \\
Tecnick & DCAE '25 & MLIC++ & $-$4.36 & $-$4.19 & 13.32 & \textbf{$-$3.25} & \textbf{$-$3.15} & \textbf{$-$2.50} & \textbf{$-$2.19} & \textbf{$-$0.76} &  \\
Tecnick & HPCM '25 & MLIC++ & $-$5.95 & $-$5.75 & 18.13 & \textbf{$-$4.80} & \textbf{$-$4.70} & \textbf{$-$4.05} & \textbf{$-$3.74} & \textbf{$-$2.29} & hardest: Minnen'20 (ChARM) \\
Tecnick & LALIC '25 & MLIC++ & $-$2.14 & $-$1.69 & 5.58 & \textbf{$-$0.78} & \textbf{$-$0.69} & \textbf{$-$0.06} & 0 & 0 &  \\
Tecnick & MSINN-VR '25 & MLIC++ & +21.04 & +21.03 & 64.61 & +20.05 & +19.95 & +19.28 & +18.96 & +17.48 &  \\
Tecnick & CMIC '26 & HPCM & $-$2.33 & $-$2.30 & 7.48 & \textbf{$-$1.38} & \textbf{$-$1.28} & \textbf{$-$0.65} & \textbf{$-$0.35} & 0 &  \\
Tecnick & ChWDTA '26 & HPCM & $-$2.05 & $-$2.05 & 6.74 & \textbf{$-$1.13} & \textbf{$-$1.04} & \textbf{$-$0.41} & \textbf{$-$0.12} & 0 &  \\
Tecnick & GLIC '26 & HPCM & $-$1.00 & $-$0.91 & 3.20 & \textbf{$-$0.06} & 0 & 0 & 0 & 0 &  \\
Tecnick & SAAF '26 & HPCM & +0.96 & +0.95 & 3.31 & +0.09 & 0.00 & 0 & 0 & 0 & Bonf.\ changes at 3, 3.30 \\
CLIC~Pro & C3 '24 & MLIC++ & +11.77 & +10.57 & 21.29 & +9.08 & +8.93 & +7.91 & +7.42 & +5.16 &  \\
CLIC~Pro & CCA '24 & MLIC++ & +1.99 & +1.16 & 2.67 & 0 & 0 & 0 & 0 & 0 &  \\
CLIC~Pro & FTIC '24 & MLIC++ & +2.95 & +2.14 & 4.63 & +0.75 & +0.61 & 0 & 0 & 0 &  \\
CLIC~Pro & SegPIC '24 & MLIC++ & +5.06 & +4.46 & 9.17 & +3.00 & +2.85 & +1.84 & +1.37 & 0 &  \\
CLIC~Pro & DCAE '25 & MLIC++ & $-$3.42 & $-$3.35 & 7.03 & \textbf{$-$1.92} & \textbf{$-$1.78} & \textbf{$-$0.79} & \textbf{$-$0.33} & 0 &  \\
CLIC~Pro & HPCM '25 & MLIC++ & $-$6.39 & $-$6.32 & 12.92 & \textbf{$-$4.85} & \textbf{$-$4.71} & \textbf{$-$3.69} & \textbf{$-$3.22} & \textbf{$-$0.98} &  \\
CLIC~Pro & LALIC '25 & MLIC++ & $-$2.27 & $-$2.11 & 4.60 & \textbf{$-$0.74} & \textbf{$-$0.60} & 0 & 0 & 0 &  \\
CLIC~Pro & LotteryCodec '25 & MLIC++ & +7.68 & +6.72 & 13.60 & +5.24 & +5.09 & +4.07 & +3.59 & +1.33 &  \\
CLIC~Pro & MSINN-VR '25 & MLIC++ & +18.02 & +16.65 & 32.89 & +15.13 & +14.98 & +13.93 & +13.44 & +11.13 &  \\
CLIC~Pro & CMIC '26 & HPCM & +0.45 & +0.25 & 0.62 & 0 & 0 & 0 & 0 & 0 &  \\
CLIC~Pro & ChWDTA '26 & HPCM & +1.08 & +0.87 & 2.06 & 0 & 0 & 0 & 0 & 0 &  \\
CLIC~Pro & Cool-chic 5 '26 & HPCM & +7.17 & +6.67 & 13.55 & +5.20 & +5.05 & +4.03 & +3.55 & +1.30 &  \\
CLIC~Pro & GLIC '26 & HPCM & +0.43 & +0.17 & 0.44 & 0 & 0 & 0 & 0 & 0 &  \\
CLIC~Pro & SAAF '26 & HPCM & +2.98 & +2.49 & 5.31 & +1.08 & +0.94 & 0 & 0 & 0 &  \\
\bottomrule
\end{tabular}
\end{table}

\begin{table}[htbp]\centering\footnotesize
\caption{Coverage-preserving variants of the 44 frontier tests of 2024--2026: one interior operating point removed from either curve (PCHIP; Akima where five points remain), and $W_d$ shrunk by 0.1\,dB at either or both ends or widened by 0.1\,dB where both curves span it. Every variant spans its window, so all are valid. Range: smallest and largest direct BD-rate (\%) over the variants; $\sigma^\star$ from the smallest-magnitude value; $^\ast$: status differs from the frozen decision (last column), which uses the predeclared family of Tab.~\mref{tab:frontier} of the main paper.}\label{tab:s_interior}
\begin{tabular}{@{}lllrrcrll@{}}
\toprule Dataset & Method & vs. & Variants & Canonical & Range & $\sigma^\star$ & Status & Frozen \\ \midrule
Kodak & QARV & MLIC++ & 41 & +17.05 & [+16.74, +17.84] & 41.83 & robust & robust \\
Kodak & FTIC & MLIC++ & 21 & $-$0.11 & [$-$0.38, +0.57] & -- & unstable$^\ast$ & unresolved \\
Kodak & CCA & MLIC++ & 20 & +0.72 & [+0.45, +1.36] & 1.12 & unresolved & unresolved \\
Kodak & WeConvene & MLIC++ & 21 & +6.87 & [+6.42, +7.60] & 16.04 & robust & robust \\
Kodak & SegPIC & MLIC++ & 21 & +6.00 & [+5.12, +6.64] & 12.80 & robust & robust \\
Kodak & C3 & MLIC++ & 37 & +22.16 & [+21.82, +23.01] & 54.53 & robust & robust \\
Kodak & DCAE & FTIC & 21 & $-$2.29 & [$-$2.68, $-$2.10] & 5.24 & unresolved & unresolved \\
Kodak & LALIC & FTIC & 21 & $-$0.60 & [$-$1.03, $-$0.40] & 0.99 & unresolved & unresolved \\
Kodak & HPCM & FTIC & 21 & $-$5.69 & [$-$5.99, $-$5.50] & 13.75 & robust & robust \\
Kodak & MSINN-VR & FTIC & 33 & +14.78 & [+14.62, +15.01] & 36.55 & robust & robust \\
Kodak & ShiftLIC & FTIC & 25 & +16.75 & [+16.41, +16.99] & 41.02 & robust & robust \\
Kodak & LotteryCodec & FTIC & 21 & +19.14 & [+17.52, +19.44] & 43.79 & robust & robust \\
Kodak & S2CFormer & FTIC & 21 & $-$1.36 & [$-$1.58, $-$1.16] & 2.91 & unresolved & unresolved \\
Kodak & Cool-chic 5 & HPCM & 16 & +16.76 & [+14.96, +17.20] & 37.39 & robust & robust \\
Kodak & CMIC & HPCM & 23 & +2.24 & [+1.95, +2.57] & 4.87 & unresolved & unresolved \\
Kodak & GLIC & HPCM & 21 & +0.43 & [+0.24, +0.76] & 0.60 & unresolved & unresolved \\
Kodak & SAAF & HPCM & 21 & +3.23 & [+2.86, +3.56] & 7.15 & resolved & resolved \\
Kodak & ChWDTA & HPCM & 21 & +2.04 & [+1.77, +2.37] & 4.43 & unresolved & unresolved \\
Tecnick & QARV & MLIC++ & 40 & +17.14 & [+16.84, +17.44] & 85.94 & robust & robust \\
Tecnick & FTIC & MLIC++ & 20 & +1.56 & [+1.38, +1.73] & 7.06 & resolved & resolved \\
Tecnick & CCA & MLIC++ & 19 & +2.28 & [+1.98, +2.43] & 10.13 & resolved$^\ast$ & robust \\
Tecnick & WeConvene & MLIC++ & 20 & +9.65 & [+9.11, +9.97] & 46.50 & robust & robust \\
Tecnick & DCAE & MLIC++ & 20 & $-$4.36 & [$-$4.75, $-$4.10] & 20.94 & robust & robust \\
Tecnick & LALIC & MLIC++ & 20 & $-$2.14 & [$-$2.59, $-$1.70] & 8.65 & resolved & resolved \\
Tecnick & HPCM & MLIC++ & 20 & $-$5.95 & [$-$6.10, $-$5.86] & 29.92 & robust & robust \\
Tecnick & MSINN-VR & MLIC++ & 32 & +21.04 & [+20.74, +21.35] & 105.79 & robust & robust \\
Tecnick & CMIC & HPCM & 23 & $-$2.33 & [$-$2.42, $-$2.17] & 11.07 & robust & robust \\
Tecnick & GLIC & HPCM & 21 & $-$1.00 & [$-$1.09, $-$0.84] & 4.27 & unresolved & unresolved \\
Tecnick & SAAF & HPCM & 21 & +0.96 & [+0.58, +1.23] & 2.97 & unresolved & unresolved \\
Tecnick & ChWDTA & HPCM & 21 & $-$2.05 & [$-$2.25, $-$1.89] & 9.63 & resolved & resolved \\
CLIC Pro & FTIC & MLIC++ & 21 & +2.95 & [+2.72, +3.34] & 8.88 & resolved & resolved \\
CLIC Pro & CCA & MLIC++ & 21 & +1.99 & [+1.54, +2.38] & 5.04 & unresolved & unresolved \\
CLIC Pro & SegPIC & MLIC++ & 21 & +5.06 & [+4.27, +5.66] & 13.94 & robust & robust \\
CLIC Pro & C3 & MLIC++ & 37 & +11.77 & [+11.52, +12.20] & 37.63 & robust & robust \\
CLIC Pro & DCAE & MLIC++ & 21 & $-$3.42 & [$-$4.08, $-$3.03] & 9.90 & resolved$^\ast$ & robust \\
CLIC Pro & LALIC & MLIC++ & 21 & $-$2.27 & [$-$2.78, $-$1.85] & 6.03 & unresolved$^\ast$ & resolved \\
CLIC Pro & HPCM & MLIC++ & 21 & $-$6.39 & [$-$6.60, $-$6.03] & 19.70 & robust & robust \\
CLIC Pro & MSINN-VR & MLIC++ & 33 & +18.02 & [+17.71, +18.43] & 57.85 & robust & robust \\
CLIC Pro & LotteryCodec & MLIC++ & 16 & +7.68 & [+5.62, +8.19] & 18.37 & robust & robust \\
CLIC Pro & Cool-chic 5 & HPCM & 16 & +7.17 & [+6.20, +7.41] & 20.27 & robust & robust \\
CLIC Pro & CMIC & HPCM & 23 & +0.45 & [+0.29, +0.66] & 0.94 & unresolved & unresolved \\
CLIC Pro & GLIC & HPCM & 21 & +0.43 & [+0.29, +0.64] & 0.93 & unresolved & unresolved \\
CLIC Pro & SAAF & HPCM & 21 & +2.98 & [+2.46, +3.29] & 8.03 & resolved & resolved \\
CLIC Pro & ChWDTA & HPCM & 23 & +1.08 & [+0.83, +1.41] & 2.71 & unresolved & unresolved \\
\bottomrule
\end{tabular}
\end{table}

\section{Image-to-image variability and evidence}
\label{sec:s_boot}
\Cref{tab:s_strata} summarizes the heterogeneity of every released pair of codecs with per-image results.
It grows with the size of an effect: cross-architecture pairs with effects below 12\% have $\sigma_{\rm pair}\le\sigmaCrossSmallMax$, whereas larger steps reach \sigmaCrossMax\ (Fig.~\mref{fig:sampling}).

\begin{table}[htbp]
\centering
\caption{Heterogeneity of released per-image pairs, by class ($\sigma_{\rm pair}$: bootstrap SD at the full image count times $\sqrt n$). Recent pairs: Cool-chic~5.0 and ShiftLIC on Kodak (one-point-inset window). $\sigma_{\rm ref}$ is the median of the reference pairs' slopes, so the reference row differs slightly from it.}
\label{tab:s_strata}
\small
\begin{tabular}{@{}P{8.2cm}rrcrr@{}}
\toprule
Stratum & Pairs & Median $\sigma_{\text{pair}}$ & Range & $\le\sigma_{\text{ref}}$ (\%) & $\le\sigma_{\text{hi}}$ (\%) \\
\midrule
Codec vs.\ VTM anchor & 13 & 12.76 & 6.06--43.16 & 8 & 31 \\
LIC vs.\ LIC, held out & 6 & 8.65 & 5.71--13.55 & 17 & 67 \\
LIC vs.\ hyperprior model & 5 & 13.81 & 11.77--18.79 & 0 & 0 \\
Reference pairs (define $\sigma_{\text{ref}}$) & 4 & 6.55 & 5.53--6.63 & 25 & 100 \\
Within family (Cool-chic versions) & 49 & 5.17 & 0.05--10.91 & 71 & 98 \\
Recent, cross-architecture (Cool-chic 5.0 vs.\ ShiftLIC) & 3 & 6.50 & 5.92--7.57 & 33 & 100 \\
Recent, within ShiftLIC & 3 & 5.35 & 2.04--6.81 & 67 & 100 \\
\bottomrule
\end{tabular}

\end{table}

\subsection{Exact-null calibration on released pairs}
\label{sec:s_null}
For each of the \nNullPairs\ released per-image pairs of two learned codecs, let $\beta$ be the full-sample BD-rate of codec $B$ against $A$ on the pair's one-point-inset window. Dividing every released rate of $B$ by $1+\beta$ creates a partner $B_0$ with exactly zero effect on the full sample (BD-rate is equivariant to a common rate factor), while every resampled test set keeps the real pair's fluctuation; its scale is $\sigma_0=\sigma_{\rm pair}/(1+\beta)$.
We draw \nullR\ test sets per pair, of 24 images and of the released image count, and resolve a draw if $|\hat\Delta|>1.96\,\sigma/\sqrt n$ (\cref{tab:s_null}).
False resolutions stay at or below 5\% when $\sigma_0$ does not exceed the assumed $\sigma$ and rise beyond it; transporting a scale to recent pairs therefore assumes that they are no more heterogeneous than these, which only matched per-image results can confirm (\cref{sec:s_r19_scale}).
The matched percentile bootstrap, which needs no assumed scale, resolves \nullNonRefBoot\% of the nulls (pooled), somewhat above its nominal 5\%.

\begin{table}[htbp]
\centering
\caption{False resolutions on exact nulls (\%, Wilson 95\% interval), pooled over the pairs of each kind. Rule: $|\hat\Delta|>1.96\,\sigma/\sqrt n$ with the frozen $\sigma_{\rm ref}=\sigmaRef$ or $\sigma_{\rm hi}=\sigmaHi$, or with the null's own $\sigma_0$; matched bootstrap: the percentile interval excludes zero.}
\label{tab:s_null}
\footnotesize
\begin{tabular}{@{}lrcccccrr@{}}
\toprule
 & & & \multicolumn{4}{c}{False resolutions, released image count (\%)} & \multicolumn{2}{c}{$n=24$ (\%)} \\
\cmidrule(lr){4-7}\cmidrule(lr){8-9}
Pairs (exact null) & $N$ & $\sigma_0$ & rule, $\sigma_{\text{ref}}$ & rule, $\sigma_{\text{hi}}$ & rule, $\sigma_0$ & matched bootstrap & $\sigma_{\text{ref}}$ & $\sigma_{\text{hi}}$ \\
\midrule
Cross-architecture, held out & 6 & 7.0--16.7 & 24.4 [22.7, 26.2] & 7.9 [6.9, 9.0] & 5.2 & 9.5 [7.4, 12.1] & 23.0 & 6.9 \\
LIC vs.\ hyperprior & 5 & 17.2--33.7 & 59.8 [57.6, 61.9] & 35.6 [33.6, 37.8] & 5.8 & 7.4 [5.4, 10.0] & 60.4 & 36.5 \\
Within family (Cool-chic) & 13 & 0.0--8.1 & 2.1 [1.7, 2.5] & 0.0 [0.0, 0.1] & 5.2 & 7.3 [6.0, 8.9] & 1.8 & 0.0 \\
\midrule
All held out, $\sigma_0\le\sigma_{\text{hi}}$ & 17 & 0.0--9.2 & 4.9 [4.4, 5.4] & 0.5 [0.4, 0.7] & 5.2 & 8.4 [7.2, 9.8] & 4.2 & 0.5 \\
All held out, $\sigma_0>\sigma_{\text{hi}}$ & 7 & 16.6--33.7 & 55.6 [53.8, 57.5] & 31.0 [29.3, 32.8] & 5.5 & 6.6 [5.0, 8.7] & 56.0 & 30.9 \\
\midrule
Reference pairs (set $\sigma_{\text{ref}}$) & 4 & 5.6--7.3 & 7.6 [6.4, 9.0] & 0.4 [0.2, 0.9] & 5.8 & 12.0 [9.2, 15.6] & 7.1 & 0.4 \\
\bottomrule
\end{tabular}

\end{table}

\subsection{Scales measured on the matched releases}
\label{sec:s_r19_scale}
For the six matched 2025--26 releases, the \rnSigN\ pairwise $\sigma_{\rm pair}$ values range from \rnSigMinV\ to \rnSigMaxV\ (median \rnSigMedV, 90th percentile \rnSigPninety), all below $\sigma_{\rm ref}$.
To test the scale rules where the answer is known, we drew \rnCalOuter\ test sets of \rnCalM\ images (the size of Kodak) for each of the \rnCalPairs\ frontier pairs on Tecnick and CLIC~Pro and built exact nulls as above.
Fixed-scale rules are evaluated on every test set, as they are used for aggregate curves, and, for comparison, on those whose inner bootstrap passes the coverage gate (only \rnCalGatePass\% do), which flatters them.
A scale of 3, close to the median, is anti-conservative for an unmeasured pair; $\hat\sigma_{90}$ errs on the conservative side (\cref{tab:s_r19_calibration}), and we use it in Tab.~\mref{tab:frontier}.

\paragraph{Does a borrowed scale transfer to a new codec?}
The \hoPairs\ comparisons reuse six codecs, so they are not independent, and a scale estimated from them could fit these codecs but not a new one.
We therefore held out each codec in turn: for each dataset, the scale is estimated from the 10 pairs that do not involve the held-out codec and applied to the 5 pairs that do, and the conditional decision ($|\Delta|>1.96\,\sigma/\sqrt{n_d}$) is compared with the observed paired decision.
Every pair is held out twice, once with each of its codecs (\hoN\ evaluations).
The 90th percentile of the other pairs makes \hoPninetyFalse\ false and \hoPninetyMiss\ missed resolutions, whereas $\sigma=3$ makes \hoThreeFalse\ false resolutions (\cref{tab:s_holdout}).
The borrowed 90th percentile covers \hoPninetyCov\% of the held-out scales, somewhat below the \hoExchLo--\hoExchHi\% expected for exchangeable codecs; coverage is lowest for \hoLowCodec\ (\hoLowCov\ of \hoLowN), yet each of its held-out comparisons is decided as its paired interval decides it, because its differences are also the largest (median $|\Delta|$ \hoLowMedEff\ against \hoRestMedEff\ points).
A new codec can therefore be more heterogeneous than the codecs a scale was estimated from; only its per-image records show whether it is.

\begin{table}[htbp]\centering\small
\caption{Transfer of scale rules to held-out codecs and held-out pairs among the six 2025--26 releases (number of held-out evaluations in parentheses). Codec held out: scale from the 10 pairs of the dataset without the held-out codec, applied to its 5 pairs; pair held out: scale from the other 14 pairs of the dataset. False: resolved under the scale model but not by the paired interval; missed: the reverse. Covered: the held-out pair's measured $\sigma_{\rm pair}$ does not exceed the scale used; with the 90th percentile, the pairs covered of 15 per held-out codec are \hoByCodec.}\label{tab:s_holdout}
\begin{tabular}{@{}l c r r r @{\hspace{14pt}} r r@{}}
\toprule
 & & \multicolumn{3}{c}{Codec held out (90)} & \multicolumn{2}{c}{Pair held out (45)} \\
\cmidrule(lr){3-5}\cmidrule(l){6-7}
Scale & range used & covered (\%) & false & missed & false & missed \\
\midrule
median of the other pairs & 1.82--4.65 & 51 & 1 & 6 & 1 & 3 \\
90th percentile of the other pairs & 2.49--5.55 & 80 & 0 & 14 & 0 & 8 \\
maximum of the other pairs & 2.60--6.05 & 90 & 0 & 17 & 0 & 9 \\
$\sigma=3$ & 3.00 & 47 & 4 & 4 & 2 & 2 \\
$\sigma_{\rm ref}$ & 6.35 & 100 & 0 & 28 & 0 & 14 \\
$\sigma_{\rm hi}$ & 10.91 & 100 & 0 & 48 & 0 & 24 \\
\bottomrule
\end{tabular}

\end{table}

\begin{table}[htbp]\centering\small
\caption{Coverage and false resolutions of interval rules on \rnCalM-image test sets of Tecnick and CLIC~Pro (\rnCalPairs\ pairs, \rnCalOuter\ test sets each; exact nulls as in \cref{sec:s_null}).}\label{tab:s_r19_calibration}
\begin{tabular}{@{}l rr rr@{}}
\toprule
 & \multicolumn{2}{c}{Coverage of the full-set effect (\%)} & \multicolumn{2}{c}{False resolutions on exact nulls (\%)} \\
\cmidrule(lr){2-3}\cmidrule(lr){4-5}
Rule & gate passes & every subsample & gate passes & every subsample \\
\midrule
Percentile bootstrap interval & 96.0 & -- & 4.0 & -- \\
Normal, bootstrap SD & 94.3 & -- & 5.7 & -- \\
Normal, pair's own full-set $\sigma_{\rm pair}$ & 99.0 & 95.7 & 0.9 & 3.9 \\
Normal, $\sigma=3$ & 90.6 & 87.1 & 8.5 & 12.0 \\
Normal, measured median 3.30 & 93.4 & 90.0 & 6.2 & 9.3 \\
Normal, measured 90th percentile 5.37 & 99.7 & 98.2 & 0.2 & 1.6 \\
Normal, $\sigma_{\rm ref}=6.35$ & 100.0 & 99.3 & 0.0 & 0.6 \\
Normal, $\sigma_{\rm hi}=10.91$ & 100.0 & 100.0 & 0.0 & 0.0 \\
\bottomrule
\end{tabular}

\end{table}

\subsection{Observed evidence and further controls}
\label{sec:s_evidence}
\begin{figure}[htbp]
  \centering
  \includegraphics[width=.8\linewidth]{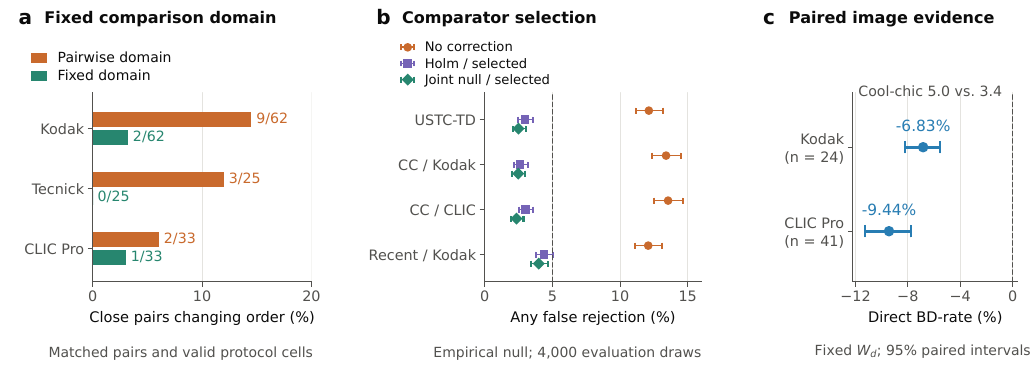}
  \caption{\textbf{Three controls of the evidence rules.}
  \textbf{a},~Changing only the quality-domain rule on matched valid protocol cells reduces close-pair order reversals.
  \textbf{b},~Empirical global-null controls select the best codec on each draw, then test the others against it (bars: 95\% Monte Carlo intervals; dashed: 5\%; CC: Cool-chic; Recent: Cool-chic/ShiftLIC).
  \textbf{c},~Cool-chic~5.0 vs.~3.4: observed paired-bootstrap intervals on the frozen windows.}
  \label{fig:s_evidence_controls}
\end{figure}

Where per-image records exist, the tool decides from them: images are resampled with replacement, the same images for both codecs, and the direct BD-rate is recomputed on the frozen $W_d$; with more than 1\% of replicates failing coverage, it makes no decision.
Before any statistic, the records must name their images and operating points, cover the same images at every point, and reproduce every released point.
Cool-chic~5.0 and three ShiftLIC models release such records on Kodak; on $W_d$ all their pairs exceed the 1\% limit and abstain, which shows that the rule does not force a decision.
\Cref{fig:s_evidence_controls} summarizes three further controls, each on its own cohort.
(a) To isolate the integration domain from the set of valid protocol cells, we match each method pair, VTM run, interpolant, and point rule and keep a cell only if it is valid under both domains; the fixed window removes most order flips (\cref{tab:s_component}).
(b) On four empirical families with released per-image records (six USTC-TD codecs, the Cool-chic versions on Kodak and on CLIC~Pro, and Cool-chic~5.0 with ShiftLIC), rates are rescaled to an exact global null; selecting the best codec on each test set and testing the others against it falsely resolves \evRawMin--\evRawMax\% of draws without adjustment and \evSelHolmMin--\evSelHolmMax\% with Holm.
(c) Cool-chic~5.0 against 3.4 is resolved from per-image records on both frozen windows without coverage failures.

\begin{table}[htbp]\centering\small
\caption{Matched-cell domain ablation. A method pair enters both arms on exactly the same valid anchor, interpolant and point-rule cells. Counts are pairwise-domain / fixed-domain order changes. Close means a canonical-score difference below two points. The final column counts method--dataset cells documenting both rate accounting and aggregation; documentation is not proof of every measurement detail.}\label{tab:s_component}
\begin{tabular}{@{}lrrrrr@{}}
\toprule Dataset & All pairs & Flips (all) & Close pairs & Flips (close) & Documented \\ \midrule
Kodak & 496 & 22 / 2 & 62 & 9 / 2 & 3 \\
Tecnick & 190 & 6 / 0 & 25 & 3 / 0 & 2 \\
CLIC Pro & 276 & 4 / 1 & 33 & 2 / 1 & 1 \\
\bottomrule
\end{tabular}
\end{table}

\section{Matched measurement of public checkpoints}
\label{sec:s_batches}
Public checkpoints were measured in four batches of cluster jobs (\cref{tab:s_batches}).
Each run encodes every image of a test set with one checkpoint, writes the entropy-coded streams to a self-describing file, reads the file back, and decodes it.
Every record stores the entropy-payload and file bytes, the entropy model's rate estimate, PSNR before and after 8-bit rounding, and encoding and decoding times; batch~4 adds MS-SSIM.
The primary comparison counts payload bytes over the original image area and computes 8-bit RGB PSNR on the reconstruction cropped to the original size, with each codec's own padding; the other quantities identify how released curves were measured.
All batches share the stream container and metric code, and every run specification records the SHA-256 of the checkpoint, the image pixels, and the runner code.
\Cref{tab:s_releases} lists the values drawn in Fig.~\mref{fig:matched} for Kodak, Tecnick, and CLIC~Pro, together with the two padding controls and the measurement profiles of Tab.~\mref{tab:releases}.

\begin{table}[htbp]
\centering\small
\caption{The four measurement batches (NVIDIA H100, MIG slices; GPU-hours are job wall times). Native: each release's own padding; shared: HPCM-Large's padding for both codecs; replicate: right/bottom replication to multiples of 256 for both.}
\label{tab:s_batches}
\begin{tabular}{@{}l P{3.2cm} P{2.4cm} P{4.0cm} l r r@{}}
\toprule
Batch & Releases & Test sets & Paddings and content & Slice & Records & GPU-h \\
\midrule
1 & HPCM-Large, CMIC\_AuxT & Kodak, Tecnick & native, shared & 2g.20gb & \rmRecords & \rmGpuHours \\
2 & HPCM-Large, CMIC\_AuxT & Tecnick, CLIC~Pro & Tecnick: replicate; CLIC~Pro: native, shared, replicate & 3g.40gb & \rmRecordsNew & \rmGpuHoursNew \\
3 & GLIC, SAAF, DCAE, LALIC, TCM-S & Kodak, Tecnick, CLIC~Pro & native, shared, replicate & 3g.40gb & \rrRecords & \rrGpuHours \\
4 & MLIC++, TCM-L, and the releases of batches 1--3 except LALIC & Kodak, Tecnick, CLIC~Pro, DIV2K & native; MS-SSIM; timing on all test sets & 3g.40gb & \cfRecords & \cfGpuHours \\
\midrule
All & eight releases and TCM-L & four test sets & & & \cfRecordsAll & \cfGpuHoursAll \\
\bottomrule
\end{tabular}
\end{table}

\begin{table}[htbp]
\centering
\caption{Measurement profiles and matched comparisons of the eight releases (the values of Fig.~\mref{fig:matched}a--c). Left: the best-fitting tested measurement rules; no tested rule reproduces TCM-S, and the public MLIC++ weights outperform its released curves. Right: direct BD-rate (\%) against HPCM-Large; positive values mean more rate. Rel.: released curves. Own: common rate and PSNR definitions with each codec's native padding. Sh./Rep.: shared HPCM/replicate padding (--: not measured). Bold: the paired 95\% interval on $W_d$ excludes zero, from the image bootstrap or, marked $^{\rm J}$, from the delete-one jackknife where more than 1\% of bootstrap replicates fail coverage. $^a$No release states all rules; LALIC states actual-bitstream rates. $^b$Bytes on Kodak. $^c$See \cref{sec:s_cf}. $^d$Author-reported parameters (millions); --: no author count for the tested variant. $^e$Zero or replicate padding, centered (c) or right/bottom (rb), to a multiple of 64--256. $^f$Median decoding seconds per CLIC~Pro image on one H100 MIG 3g.40gb slice; entropy coding on CPU.}
\label{tab:s_releases}
\scriptsize
\setlength{\tabcolsep}{2.6pt}
\begin{tabular}{@{}l r l l l r rr rrc rrc@{}}
\toprule
 & & \multicolumn{3}{c}{Best-fitting tested rules$^a$} & & \multicolumn{2}{c}{Kodak} & \multicolumn{3}{c}{Tecnick} & \multicolumn{3}{c}{CLIC~Pro} \\
\cmidrule(lr){3-5}\cmidrule(lr){7-8}\cmidrule(lr){9-11}\cmidrule(lr){12-14}
Release & M$^d$ & Bits & PSNR & Padding$^e$ & Dec.$^f$ & Rel. & Own & Rel. & Own & Sh./Rep. & Rel. & Own & Sh./Rep. \\
\midrule
HPCM-Large~\cite{li2025hpcm} '25 & 89.71 & written & 8-bit & zero c 256 & 0.85 & \multicolumn{8}{c}{reference of every comparison} \\
CMIC\_AuxT~\cite{chen2026cmic} '26 & -- & estimate$^b$ & unrounded & repl. rb 64 & 0.96 & +2.24 & \textbf{+2.68} & $-$2.33 & \textbf{$-$1.28}$^{\rm J}$ & \textbf{+0.95}$^{\rm J}$/\textbf{+1.24}$^{\rm J}$ & +0.45 & \textbf{+1.57} & \textbf{+2.34}/\textbf{+3.09} \\
GLIC~\cite{chen2026glic} '26 & 67.20 & estimate & unrounded & repl. rb 128 & 1.07 & +0.43 & \textbf{+2.29} & $-$1.00 & \textbf{+1.43}$^{\rm J}$ & +1.01$^{\rm J}$/\textbf{+1.58}$^{\rm J}$ & +0.43 & \textbf{+2.81} & \textbf{+2.70}/\textbf{+3.42} \\
SAAF~\cite{ma2026saaf} '26 & 123 & estimate & unrounded & zero c 128 & 0.71 & +3.23 & \textbf{+3.90} & +0.96 & \textbf{+1.67}$^{\rm J}$ & \textbf{+1.67}$^{\rm J}$/\textbf{+2.46}$^{\rm J}$ & +2.98 & \textbf{+3.77} & \textbf{+4.60}/\textbf{+6.49} \\
DCAE~\cite{lu2025dcae} '25 & 119.2 & estimate & unrounded & zero c 128 & 0.71 & +3.60 & \textbf{+4.45} & +1.69 & \textbf{+2.40}$^{\rm J}$ & \textbf{+2.40}$^{\rm J}$/\textbf{+3.19}$^{\rm J}$ & +3.18 & \textbf{+4.02} & \textbf{+4.65}/\textbf{+6.79} \\
LALIC~\cite{feng2025lalic} '25 & 63.24 & written & unrounded & zero c 128 & 0.83 & +5.40 & \textbf{+5.71} & +4.05 & \textbf{+4.61}$^{\rm J}$ & \textbf{+4.61}$^{\rm J}$/\textbf{+5.08}$^{\rm J}$ & +4.41 & \textbf{+4.94} & \textbf{+5.49}/\textbf{+6.47} \\
MLIC++~\cite{jiang2023mlicppws} '23 & -- & \multicolumn{2}{l}{none matches$^c$} & zero rb 64 & 0.74 & +6.15 & \textbf{+4.85} & +6.33 & \textbf{+4.39}$^{\rm J}$ & -- & +6.83 & \textbf{+4.96}$^{\rm J}$ & -- \\
TCM-S~\cite{liu2023tcm} '23 & 44.96 & \multicolumn{2}{l}{none matches$^c$} & zero c 128 & 0.51 & +15.99 & \textbf{+16.33}$^{\rm J}$ & +18.13 & \textbf{+18.34}$^{\rm J}$ & \textbf{+18.34}$^{\rm J}$/\textbf{+18.13}$^{\rm J}$ & +16.04 & \textbf{+15.53}$^{\rm J}$ & \textbf{+16.21}$^{\rm J}$/\textbf{+18.08}$^{\rm J}$ \\
\bottomrule
\end{tabular}

\end{table}

\subsection{Evaluation code and measurement rules}
\label{sec:s_code_identity}
A verified curve does not identify the checkpoint or the convention that produced it, so we read the released test scripts at the pinned commits (\cref{tab:s_code_identity}) and measured the public weights under every rule.
HPCM's test script rounds reconstructed pixels to 8 bits before PSNR and pads with zeros, centered, to a multiple of 256; CMIC's evaluates unrounded pixels and replicates at the right and bottom to a multiple of 64.

\begin{table}[htbp]
 \centering\small
 \caption{Pinned released evaluation code and checkpoint descriptions of HPCM and CMIC. These facts concern executable releases; which conventions produced the released curves is identified by measurement (\cref{tab:s_rm_ident,tab:s_rm_ident_clic}).}
 \label{tab:s_code_identity}
 \begin{tabular}{@{}lp{5.3cm}p{5.8cm}@{}}
\toprule Item & HPCM released evaluation & CMIC released evaluation \\ \midrule
PSNR & Clamp and round to 8-bit RGB & Floating RGB after clamp \\
Padding & Centered zero padding, multiple of 256 & Right/bottom replication, default multiple of 64 \\
Public model & HPCM-Large checkpoint links & README labels checkpoints CMIC\_AuxT \\
Rates & Actual compress/decompress payload & Forward likelihood and actual payload reported separately \\
\bottomrule
\end{tabular}

\end{table}
Identification is by exclusion: the rule shown reproduces all released points within the rounding of their printed decimals, and every alternative rate rule (payload, rate estimate, file), PSNR definition (8-bit, unrounded), padding, or PSNR of the mean MSE misses by the margins listed (\cref{tab:s_rm_ident,tab:s_rm_ident_clic,tab:s_r16_ident}).
HPCM-Large's curves are reproduced by payload bytes and 8-bit PSNR; CMIC's Tecnick and CLIC~Pro curves by the rate estimate and unrounded PSNR, and its Kodak curve, printed with fewer decimals, by payload bytes and unrounded PSNR.
LALIC counts written bytes with unrounded PSNR, and GLIC, SAAF, and DCAE report the rate estimate with unrounded PSNR.
No tested rule reproduces TCM-S or MLIC++ (\cref{sec:s_cf}).

\begin{table}[htbp]
\centering\scriptsize\setlength{\tabcolsep}{3.5pt}
\caption{Measurement rule behind each released HPCM-Large and CMIC curve on Kodak and Tecnick (each release's own padding): largest error over the released points for the identified rule and for any other rule; BD: re-measured against released curve on $W_d$; scorer class: the measurement gate applied with the identified conventions stated.}
\label{tab:s_rm_ident}
\begin{tabular}{@{}l l l r r r r r r l@{}}
\toprule
 & \multicolumn{2}{c}{Identified} & \multicolumn{2}{c}{Largest error} & \multicolumn{3}{c}{Any other rule misses by $\ge$} & BD vs. & Scorer class \\
\cmidrule(lr){2-3}\cmidrule(lr){4-5}\cmidrule(lr){6-8}
Released curve & rate & PSNR & dB & units & rate (units) & PSNR (dB) & mean-MSE (dB) & released & if stated \\
\midrule
HPCM-Large, Kodak & payload & 8-bit & 0.0002 & 1.07 & 7.6 & 0.034 & 0.55 & +0.01\% & documented \\
HPCM-Large, Tecnick & payload & 8-bit & 0.0001 & 0.47 & 4.8 & 0.042 & 0.89 & $-$0.01\% & documented \\
CMIC, Kodak & payload & floating & 0.0048 & 0.73 & 3.3 & 0.036 & 0.53 & +0.14\% & provisional \\
CMIC, Tecnick & likelihood & floating & 0.0019 & 0.39 & 22.2 & 0.045 & 0.88 & 0.00\% & separate rate class \\
\bottomrule
\end{tabular}

\end{table}
\begin{table}[htbp]
\centering\scriptsize\setlength{\tabcolsep}{3.5pt}
\caption{The same identification for the released CLIC~Pro curves.}
\label{tab:s_rm_ident_clic}
\begin{tabular}{@{}l l l r r r r r r l@{}}
\toprule
 & \multicolumn{2}{c}{Identified} & \multicolumn{2}{c}{Largest error} & \multicolumn{3}{c}{Any other rule misses by $\ge$} & BD vs. & Scorer class \\
\cmidrule(lr){2-3}\cmidrule(lr){4-5}\cmidrule(lr){6-8}
Released curve & rate & PSNR & dB & units & rate (units) & PSNR (dB) & mean-MSE (dB) & released & if stated \\
\midrule
HPCM-Large, CLIC Pro & payload & 8-bit & 0.0002 & 0.23 & 5.6 & 0.044 & 1.07 & +0.01\% & documented \\
CMIC, CLIC Pro & likelihood & floating & 0.0009 & 0.63 & 21.0 & 0.047 & 1.05 & +0.02\% & separate rate class \\
\bottomrule
\end{tabular}

\end{table}
\begin{table}[htbp]\centering
\caption{Identification of the measurement rule behind each released curve of the five releases of batch~3 (release padding). Largest error over the six released points as a relative rate error and in dB, and in units of the last released decimal; ``next rule'': the smallest largest error of every other rate rule (relative) and of the other PSNR rounding (dB); BD: re-measured curve under the identified rule against the released curve on $W_d$ (\%). For TCM-S, which no rule reproduces, the rule with the smallest error in units of the last released decimal is listed.}\label{tab:s_r16_ident}
\begin{tabular}{@{}llllrrrrrr@{}}
\toprule
Release & Data & Bits & PSNR & Rate err. & PSNR err. & Units & Next rate & Next PSNR & BD \\
\midrule
GLIC & Kodak & estimate & unrounded & 1.307\% & 0.0180 & 21.9/179.8 & 5.83\% & 0.034 & +0.16 \\
GLIC & Tecnick & estimate & unrounded & 1.214\% & 0.0140 & 17.2/139.9 & 6.34\% & 0.045 & +0.20 \\
GLIC & CLIC Pro & estimate & unrounded & 1.179\% & 0.0234 & 16.0/233.5 & 6.91\% & 0.046 & +0.14 \\
SAAF & Kodak & estimate & unrounded & 0.003\% & 0.0002 & 0.8/16.9 & 0.47\% & 0.036 & 0.00 \\
SAAF & Tecnick & estimate & unrounded & 0.003\% & 0.0001 & 0.6/6.5 & 0.32\% & 0.045 & 0.00 \\
SAAF & CLIC Pro & estimate & unrounded & 0.004\% & 0.0001 & 0.4/10.6 & 0.36\% & 0.047 & 0.00 \\
DCAE & Kodak & estimate & unrounded & 0.412\% & 0.0038 & 0.4/0.4 & 1.08\% & 0.038 & +0.11 \\
DCAE & Tecnick & estimate & unrounded & 0.354\% & 0.0129 & 0.7/1.3 & 0.56\% & 0.056 & $-$0.16 \\
DCAE & CLIC Pro & estimate & unrounded & 0.533\% & 0.0094 & 0.9/0.9 & 0.61\% & 0.044 & $-$0.07 \\
LALIC & Kodak & written & unrounded & 0.023\% & 0.0033 & 136.0/3298.0 & 1.55\% & 0.034 & 0.00 \\
LALIC & Tecnick & written & unrounded & 0.011\% & 0.0016 & 18.2/1571.4 & 0.73\% & 0.043 & 0.00 \\
LALIC & CLIC Pro & written & unrounded & 0.015\% & 0.0015 & 39.6/1467.7 & 0.73\% & 0.045 & $-$0.01 \\
TCM-S & Kodak & estimate & 8-bit & 3.619\% & 0.2784 & 22.7/27.8 & 3.99\% & 0.286 & $-$0.09 \\
TCM-S & Tecnick & estimate & 8-bit & 4.203\% & 0.1846 & 17.9/18.5 & 4.50\% & 0.197 & $-$0.13 \\
TCM-S & CLIC Pro & estimate & 8-bit & 3.556\% & 0.2209 & 12.1/22.1 & 3.34\% & 0.233 & $-$0.77 \\
\bottomrule
\end{tabular}
\end{table}

\subsection{The five further releases}
\label{sec:s_r16}
We measured every further published release in the record whose public code and at least six public MSE weights run in our environment: GLIC~\cite{chen2026glic}, SAAF~\cite{ma2026saaf}, DCAE~\cite{lu2025dcae}, LALIC~\cite{feng2025lalic}, and the small TCM~\cite{liu2023tcm}, the only TCM model with a public weight for every rate.
Not measured: FTIC~\cite{li2024ftic}, whose Python entropy coder is too slow for the full test sets and whose batch-4 job failed; CCA~\cite{han2024cca}, with four public weights; and the preprint ChWDTA~\cite{fu2026chwdta}.
Each release runs at the commit of its curve with its own model constructor, weight-loading code, and padding; the \rrCkpts\ weight files are checked by SHA-256, every loaded tensor is compared with its file, and every software adjustment (renamed parameters, a PyTorch dispatch call in LALIC's CUDA operator, an initializer import) is recorded in the run specification and changes no evaluated operator.
Repeated Kodak runs are bitwise identical, and all \rrStreams\ encoded files were re-read.
Because all curves span $W_d$ under one interpolant, the released direct BD-rate of each release against HPCM-Large splits exactly into measurement steps (\cref{tab:s_r16_attr}): the rate estimate and unrounded PSNR together favor GLIC, SAAF, and DCAE by \atEstConvMin--\atEstConvMax\ points, and unrounded PSNR alone favors LALIC by \atLalicConvMin--\atLalicConvMax\ points.

\begin{table}[htbp]\centering
\caption{Exact decomposition of the released direct BD-rate of each release against HPCM-Large (\%) through the re-measured curves on $W_d$ (Property~\mref{prop:fixed}); each step is the BD-rate it would cause alone, and the steps multiply to the released value. Residuals: released against re-measured curve of the codec and of HPCM-Large; rate and PSNR: the codec's rule replaced by payload bytes and 8-bit PSNR.}\label{tab:s_r16_attr}
\begin{tabular}{@{}llrrrrrr@{}}
\toprule
Release & Data & Released & Codec residual & Rate rule & PSNR rule & Matched (own) & HPCM residual \\
\midrule
GLIC & Kodak & +0.43 & $-$0.16 & $-$1.37 & $-$0.30 & +2.29 & +0.01 \\
GLIC & Tecnick & $-$1.00 & $-$0.20 & $-$1.68 & $-$0.52 & +1.43 & $-$0.01 \\
GLIC & CLIC Pro & +0.43 & $-$0.14 & $-$1.67 & $-$0.52 & +2.81 & +0.01 \\
SAAF & Kodak & +3.23 & 0.00 & $-$0.36 & $-$0.31 & +3.90 & +0.01 \\
SAAF & Tecnick & +0.96 & 0.00 & $-$0.17 & $-$0.52 & +1.67 & $-$0.01 \\
SAAF & CLIC Pro & +2.98 & 0.00 & $-$0.25 & $-$0.52 & +3.77 & +0.01 \\
DCAE & Kodak & +3.60 & $-$0.11 & $-$0.41 & $-$0.31 & +4.45 & +0.01 \\
DCAE & Tecnick & +1.69 & +0.16 & $-$0.31 & $-$0.54 & +2.40 & $-$0.01 \\
DCAE & CLIC Pro & +3.18 & +0.07 & $-$0.34 & $-$0.55 & +4.02 & +0.01 \\
LALIC & Kodak & +5.40 & 0.00 & 0.00 & $-$0.30 & +5.71 & +0.01 \\
LALIC & Tecnick & +4.05 & 0.00 & 0.00 & $-$0.52 & +4.61 & $-$0.01 \\
LALIC & CLIC Pro & +4.41 & +0.01 & 0.00 & $-$0.53 & +4.94 & +0.01 \\
TCM-S & Kodak & +15.99 & +0.09 & $-$0.39 & 0.00 & +16.33 & +0.01 \\
TCM-S & Tecnick & +18.13 & +0.13 & $-$0.30 & 0.00 & +18.34 & $-$0.01 \\
TCM-S & CLIC Pro & +16.04 & +0.78 & $-$0.34 & 0.00 & +15.53 & +0.01 \\
\bottomrule
\end{tabular}
\end{table}

\subsection{All pairs, joint intervals, and rate regimes}
\label{sec:s_r19}
\paragraph{When the bootstrap abstains.}
The paired bootstrap decides only if at most 1\% of its replicates fail to span $W_d$ for either codec.
It abstains in \rnPairGateAbst\ of the \rnPairCases\ comparisons of \cref{tab:s_r19_pairs}: on Tecnick, HPCM-Large's highest mean point lies only \rnTecMargin\,dB above the upper window edge, so \rnTecFailObs\% of the replicates fail for every pair with it, and on Kodak and CLIC~Pro, TCM-S's lowest point lies near the lower edge.
There we use the delete-one jackknife~\cite{efron1981jackknife}: the BD-rate on the $n$ leave-one-out mean curves gives $\widehat{\rm SE}^2=\frac{n-1}{n}\sum_i(\hat\theta_{(i)}-\bar\theta)^2$ and a $t_{n-1}$ interval, with no decision if any leave-one-out curve fails to span $W_d$.
Where the bootstrap passes its gate, the two standard errors agree closely, and the jackknife resolves all \rnPairAbstJkResolved\ comparisons in which the bootstrap abstains.

\paragraph{Simultaneous intervals.}
The 15 contrasts among the six 2025--26 releases share the same resampled images, so a simultaneous interval uses the 95\% quantile of $\max_k|\hat\theta^\ast_k-\hat\theta_k|/\mathrm{SD}_k$~\cite{westfall1993resampling}; on Tecnick, the jackknife version with a multivariate $t$ quantile is primary (\cref{tab:s_r19_joint}).

\paragraph{Rate regimes.}
A band next to $W_d$ is fixed from where the curves end, not from any effect: the low band starts at the lowest 0.1-dB value that each release reaches in at least 99.5\% of the resamples, and bands narrower than 0.3\,dB are not evaluated.
The public checkpoints end only \rnLowGapMin--\rnLowGapMax\,dB below the windows, so only the Kodak high and Tecnick low bands qualify (\cref{tab:s_r19_regimes}); on the latter, \rnBandReversals\ orderings that are resolved in both windows reverse.

\begin{table}[htbp]\centering\small
\caption{Paired-bootstrap comparisons on the two admissible rate bands ($B=10{,}000$; every comparison passes the coverage gate). Bold: excludes zero.}\label{tab:s_r19_regimes}
\begin{tabular}{@{}ll rl rl@{}}
\toprule
 & & \multicolumn{2}{c}{Kodak high band 36.6--37.0\,dB} & \multicolumn{2}{c}{Tecnick low band 32.2--32.5\,dB} \\
\cmidrule(lr){3-4}\cmidrule(lr){5-6}
A & B & $\Delta$ & 95\% & $\Delta$ & 95\% \\
\midrule
HPCM & CMIC & +0.83 & \textbf{[+0.29, +1.55]} & $-$0.81 & \textbf{[$-$1.74, $-$0.25]} \\
HPCM & GLIC & $-$0.37 & [$-$0.89, +0.36] & +6.18 & \textbf{[+4.87, +7.34]} \\
HPCM & SAAF & +1.32 & \textbf{[+0.86, +2.01]} & +4.12 & \textbf{[+1.80, +5.58]} \\
HPCM & DCAE & +2.70 & \textbf{[+2.23, +3.41]} & +1.70 & \textbf{[+0.73, +2.57]} \\
HPCM & LALIC & +3.92 & \textbf{[+2.99, +5.07]} & +9.60 & \textbf{[+8.33, +10.56]} \\
CMIC & GLIC & $-$1.18 & \textbf{[$-$1.48, $-$0.91]} & +7.04 & \textbf{[+5.46, +8.90]} \\
CMIC & SAAF & +0.49 & \textbf{[+0.14, +0.88]} & +4.96 & \textbf{[+3.30, +6.18]} \\
CMIC & DCAE & +1.85 & \textbf{[+1.32, +2.55]} & +2.53 & \textbf{[+1.79, +3.38]} \\
CMIC & LALIC & +3.07 & \textbf{[+2.42, +3.81]} & +10.49 & \textbf{[+8.99, +11.83]} \\
GLIC & SAAF & +1.70 & \textbf{[+1.25, +2.21]} & $-$1.94 & [$-$4.79, +0.27] \\
GLIC & DCAE & +3.08 & \textbf{[+2.49, +3.81]} & $-$4.22 & \textbf{[$-$5.76, $-$2.71]} \\
GLIC & LALIC & +4.30 & \textbf{[+3.42, +5.29]} & +3.22 & \textbf{[+2.23, +3.97]} \\
SAAF & DCAE & +1.36 & \textbf{[+0.99, +1.85]} & $-$2.32 & \textbf{[$-$3.13, $-$0.95]} \\
SAAF & LALIC & +2.57 & \textbf{[+1.93, +3.26]} & +5.27 & \textbf{[+2.98, +8.03]} \\
DCAE & LALIC & +1.19 & \textbf{[+0.33, +1.90]} & +7.77 & \textbf{[+6.06, +9.32]} \\
\bottomrule
\end{tabular}

\end{table}

\paragraph{A method without released numbers.}
MLICv2~\cite{jiang2025mlicv2} reports BD-rates against VTM-17.0 for MLICv2, for MLICv2$^+$ with test-time latent refinement, and for MLIC++, so the authors' own comparison gives MLICv2 against MLIC++, which we bridge to the canonical scale through MLIC++'s released curve (\cref{tab:s_r19_mlicv2}).
This shows what the numeric-release filter can hide, assuming that the authors' and our MLIC++ measurements agree; it is not a score.

\begin{table}[htbp]\centering\small
\caption{MLICv2 bridged to the canonical scale through its authors' MLIC++ claims (\%). Negative: less rate than HPCM (released curves).}\label{tab:s_r19_mlicv2}
\begin{tabular}{@{}l rrr@{}}
\toprule
 & Kodak & Tecnick & CLIC~Pro \\
\midrule
MLIC++ claim vs.\ VTM-17.0 (MLICv2 paper) & $-$13.39 & $-$17.59 & $-$13.08 \\
MLIC++ canonical score (released curve) & $-$13.96 & $-$16.41 & $-$13.74 \\
HPCM canonical score (released curve) & $-$18.95 & $-$21.38 & $-$19.26 \\
\midrule
MLICv2: claim vs.\ VTM-17.0 & $-$16.54 & $-$21.61 & $-$16.05 \\
\quad direct vs.\ MLIC++ from the claims & $-$3.64 & $-$4.88 & $-$3.42 \\
\quad bridged direct vs.\ HPCM & +2.29 & +1.14 & +3.18 \\
MLICv2$^+$ (test-time latent refinement): claim vs.\ VTM-17.0 & $-$20.46 & $-$24.35 & $-$19.14 \\
\quad direct vs.\ MLIC++ from the claims & $-$8.16 & $-$8.20 & $-$6.97 \\
\quad bridged direct vs.\ HPCM & $-$2.52 & $-$2.40 & $-$0.62 \\
\bottomrule
\end{tabular}

\end{table}

\subsection{Confirmation runs: MLIC++, TCM-L, DIV2K, MS-SSIM, and timing}
\label{sec:s_cf}
Batch~4 tests whether the conclusions hold for a widely used older baseline, on a test set used for nothing else, under a second quality measure, and for timing on every test set.
Eight of its ten jobs completed all \cfRuns\ runs. The FTIC and LALIC jobs stopped with a runtime error of our cluster environment before writing any record; FTIC therefore remains unmeasured, and LALIC enters with its batch-3 records only, so it is absent from DIV2K and MS-SSIM.
Re-encoding six earlier releases on the same images reproduced \cfIdentStream\ of \cfIdentCompared\ streams exactly; the others, CMIC\_AuxT records first measured on the smaller slice of batch~1, differ by at most \cfIdentMaxRel\% in bytes and \cfIdentMaxDb\,dB.

\paragraph{MLIC++.}
We ran the six public MSE weights that the MLIC++ README lists for its 2023 model, with the older modules it prescribes for them and its own padding.
No tested rule reproduces the released curves: with payload bytes and 8-bit PSNR, rates agree within \cfMlicRateErrMax\%, but PSNR is higher at every point, by up to \cfMlicDbMax\,dB, so the public weights need \cfMlicBetterMin--\cfMlicBetterMax\% less rate than the released curves state.
Against HPCM-Large, MLIC++ needs \cfMlicOwnK\%, \cfMlicOwnT\%, and \cfMlicOwnC\% more rate on Kodak, Tecnick, and CLIC~Pro, all resolved.
Released curves put CMIC\_AuxT, GLIC, SAAF, and DCAE \cfMlicRelLeadMin--\cfMlicRelLeadMax\% ahead of MLIC++, and matched measurement reduces each lead by \cfMlicShiftMin--\cfMlicShiftMax\ points; LALIC, \cfLalicRelMin--\cfLalicRelMax\% ahead on released curves, is level with MLIC++ on all three test sets (\cref{tab:s_cf_mlic_pairs}).

\begin{table}[htbp]
\centering\small
\caption{Direct BD-rate (\%) of MLIC++ against every other measured release on the frozen windows; positive: MLIC++ needs more rate. Released: released curves of both methods. Matched: payload bytes, 8-bit PSNR, native padding. Bold: the paired 95\% interval excludes zero (image bootstrap, or the delete-one jackknife, $^{\rm J}$). $^\dagger$LALIC records from batch~3 (same GPU type).}
\label{tab:s_cf_mlic_pairs}
\begin{tabular}{@{}l rr rr rr@{}}
\toprule
 & \multicolumn{2}{c}{Kodak} & \multicolumn{2}{c}{Tecnick} & \multicolumn{2}{c}{CLIC~Pro} \\
\cmidrule(lr){2-3}\cmidrule(lr){4-5}\cmidrule(l){6-7}
MLIC++ against & released & matched & released & matched & released & matched \\
\midrule
HPCM-Large '25 & +6.15 & \textbf{+4.85} & +6.33 & \textbf{+4.39}$^{\rm J}$ & +6.83 & \textbf{+4.96}$^{\rm J}$ \\
CMIC\_AuxT '26 & +3.82 & \textbf{+2.11} & +8.86 & \textbf{+5.75}$^{\rm J}$ & +6.35 & \textbf{+3.34}$^{\rm J}$ \\
GLIC '26 & +5.69 & \textbf{+2.50} & +7.40 & \textbf{+2.92}$^{\rm J}$ & +6.38 & +2.10$^{\rm J}$ \\
SAAF '26 & +2.83 & +0.91 & +5.31 & \textbf{+2.67}$^{\rm J}$ & +3.74 & +1.15$^{\rm J}$ \\
DCAE '25 & +2.46 & +0.38 & +4.56 & \textbf{+1.94}$^{\rm J}$ & +3.54 & +0.91$^{\rm J}$ \\
LALIC '25$^\dagger$ & +0.71 & $-$0.82 & +2.18 & $-$0.20$^{\rm J}$ & +2.32 & +0.02$^{\rm J}$ \\
TCM-S '23 & $-$8.49 & \textbf{$-$9.87}$^{\rm J}$ & $-$9.99 & \textbf{$-$11.78}$^{\rm J}$ & $-$7.93 & \textbf{$-$9.15}$^{\rm J}$ \\
\bottomrule
\end{tabular}

\end{table}

\paragraph{TCM-L.}
The single public TCM-L weight ($\lambda=0.05$) reproduces the top point of the released curves within \cfTcmlRateErrMax\% in rate and \cfTcmlDbMax\,dB, and needs \cfTcmlSaveMin--\cfTcmlSaveMax\% less rate than TCM-S at the same $\lambda$ with \cfTcmlGainMin--\cfTcmlGainMax\,dB higher PSNR. One point defines no curve, so TCM-L enters no BD-rate comparison.

\paragraph{DIV2K.}
The 100 validation images of DIV2K~\cite{agustsson2017div2k} (2.8 megapixels on average) enter no other analysis of ours. The DIV2K training images are a common training set for LIC, and we do not know whether every release excluded the validation images from training or model selection, so DIV2K is an additional test set rather than a guaranteed unseen one.
Their window is fixed by the rule of the resampling analyses before any comparison: the common range of the seven measured curves, each inset by one operating point and rounded inward to 0.1\,dB (\cfDivLo--\cfDivHi\,dB).
HPCM-Large ranks first in \cfDivFirst\% of the bootstrap replicates, and \cfDivPairsRes\ of the \cfDivPairs\ pairs are resolved (\cref{tab:s_cf_div2k}).

\begin{table}[htbp]
\centering\small
\caption{DIV2K (100 images): direct BD-rate (\%) against HPCM-Large on the inset window with its paired 95\% interval, the same comparison on the full intersection of the mean curves (sensitivity), and the share of bootstrap replicates in which each release ranks first. Bold: excludes zero; $^{\rm J}$: jackknife; $^{\rm n}$: no decision.}
\label{tab:s_cf_div2k}
\begin{tabular}{@{}l r l r r@{}}
\toprule
Against HPCM-Large & inset window & 95\% interval & full intersection & ranked first (\%) \\
\midrule
HPCM-Large '25 & \multicolumn{3}{c}{reference} & 95.2 \\
CMIC\_AuxT '26 & \textbf{+1.00} & 0.28 to 1.64 & \textbf{+0.78}$^{\rm J}$ & 0.0 \\
GLIC '26 & +0.97 & $-$0.17 to 2.19 & +0.85$^{\rm J}$ & 4.8 \\
SAAF '26 & \textbf{+5.39} & 4.20 to 6.76 & \textbf{+5.02}$^{\rm J}$ & 0.0 \\
DCAE '25 & \textbf{+3.37} & 2.56 to 4.22 & \textbf{+3.01}$^{\rm J}$ & 0.0 \\
MLIC++ '23 & \textbf{+5.46} & 4.57 to 6.42 & +5.65$^{\rm n}$ & 0.0 \\
TCM-S '23 & \textbf{+15.38} & 12.94 to 18.24 & \textbf{+14.93}$^{\rm J}$ & 0.0 \\
\bottomrule
\end{tabular}

\end{table}

\paragraph{MS-SSIM.}
MS-SSIM~\cite{wang2003msssim} is computed on the same decoded images of the MSE-optimized checkpoints, as in CompressAI's evaluation, with windows set by the same inset rule; no heterogeneity scale is calibrated for it, so only paired intervals are reported.
Against HPCM-Large, \cfMsSame\ of \cfMsTot\ comparisons keep their PSNR sign, but the gaps of CMIC\_AuxT, GLIC, SAAF, and DCAE shrink from a median of \cfMsMedPsnr\ to \cfMsMedMs\ points; on CLIC~Pro and DIV2K, HPCM-Large, CMIC\_AuxT, and DCAE are pairwise unresolved (\cref{tab:s_cf_msssim}).

\begin{table}[htbp]
\centering\small
\caption{Matched comparisons against HPCM-Large with PSNR and with MS-SSIM as the quality axis, on the same decoded images (payload bytes, native padding). Bold: the paired 95\% MS-SSIM interval excludes zero ($^{\rm J}$: jackknife). First rank: share of bootstrap replicates (MS-SSIM) for every release ranked first in at least 20\%.}
\label{tab:s_cf_msssim}
\begin{tabular}{@{}l rr rr rr rr@{}}
\toprule
 & \multicolumn{2}{c}{Kodak} & \multicolumn{2}{c}{Tecnick} & \multicolumn{2}{c}{CLIC~Pro} & \multicolumn{2}{c}{DIV2K} \\
\cmidrule(lr){2-3}\cmidrule(lr){4-5}\cmidrule(lr){6-7}\cmidrule(l){8-9}
Against HPCM-Large & PSNR & MS-SSIM & PSNR & MS-SSIM & PSNR & MS-SSIM & PSNR & MS-SSIM \\
\midrule
CMIC\_AuxT '26 & +2.68 & \textbf{+1.55} & $-$1.28 & \textbf{$-$1.64} & +1.57 & 0.00 & +1.00 & +0.03 \\
GLIC '26 & +2.29 & \textbf{+1.27} & +1.43 & \textbf{+1.72} & +2.81 & \textbf{+1.62} & +0.97 & +0.66 \\
SAAF '26 & +3.90 & \textbf{+1.75} & +1.67 & +0.74 & +3.77 & +0.83 & +5.39 & \textbf{+2.32} \\
DCAE '25 & +4.45 & \textbf{+1.55} & +2.40 & +0.12 & +4.02 & +0.55 & +3.37 & +0.09 \\
MLIC++ '23 & +4.85 & +1.38 & +4.39 & \textbf{+0.99} & +4.96 & +1.75 & +5.46 & \textbf{+2.00} \\
TCM-S '23 & +16.33 & \textbf{+11.04} & +18.34 & \textbf{+11.20} & +15.53 & \textbf{+8.54}$^{\rm J}$ & +15.38 & \textbf{+9.18} \\
\midrule
First rank (\%, MS-SSIM) & \multicolumn{2}{c}{\begin{tabular}[t]{@{}c@{}}HPCM 96\end{tabular}} & \multicolumn{2}{c}{\begin{tabular}[t]{@{}c@{}}CMIC 100\end{tabular}} & \multicolumn{2}{c}{\begin{tabular}[t]{@{}c@{}}HPCM 45\\CMIC 41\end{tabular}} & \multicolumn{2}{c}{\begin{tabular}[t]{@{}c@{}}HPCM 42\\DCAE 33\\CMIC 24\end{tabular}} \\
MS-SSIM window (dB) & \multicolumn{2}{c}{13.4--18.2} & \multicolumn{2}{c}{15.0--18.3} & \multicolumn{2}{c}{14.3--18.1} & \multicolumn{2}{c}{14.7--18.8} \\
\bottomrule
\end{tabular}

\end{table}

\paragraph{Timing.}
Decoding times on CLIC~Pro reproduce those of batches 2 and 3 within \cfDecAgreeMax\% (rank correlation \cfDecSpearman).
Entropy coding runs on the host CPU and the implementations differ in optimization, so these times are context, not an efficiency ranking (\cref{tab:s_cf_timing}).

\begin{table}[htbp]
\centering\small
\caption{Median encoding and decoding time (seconds per image) on one H100 MIG 3g.40gb slice, native padding; CUDA-synchronized wall-clock times around \texttt{compress} and \texttt{decompress}, after warm-up. $^\ast$Decoding seconds per megapixel on DIV2K.}
\label{tab:s_cf_timing}
\begin{tabular}{@{}l rr rr rr rr r@{}}
\toprule
 & \multicolumn{2}{c}{Kodak} & \multicolumn{2}{c}{Tecnick} & \multicolumn{2}{c}{CLIC~Pro} & \multicolumn{2}{c}{DIV2K} & \\
\cmidrule(lr){2-3}\cmidrule(lr){4-5}\cmidrule(lr){6-7}\cmidrule(lr){8-9}
Release & enc. & dec. & enc. & dec. & enc. & dec. & enc. & dec. & s/MP$^\ast$ \\
\midrule
HPCM-Large & 0.08 & 0.11 & 0.31 & 0.44 & 0.58 & 0.84 & 0.59 & 0.85 & 0.31 \\
CMIC\_AuxT & 0.21 & 0.18 & 0.52 & 0.50 & 0.91 & 0.96 & 0.93 & 0.98 & 0.35 \\
GLIC & 0.23 & 0.19 & 0.66 & 0.62 & 1.10 & 1.08 & 1.10 & 1.09 & 0.39 \\
SAAF & 0.14 & 0.16 & 0.38 & 0.43 & 0.62 & 0.71 & 0.63 & 0.72 & 0.26 \\
DCAE & 0.13 & 0.16 & 0.36 & 0.43 & 0.60 & 0.71 & 0.60 & 0.72 & 0.26 \\
MLIC++ & 0.17 & 0.22 & 0.35 & 0.44 & 0.61 & 0.74 & 0.62 & 0.76 & 0.27 \\
TCM-S & 0.13 & 0.14 & 0.29 & 0.32 & 0.46 & 0.52 & 0.49 & 0.54 & 0.19 \\
TCM-L & 0.16 & 0.19 & 0.41 & 0.49 & 0.68 & 0.83 & 0.68 & 0.84 & 0.30 \\
\bottomrule
\end{tabular}

\end{table}
\begin{table}[htbp]\centering\scriptsize\setlength{\tabcolsep}{3.5pt}
\caption{All \rnPairs\ pairs of the seven releases of batches 1--3 on the frozen windows (native padding, payload bytes, 8-bit PSNR). $\Delta$: direct BD-rate (\%) of B against A; positive: B needs more rate. Bootstrap: paired percentile interval ($B=10{,}000$) where at most 1\% of replicates fail coverage, otherwise the failure share. Jackknife: delete-one interval ($t_{n-1}$). $\sigma_{\rm pair}=\sqrt{n}\,\mathrm{SD}_{\rm boot}$.}\label{tab:s_r19_pairs}
\begin{tabular}{@{}lll r l r l r@{}}
\toprule Dataset & A & B & $\Delta$ & Bootstrap 95\% & fail (\%) & Jackknife 95\% & $\sigma_{\rm pair}$ \\ \midrule
Kodak & HPCM & CMIC & +2.68 & \textbf{[+1.93, +3.60]} & 0.08 & \textbf{[+1.75, +3.61]} & 2.08 \\
Kodak & HPCM & GLIC & +2.29 & \textbf{[+1.10, +4.06]} & 0.10 & \textbf{[+0.66, +3.93]} & 3.71 \\
Kodak & HPCM & SAAF & +3.90 & \textbf{[+3.06, +4.76]} & 0.11 & \textbf{[+2.85, +4.96]} & 2.15 \\
Kodak & HPCM & DCAE & +4.45 & \textbf{[+3.47, +5.44]} & 0.15 & \textbf{[+3.21, +5.70]} & 2.52 \\
Kodak & HPCM & LALIC & +5.71 & \textbf{[+4.48, +7.11]} & 0.13 & \textbf{[+4.19, +7.24]} & 3.30 \\
Kodak & HPCM & TCM-S & +16.33 & gate abstains & 12.63 & \textbf{[+11.78, +20.88]} & 9.26 \\
Kodak & CMIC & GLIC & $-$0.38 & [$-$0.90, +0.54] & 0.10 & [$-$1.15, +0.40] & 1.80 \\
Kodak & CMIC & SAAF & +1.19 & \textbf{[+0.65, +1.66]} & 0.11 & \textbf{[+0.67, +1.71]} & 1.23 \\
Kodak & CMIC & DCAE & +1.72 & \textbf{[+1.10, +2.37]} & 0.15 & \textbf{[+0.99, +2.45]} & 1.58 \\
Kodak & CMIC & LALIC & +2.95 & \textbf{[+2.28, +3.75]} & 0.13 & \textbf{[+2.11, +3.80]} & 1.85 \\
Kodak & CMIC & TCM-S & +13.29 & gate abstains & 12.63 & \textbf{[+9.53, +17.05]} & 7.93 \\
Kodak & GLIC & SAAF & +1.58 & \textbf{[+0.37, +2.21]} & 0.11 & \textbf{[+0.66, +2.50]} & 2.30 \\
Kodak & GLIC & DCAE & +2.11 & \textbf{[+0.93, +3.03]} & 0.15 & \textbf{[+1.01, +3.22]} & 2.60 \\
Kodak & GLIC & LALIC & +3.34 & \textbf{[+2.35, +4.32]} & 0.13 & \textbf{[+2.27, +4.42]} & 2.48 \\
Kodak & GLIC & TCM-S & +13.72 & gate abstains & 12.63 & \textbf{[+10.10, +17.35]} & 7.96 \\
Kodak & SAAF & DCAE & +0.53 & \textbf{[+0.16, +0.97]} & 0.15 & \textbf{[+0.07, +0.99]} & 1.01 \\
Kodak & SAAF & LALIC & +1.74 & \textbf{[+0.96, +2.75]} & 0.13 & \textbf{[+0.79, +2.69]} & 2.24 \\
Kodak & SAAF & TCM-S & +11.96 & gate abstains & 12.63 & \textbf{[+8.27, +15.65]} & 7.84 \\
Kodak & DCAE & LALIC & +1.21 & \textbf{[+0.58, +2.00]} & 0.15 & \textbf{[+0.45, +1.96]} & 1.78 \\
Kodak & DCAE & TCM-S & +11.37 & gate abstains & 12.63 & \textbf{[+8.06, +14.69]} & 7.01 \\
Kodak & LALIC & TCM-S & +10.04 & gate abstains & 12.63 & \textbf{[+7.17, +12.92]} & 6.05 \\
Tecnick & HPCM & CMIC & $-$1.28 & gate abstains & 3.56 & \textbf{[$-$1.60, $-$0.96]} & 1.63 \\
Tecnick & HPCM & GLIC & +1.43 & gate abstains & 3.56 & \textbf{[+0.42, +2.44]} & 4.87 \\
Tecnick & HPCM & SAAF & +1.67 & gate abstains & 3.56 & \textbf{[+0.70, +2.65]} & 4.64 \\
Tecnick & HPCM & DCAE & +2.40 & gate abstains & 3.56 & \textbf{[+1.75, +3.05]} & 3.10 \\
Tecnick & HPCM & LALIC & +4.61 & gate abstains & 3.56 & \textbf{[+3.50, +5.71]} & 5.23 \\
Tecnick & HPCM & TCM-S & +18.34 & gate abstains & 6.82 & \textbf{[+15.56, +21.11]} & 12.85 \\
Tecnick & CMIC & GLIC & +2.75 & \textbf{[+1.95, +3.81]} & 0.00 & \textbf{[+1.81, +3.68]} & 4.77 \\
Tecnick & CMIC & SAAF & +2.99 & \textbf{[+2.13, +3.92]} & 0.02 & \textbf{[+2.09, +3.90]} & 4.53 \\
Tecnick & CMIC & DCAE & +3.73 & \textbf{[+3.21, +4.35]} & 0.50 & \textbf{[+3.16, +4.30]} & 2.87 \\
Tecnick & CMIC & LALIC & +5.96 & \textbf{[+4.95, +7.12]} & 0.06 & \textbf{[+4.86, +7.07]} & 5.53 \\
Tecnick & CMIC & TCM-S & +19.87 & gate abstains & 3.58 & \textbf{[+17.05, +22.70]} & 13.45 \\
Tecnick & GLIC & SAAF & +0.24 & [$-$0.40, +0.71] & 0.02 & [$-$0.30, +0.78] & 2.78 \\
Tecnick & GLIC & DCAE & +0.96 & \textbf{[+0.29, +1.48]} & 0.50 & \textbf{[+0.36, +1.56]} & 3.03 \\
Tecnick & GLIC & LALIC & +3.13 & \textbf{[+2.60, +3.64]} & 0.06 & \textbf{[+2.60, +3.66]} & 2.68 \\
Tecnick & GLIC & TCM-S & +16.67 & gate abstains & 3.58 & \textbf{[+14.60, +18.74]} & 10.21 \\
Tecnick & SAAF & DCAE & +0.72 & \textbf{[+0.23, +1.25]} & 0.50 & \textbf{[+0.18, +1.25]} & 2.62 \\
Tecnick & SAAF & LALIC & +2.88 & \textbf{[+2.15, +3.78]} & 0.06 & \textbf{[+2.07, +3.70]} & 4.16 \\
Tecnick & SAAF & TCM-S & +16.39 & gate abstains & 3.58 & \textbf{[+14.02, +18.75]} & 11.61 \\
Tecnick & DCAE & LALIC & +2.15 & \textbf{[+1.47, +2.93]} & 0.50 & \textbf{[+1.40, +2.90]} & 3.73 \\
Tecnick & DCAE & TCM-S & +15.56 & gate abstains & 3.76 & \textbf{[+13.26, +17.86]} & 11.06 \\
Tecnick & LALIC & TCM-S & +13.13 & gate abstains & 3.58 & \textbf{[+11.51, +14.74]} & 7.87 \\
CLIC~Pro & HPCM & CMIC & +1.57 & \textbf{[+0.89, +2.20]} & 0.34 & \textbf{[+0.85, +2.29]} & 2.16 \\
CLIC~Pro & HPCM & GLIC & +2.81 & \textbf{[+1.16, +4.85]} & 0.34 & \textbf{[+0.83, +4.78]} & 6.05 \\
CLIC~Pro & HPCM & SAAF & +3.77 & \textbf{[+2.29, +5.25]} & 0.30 & \textbf{[+2.15, +5.39]} & 4.85 \\
CLIC~Pro & HPCM & DCAE & +4.02 & \textbf{[+3.11, +4.90]} & 0.53 & \textbf{[+3.05, +4.99]} & 2.95 \\
CLIC~Pro & HPCM & LALIC & +4.94 & \textbf{[+3.49, +6.67]} & 0.40 & \textbf{[+3.16, +6.73]} & 5.31 \\
CLIC~Pro & HPCM & TCM-S & +15.53 & gate abstains & 12.97 & \textbf{[+12.10, +18.95]} & 9.79 \\
CLIC~Pro & CMIC & GLIC & +1.22 & \textbf{[+0.13, +2.90]} & 0.34 & [$-$0.20, +2.64] & 4.51 \\
CLIC~Pro & CMIC & SAAF & +2.17 & \textbf{[+1.12, +3.34]} & 0.33 & \textbf{[+0.98, +3.35]} & 3.64 \\
CLIC~Pro & CMIC & DCAE & +2.41 & \textbf{[+1.70, +3.09]} & 0.52 & \textbf{[+1.68, +3.14]} & 2.26 \\
CLIC~Pro & CMIC & LALIC & +3.32 & \textbf{[+2.33, +4.73]} & 0.39 & \textbf{[+2.03, +4.61]} & 3.96 \\
CLIC~Pro & CMIC & TCM-S & +13.74 & gate abstains & 12.96 & \textbf{[+10.78, +16.70]} & 8.59 \\
CLIC~Pro & GLIC & SAAF & +0.94 & [$-$0.54, +1.85] & 0.33 & [$-$0.27, +2.15] & 3.85 \\
CLIC~Pro & GLIC & DCAE & +1.18 & [$-$0.65, +2.51] & 0.52 & [$-$0.51, +2.88] & 5.21 \\
CLIC~Pro & GLIC & LALIC & +2.08 & \textbf{[+1.00, +3.05]} & 0.39 & \textbf{[+1.00, +3.16]} & 3.34 \\
CLIC~Pro & GLIC & TCM-S & +12.37 & gate abstains & 12.96 & \textbf{[+9.72, +15.03]} & 8.13 \\
CLIC~Pro & SAAF & DCAE & +0.24 & [$-$0.82, +1.34] & 0.52 & [$-$0.95, +1.43] & 3.55 \\
CLIC~Pro & SAAF & LALIC & +1.13 & \textbf{[+0.27, +2.45]} & 0.39 & \textbf{[+0.05, +2.20]} & 3.51 \\
CLIC~Pro & SAAF & TCM-S & +11.33 & gate abstains & 12.96 & \textbf{[+8.52, +14.13]} & 8.48 \\
CLIC~Pro & DCAE & LALIC & +0.89 & [$-$0.21, +2.42] & 0.52 & [$-$0.54, +2.31] & 4.38 \\
CLIC~Pro & DCAE & TCM-S & +11.06 & gate abstains & 12.96 & \textbf{[+8.36, +13.76]} & 7.68 \\
CLIC~Pro & LALIC & TCM-S & +10.09 & gate abstains & 12.96 & \textbf{[+7.96, +12.21]} & 6.50 \\
\bottomrule
\end{tabular}
\end{table}

\begin{table}[htbp]\centering\scriptsize\setlength{\tabcolsep}{3pt}
\caption{Simultaneous 95\% intervals over the \rnJointPairs\ pairs of the \rnJointMembers\ 2025--26 frontier releases on each frozen window. Bootstrap: max-$|t|$ over the shared replicates that span $W_d$ (critical values 2.79, 2.80, 2.85; replicates with any coverage failure: 0.15\%, 3.56\%, 0.53\%). Jackknife: max-$|T|$ for a multivariate $t_{n-1}$ with the jackknife correlation of the 15 contrasts (critical values 2.94, 2.79, 2.89). Holm: jackknife $t$ tests, Holm-adjusted over the 15 pairs. Bold: excludes zero.}\label{tab:s_r19_joint}
\begin{tabular}{@{}lll r ll r@{}}
\toprule Dataset & A & B & $\Delta$ & Bootstrap simultaneous & Jackknife simultaneous & Holm $p$ \\ \midrule
Kodak & HPCM & CMIC & +2.68 & \textbf{[+1.51, +3.86]} & \textbf{[+1.37, +4.01]} & $<$0.001 \\
Kodak & HPCM & GLIC & +2.29 & \textbf{[+0.22, +4.41]} & [$-$0.01, +4.65] & 0.022 \\
Kodak & HPCM & SAAF & +3.90 & \textbf{[+2.69, +5.14]} & \textbf{[+2.42, +5.41]} & $<$0.001 \\
Kodak & HPCM & DCAE & +4.45 & \textbf{[+3.03, +5.90]} & \textbf{[+2.70, +6.24]} & $<$0.001 \\
Kodak & HPCM & LALIC & +5.71 & \textbf{[+3.85, +7.60]} & \textbf{[+3.57, +7.90]} & $<$0.001 \\
Kodak & CMIC & GLIC & $-$0.38 & [$-$1.39, +0.64] & [$-$1.47, +0.73] & 0.322 \\
Kodak & CMIC & SAAF & +1.19 & \textbf{[+0.49, +1.89]} & \textbf{[+0.45, +1.93]} & $<$0.001 \\
Kodak & CMIC & DCAE & +1.72 & \textbf{[+0.83, +2.63]} & \textbf{[+0.69, +2.77]} & $<$0.001 \\
Kodak & CMIC & LALIC & +2.95 & \textbf{[+1.91, +4.01]} & \textbf{[+1.75, +4.16]} & $<$0.001 \\
Kodak & GLIC & SAAF & +1.58 & \textbf{[+0.28, +2.89]} & \textbf{[+0.28, +2.89]} & 0.008 \\
Kodak & GLIC & DCAE & +2.11 & \textbf{[+0.64, +3.60]} & \textbf{[+0.55, +3.70]} & 0.004 \\
Kodak & GLIC & LALIC & +3.34 & \textbf{[+1.95, +4.76]} & \textbf{[+1.83, +4.89]} & $<$0.001 \\
Kodak & SAAF & DCAE & +0.53 & [$-$0.05, +1.10] & [$-$0.12, +1.18] & 0.051 \\
Kodak & SAAF & LALIC & +1.74 & \textbf{[+0.48, +3.02]} & \textbf{[+0.40, +3.10]} & 0.005 \\
Kodak & DCAE & LALIC & +1.21 & \textbf{[+0.20, +2.22]} & \textbf{[+0.14, +2.29]} & 0.012 \\
Tecnick & HPCM & CMIC & $-$1.28 & \textbf{[$-$1.74, $-$0.82]} & \textbf{[$-$1.73, $-$0.83]} & $<$0.001 \\
Tecnick & HPCM & GLIC & +1.43 & \textbf{[+0.08, +2.80]} & \textbf{[+0.03, +2.85]} & 0.016 \\
Tecnick & HPCM & SAAF & +1.67 & \textbf{[+0.38, +2.98]} & \textbf{[+0.32, +3.05]} & 0.004 \\
Tecnick & HPCM & DCAE & +2.40 & \textbf{[+1.54, +3.27]} & \textbf{[+1.50, +3.31]} & $<$0.001 \\
Tecnick & HPCM & LALIC & +4.61 & \textbf{[+3.15, +6.08]} & \textbf{[+3.07, +6.17]} & $<$0.001 \\
Tecnick & CMIC & GLIC & +2.75 & \textbf{[+1.46, +4.05]} & \textbf{[+1.44, +4.07]} & $<$0.001 \\
Tecnick & CMIC & SAAF & +2.99 & \textbf{[+1.78, +4.22]} & \textbf{[+1.73, +4.27]} & $<$0.001 \\
Tecnick & CMIC & DCAE & +3.73 & \textbf{[+2.94, +4.52]} & \textbf{[+2.93, +4.54]} & $<$0.001 \\
Tecnick & CMIC & LALIC & +5.96 & \textbf{[+4.48, +7.46]} & \textbf{[+4.42, +7.53]} & $<$0.001 \\
Tecnick & GLIC & SAAF & +0.24 & [$-$0.54, +1.03] & [$-$0.51, +1.00] & 0.377 \\
Tecnick & GLIC & DCAE & +0.96 & \textbf{[+0.12, +1.80]} & \textbf{[+0.12, +1.80]} & 0.008 \\
Tecnick & GLIC & LALIC & +3.13 & \textbf{[+2.39, +3.88]} & \textbf{[+2.39, +3.88]} & $<$0.001 \\
Tecnick & SAAF & DCAE & +0.72 & [0.00, +1.44] & [$-$0.03, +1.46] & 0.017 \\
Tecnick & SAAF & LALIC & +2.88 & \textbf{[+1.72, +4.06]} & \textbf{[+1.74, +4.04]} & $<$0.001 \\
Tecnick & DCAE & LALIC & +2.15 & \textbf{[+1.13, +3.19]} & \textbf{[+1.11, +3.21]} & $<$0.001 \\
CLIC~Pro & HPCM & CMIC & +1.57 & \textbf{[+0.61, +2.54]} & \textbf{[+0.55, +2.60]} & $<$0.001 \\
CLIC~Pro & HPCM & GLIC & +2.81 & \textbf{[+0.17, +5.51]} & \textbf{[+0.02, +5.66]} & 0.040 \\
CLIC~Pro & HPCM & SAAF & +3.77 & \textbf{[+1.64, +5.95]} & \textbf{[+1.47, +6.12]} & $<$0.001 \\
CLIC~Pro & HPCM & DCAE & +4.02 & \textbf{[+2.71, +5.34]} & \textbf{[+2.64, +5.42]} & $<$0.001 \\
CLIC~Pro & HPCM & LALIC & +4.94 & \textbf{[+2.62, +7.32]} & \textbf{[+2.42, +7.52]} & $<$0.001 \\
CLIC~Pro & CMIC & GLIC & +1.22 & [$-$0.75, +3.22] & [$-$0.80, +3.27] & 0.447 \\
CLIC~Pro & CMIC & SAAF & +2.17 & \textbf{[+0.56, +3.80]} & \textbf{[+0.48, +3.88]} & 0.005 \\
CLIC~Pro & CMIC & DCAE & +2.41 & \textbf{[+1.41, +3.42]} & \textbf{[+1.38, +3.46]} & $<$0.001 \\
CLIC~Pro & CMIC & LALIC & +3.32 & \textbf{[+1.58, +5.09]} & \textbf{[+1.50, +5.18]} & $<$0.001 \\
CLIC~Pro & GLIC & SAAF & +0.94 & [$-$0.75, +2.66] & [$-$0.78, +2.69] & 0.492 \\
CLIC~Pro & GLIC & DCAE & +1.18 & [$-$1.12, +3.53] & [$-$1.22, +3.64] & 0.493 \\
CLIC~Pro & GLIC & LALIC & +2.08 & \textbf{[+0.60, +3.58]} & \textbf{[+0.54, +3.64]} & 0.003 \\
CLIC~Pro & SAAF & DCAE & +0.24 & [$-$1.33, +1.83] & [$-$1.45, +1.96] & 0.685 \\
CLIC~Pro & SAAF & LALIC & +1.13 & [$-$0.41, +2.69] & [$-$0.40, +2.68] & 0.235 \\
CLIC~Pro & DCAE & LALIC & +0.89 & [$-$1.04, +2.85] & [$-$1.13, +2.95] & 0.493 \\
\bottomrule
\end{tabular}
\end{table}

\section{Using LIC-Audit for a new method}
\label{sec:s_r20_guide}
LIC-Audit does not propose a new quality measure.
It fixes three aspects of the measure the field already uses, BD-rate on PSNR: a \emph{comparison rule} (direct BD-rate on one frozen window per dataset), a \emph{measurement rule} (written bits over the original image area, 8-bit RGB PSNR on the cropped reconstruction, native padding, per-image means, all stated), and an \emph{evidence rule} (paired intervals over images when per-image records exist, otherwise conditional margins at stated scales).
The output remains a BD-rate, reported with its benchmark version, measurement class, and evidence status, beside the conventional score against a paper's own anchor.
\Cref{tab:s_r20_ladder} lists what each additional release enables; the steps are voluntary and cumulative, and the first needs only a numeric curve, which \nReposUsed\ of the \nReposCloned\ repositories we cloned already provide.

\begin{table}[htbp]\centering\small
\caption{What a new method releases, what the released tools return, and which claim that supports. Each step includes the previous ones.}\label{tab:s_r20_ladder}
\begin{tabular}{@{}lP{0.25\linewidth}P{0.36\linewidth}P{0.27\linewidth}@{}}
\toprule
Step & Authors release & LIC-Audit returns & Supported claim \\
\midrule
0 & A printed BD-rate only & No score; the claim is recorded with its quote and stated anchor & A gain against that anchor, under its own window and conventions \\
1 & A numeric curve that spans $W_d$, at a pinned public commit & Canonical score, position, direct comparison with the best entry, protocol range, status, and $\sigma^\star$ (\texttt{score}) & An ordering on the frozen window, conditional on the assumed scale \\
2 & A stated measurement profile & Measurement class \emph{documented}; estimated rates or another averaging rule receive a diagnostic score only & Comparison with curves on the same measurement basis \\
3 & Per-image records at every operating point & Paired bootstrap on $W_d$ after the consistency gate (\texttt{compare}); the jackknife of the analysis code where the bootstrap abstains & An observed decision on the test images, without an assumed scale \\
4 & Checkpoint and code & Re-measurement under the common definitions, padding controls, and timing & A measurement-level comparison of the released models \\
\bottomrule
\end{tabular}
\end{table}

\paragraph{Workflow.}
A submission is a JSON file with the method, its curves, the stated conventions, and the provenance of every curve; per-image records are optional.
The scores in this paper are computed with the Python package \texttt{licledger}, named after the source-linked record (the ledger) that it scores; LIC-Audit names the benchmark and protocol.
The public reference tool in the code repository uses the same engine files, whose SHA-256 the benchmark manifest records; before any score is computed it verifies these hashes, and a curve joins the leaderboard only if its numbers occur at the stated file and commit of a public repository.
\par\medskip\noindent\begin{minipage}{\linewidth}
\begin{verbatim}
python tools/check_submission.py --verify-engine                 # engine files unchanged
python tools/check_submission.py method.json --markdown card.md  # validate, score, report
python tools/check_submission.py method.json --online           # look up the numbers
\end{verbatim}
\end{minipage}\par\medskip
On two example submissions, the per-image comparison of \texttt{licledger} resolves \rgCmpTest\ against \rgCmpBase\ from their per-image records at \rgCmpEst\% (95\%: \rgCmpLo\ to \rgCmpHi).
In a red-team test with 17 declared cases and 350 seeded source mutations, no unverified curve was ranked.
A simpler, self-contained version of the tool runs on the LIC-Audit website, linked from the code repository (\licCodeLink): it produces this report in the browser from a method's curves or per-image records, offers a prompt with which an AI assistant can extract the curves and their sources from papers and repositories, and ranks a curve only after its numbers have been found at the stated source, as above.

\paragraph{How large a gain can a test set resolve?}
The conditional margin is zero whenever $|\Delta_{\rm least}|\le1.96\,\sigma/\sqrt{n_d}$, and the same rule resolves a true difference of $2.80\,\sigma/\sqrt{n_d}$ with 80\% probability (\cref{tab:s_r20_limits}).
\Cref{tab:s_r20_samples} inverts this into the number of test images: a one-point gain needs \rgImgOne\ images for 80\% probability at $\hat\sigma_{90}$ and \rgImgOneRef\ at $\sigma_{\rm ref}$, against 24 in Kodak.
These numbers transport a scale measured on six recent releases; a new test set or a very different pair can be more heterogeneous, which only per-image records reveal.

\paragraph{What per-image records add.}
\Cref{tab:s_r20_records} applies the limit to the \rgNcmp\ frozen-window comparisons among the six 2025--26 releases, each with its own paired 95\% interval (the percentile bootstrap, or the delete-one jackknife where the bootstrap abstains).
At $\hat\sigma_{90}$, \rgNbelow\ differences lie below the limit, so aggregate curves alone would leave them unresolved; their individual paired intervals exclude zero for \rgNbelowRes\ of them.
These are nominal 95\% intervals: under the simultaneous intervals over the \rnJointPairs\ pairs of each dataset (\cref{tab:s_r19_joint}; bootstrap max-$|t|$ on Kodak and CLIC~Pro, the jackknife on Tecnick), \simSim\ of the \rgNbelowRes\ still exclude zero, and \simHolm\ under Holm-adjusted jackknife tests.
For example, CMIC\_AuxT needs \rgExEst\% more rate than HPCM-Large on CLIC~Pro, below the \rgLimC\% limit, yet its paired interval is \rgExLo\ to \rgExHi.
Every difference above the limit at $\hat\sigma_{90}$ or $\sigma_{\rm ref}$ is also resolved by its paired interval.

\begin{table}[htbp]\centering\small
\caption{Smallest resolvable difference ($1.96\,\sigma/\sqrt{n_d}$) and the difference resolved with 80\% probability ($2.80\,\sigma/\sqrt{n_d}$), in BD-rate points.}\label{tab:s_r20_limits}
\begin{tabular}{@{}lrrrrrrrrr@{}}
\toprule
 & & \multicolumn{2}{c}{$\sigma=\rgSigMed$ (median)} & \multicolumn{2}{c}{$\hat\sigma_{90}=\rgSigNinety$} & \multicolumn{2}{c}{$\sigma_{\rm ref}=\rgSigRef$} & \multicolumn{2}{c}{$\sigma_{\rm hi}=\rgSigHi$} \\
\cmidrule(lr){3-4}\cmidrule(lr){5-6}\cmidrule(lr){7-8}\cmidrule(l){9-10}
Dataset & $n_d$ & limit & 80\% & limit & 80\% & limit & 80\% & limit & 80\% \\
\midrule
Kodak & 24 & 1.32 & 1.89 & 2.15 & 3.07 & 2.54 & 3.63 & 4.36 & 6.24 \\
Tecnick & 100 & 0.65 & 0.93 & 1.05 & 1.50 & 1.24 & 1.78 & 2.14 & 3.06 \\
CLIC~Pro & 41 & 1.01 & 1.45 & 1.64 & 2.35 & 1.94 & 2.78 & 3.34 & 4.77 \\
\bottomrule
\end{tabular}

\end{table}
\begin{table}[htbp]\centering\small
\caption{Number of test images for which a target gain is resolvable or resolved with 80\% probability, under the working model with the stated scale.}\label{tab:s_r20_samples}
\begin{tabular}{@{}lrrrr@{}}
\toprule
 & \multicolumn{2}{c}{$\hat\sigma_{90}=\rgSigNinety$} & \multicolumn{2}{c}{$\sigma_{\rm ref}=\rgSigRef$} \\
\cmidrule(lr){2-3}\cmidrule(l){4-5}
Target gain & resolvable & 80\% & resolvable & 80\% \\
\midrule
0.5\% & 444 & 906 & 619 & 1{,}265 \\
1.0\% & 111 & 227 & 155 & 317 \\
2.0\% & 28 & 57 & 39 & 80 \\
3.0\% & 13 & 26 & 18 & 36 \\
\bottomrule
\end{tabular}

\end{table}
\begin{table}[htbp]\centering\small
\caption{The \rgNcmp\ frozen-window comparisons among the six 2025--26 releases. Resolved: the paired 95\% interval excludes zero. For each scale, ``below'' counts the differences at or below its limit that paired intervals resolve, and ``above'' the differences above its limit that they leave unresolved.}\label{tab:s_r20_records}
\begin{tabular}{@{}lrrrrrrrr@{}}
\toprule
 & & & \multicolumn{2}{c}{$\sigma=\rgSigMed$ (median)} & \multicolumn{2}{c}{$\hat\sigma_{90}=\rgSigNinety$} & \multicolumn{2}{c}{$\sigma_{\rm ref}=\rgSigRef$} \\
\cmidrule(lr){4-5}\cmidrule(lr){6-7}\cmidrule(l){8-9}
Dataset & comparisons & resolved & below & above & below & above & below & above \\
\midrule
Kodak & 15 & 14 & 3 of 4 & 0 of 11 & 7 of 8 & 0 of 7 & 8 of 9 & 0 of 6 \\
Tecnick & 15 & 14 & 0 of 1 & 0 of 14 & 2 of 3 & 0 of 12 & 2 of 3 & 0 of 12 \\
CLIC~Pro & 15 & 11 & 0 of 3 & 1 of 12 & 3 of 7 & 0 of 8 & 3 of 7 & 0 of 8 \\
\midrule
All & 45 & 39 & 3 of 8 & 1 of 37 & 12 of 18 & 0 of 27 & 13 of 19 & 0 of 26 \\
\bottomrule
\end{tabular}

\end{table}

\paragraph{Stating a claim.}
Report the frozen-window value beside the conventional score, with the anchor's version and configuration.
A claim to improve on every earlier method or every dataset holds only if each comparison holds at the 5\% level~\cite{berger1982iut}; a full ordering needs simultaneous intervals.
A difference whose interval includes zero, or whose conditional margin is zero, is reported as unresolved; it may still matter for complexity or speed.
A new window, for example at lower rates, defines a new benchmark version; earlier versions stay archived and verifiable with their own hashes.

\subsection{Reporting checklist}
\label{sec:s_card}
\begin{description}[leftmargin=2.4em,labelwidth=1.8em,labelsep=0.6em,itemsep=1pt,parsep=0pt,topsep=3pt,font=\normalfont]
\item[R1] Per-image results at every operating point: bits from the actual bitstream (all side information included), PSNR of the 8-bit RGB reconstruction, image size. The anchor curve actually used, with version, configuration file, chroma format, bit depth, and color conversion.
\item[R2] The canonical score on the frozen windows of a named benchmark version (Kodak \winK, Tecnick \winT, CLIC~Pro \winC\,dB) next to the conventional pairwise-overlap value; the interpolant (PCHIP or Akima).
\item[R3] A paired bootstrap interval over images; differences whose interval contains zero are reported as unresolved. Without per-image results, explicitly conditional sensitivity to assumed image variability and the break-even heterogeneity $\sigma^\star$.
\item[R4] Results on several test sets (Tecnick, CLIC~Pro, or larger, alongside Kodak); low/mid/high-rate BD-rates.
\item[R5] PSNR color space; per-image vs.\ dataset averaging of PSNR and bpp; estimated vs.\ measured rates; rounding, native padding, and cropping. Shared padding should be identified as a diagnostic control.
\item[R6] For every baseline curve: source (paper, repository and commit, or own run), checkpoint, and tier. Report model complexity with its measurement conditions.
\end{description}

\subsection{Reproducibility}
\label{sec:s_repro}
The public code repository (\licCodeLink) contains the benchmark data (every released curve with its source, the published claims, and the per-image records of the matched measurements), the reference scoring tool, the LIC-Audit website, and tests that compare the scores of the website with those of the reference tool.
Values of curves whose source repositories carry no standard license are not copied; the repository gives their source file and commit instead.
Independent scripts recounted the key results, including the conditional margins, the Holm decisions, and the matched comparisons.

\end{document}